\documentclass[journal=mamobx,manuscript=article]{achemso}
\usepackage{achemso}
\setkeys{acs}{usetitle = true}
\usepackage[version=3]{mhchem}
\usepackage[T1]{fontenc}
\usepackage{type1cm}
\usepackage{graphicx}
\usepackage{dcolumn}

\usepackage{bm}
\usepackage{units}
\usepackage{color}
\usepackage{xcolor}
\usepackage{latexsym}
\usepackage{subfig}
\usepackage{amsmath,amssymb}
\SectionNumbersOn

\title{Computational Investigation of Microgels: \\ Synthesis and Effect of the Microstructure on the Deswelling Behavior}

\author{Angel J. Moreno}
\affiliation{Centro de F\'isica de Materiales (CSIC, UPV/EHU) and Materials Physics Center MPC, Paseo Manuel de Lardizabal 5, 20018 San Sebasti\'an, Spain}
\alsoaffiliation{Donostia International Physics Center, Paseo Manuel de Lardizabal 4, 20018 San Sebastian, Spain}
\email{angeljose.moreno@ehu.es}

\author{Federica Lo Verso}
\affiliation{Department of Physics, Chemistry and Pharmacy, University of Southern Denmark, Campusvej 55, Odense M 5230, Denmark}
\alsoaffiliation{Donostia International Physics Center, Paseo Manuel de Lardizabal 4, 20018 San Sebastian, Spain}

\begin{document}

\newpage

\begin{abstract}
%Microgels represent unique systems due to their tunability in terms of architecture, softness, permeability and deformability. Their conformational changes in solution can be e.g., achieved by changing temperature and pH. The importance of these molecules relies on their huge amount of possible applications such as drug delivery, photonic crystals or purification technologies, just to name a few.  
We present computer simulations of a realistic model of microgels.
Unlike the regular network frameworks usually assumed in the simulation literature, we model and simulate a realistic and efficient synthesis route, mimicking cross-linking of functionalized chains inside a cavity. This model is inspired, e.g., by microfluidic fabrication of microgels from macromolecular precursors and is different from standard polymerization routes. 
The assembly of the chains is mediated by a low fraction of interchain crosslinks. The microgels are polydisperse in size and shape but globally spherical objects.
In order to deeply understand the microgel structure and eventually improve the synthesis protocol we characterize their conformational properties and deswelling kinetics, and compare them with the results found for microgels obtained via underlying regular (diamond-like) structures. The specific microstructure of the microgel has no significant effect on the locus of the volume phase transition (VPT).
However, it strongly affects the deswelling kinetics, as revealed by a consistent analysis of the domain growth during the microgel collapse. Though both the disordered and the regular networks exhibit a similar early growth of the domains, an acceleration is observed
in the regular network at the late stage of the collapse. 
Similar trends are found for the dynamic correlations coupled to the domain growth. As a consequence,
the fast late processes for the domain growth and the dynamic correlations in the regular network are compensated,
and the dynamic correlations follow a power-law dependence on the growing length scale 
that is independent of the microgel microstructure.

\end{abstract}

\newpage
\maketitle

\section{Introduction}
Microgels are one of the most popular systems in the soft matter community \cite{Microgelsbook}.
These cross-linked polymer networks, of size typically in the range 100 nm - 100 $\mu$m, are unique due to their tunability in terms of architecture, softness, permeability and deformability. Their conformational changes in solution and in particular their swelling/deswelling behavior can be achieved, e.g., by changing the temperature, the ionic strength or the pH. The importance of microgels relies on their huge amount of applications \cite{microgelDas2006, Plamper2017}
such as drug delivery \cite{Kwon2008,Klinger2012}, photonic crystals \cite{Zhang2014,Islam2014}, sensors \cite{Ancla2011,Zhang2016}, 
or purification technologies \cite{Parasumaran2012, Jin2015}, just to name a few.

From a synthetic perspective and commercial applications, hot areas encompass
nanoparticle and drug upload, non-polymerization routes, control of composition and stability for efficacy in biological applications, etc. \cite{appl1,Murthy4995,appl3,doi:10.1002/adfm.200304338}
%\textcolor{blue}{To discuss: different procedures of computational synthesis, with reference to regular structures, and experiments. What should be clarified..open problems...}
%In this manuscript we systematically analyse the deswelling properties of microgels and determine structural and dynamical changes during the collapse.
Control of the local network microstructure is highly desirable since this directly influences  the efficiency of many of the former applications. 
Thus, the microstructure may affect both the diffusion of particles in and out of the microgel, as well as its swelling/deswelling time scales.
The characterization of the microgel global structure is usually addressed through scattering techniques.
Regarding the local microstructure of the network, a highly detailed characterization is still
hardly accessible in experiments, despite noteworthy recent advances in super-resolution optical microscopies \cite{Gelissen2016,Conley2016,Conley2017,Bergmann2018}. 

Computer simulations constitute a powerful tool for accessing information not directly provided by experiments, for understanding
the factors controlling the formation of specific microstructures, and for envisaging routes to tune and improve them.
However, most of the computational investigations of microgels in the literature do not implement 
a realistic synthesis procedure. They instead construct the microgels
as spherical cross-linked networks whose dynamics exhibit the expected broad internal fluctuations, but whose topology is equivalent to that of a 
crystal lattice with nodes connected by polymer strands (regular network) \cite{olvera2011,winkler2014,winkler2016,Ahualli2017,Sean2018}. 
Some works have introduced more realistic networks with spatially disordered distributions of the cross-linked sites (i.e., with a distribution of the strand length) \cite{Kamerlin2016}. Still, both approaches lack of defects 
in the network topology (loops, knots, concatenations, etc) which should be 
generally present in real microgels synthesized with conventional routes.
In an effort to construct realistic {\it in silico} microgels, Zaccarelli and co-workers \cite{Gnan2017,Rovigatti2018}
have very recently reported a detailed computational investigation of microgels inspired by the usual polymerization routes. 
By using 2- and 4-patchy particles both chain growth and branching are implemented, producing cross-linked networks
whose structural properties are consistent with the standard fuzzy sphere model widely 
used in the analysis of scattering experiments \cite{Stieger2004,Eckert2008}. 

In this article we present simulations of an alternative, realistic model of microgels. Instead of following a polymerization
route, the microgels are synthesized {\it in silico} through cross-linking, within a cavity, of pre-existing polymer chains functionalized with 
reactive groups. This procedure is inspired, e.g., by microfluidic fabrication of microgels from macromolecular precursors
\cite{Tumarkin2009,Weitz2010,Seiffert2010,Seiffert2011}. 
We characterize the global and internal structure of the generated microgels, and discuss their effect on the thermodynamic properties
by comparing results with those of a regular (diamond-like) network model. For the same molecular weight and effective degree of cross-linking,
the specific microstructure of the microgel 
has a negligible effect on the volume phase transition (VPT).
We find, however, that the microstructure has a significant effect on the deswelling kinetics. 
The collapse is more heterogeneous in the disordered microgels. We characterize the coarsening kinetics of the microgel by monitoring the growth of the domains during the collapse.  The observed behavior is apparantely intermediate between the cases of the coil-to-globule transition for linear polymers and the spinodal decomposition of a liquid-gas system.
At early times the domains grow up following a power-law that is almost independent of the network structure.
In the disordered microgels the same power-law persists until the end of the collapse. However, an acceleration is observed at late times for the regular networks. This unusual behavior is related to the fast merging
of the regularly distributed nucleating centers. Similar trends are found for the dynamic correlations coupled to the domain growth. The fast late processes for the domain growth and the dynamic correlations in the regular network are compensated. As a consequence the dynamic correlations follow a power-law dependence on the growing length scale that is independent of the specific microstructure.

The paper is organised as follows.
In \ref{sec2} the main details about the model and methods are discussed.
In \ref{sec3} conformational and structural properties in good solvent conditions are studied. \ref{sec4} presents an analysis of the 
thermodynamics and kinetics of the microgel collapse.
\ref{sec5} summarizes our conclusions.

\section{Model and simulation details}\label{sec2}

The experimental procedure we modeled is the synthesis of a microgel in a spherical cavity or droplet, via inter- and intra-molecular irreversible association of linear polymer chains.
In our model a single polymer chain consists of a linear backbone of $N$ beads. A number $N_{\rm l}$ of these beads are the reactive groups, randomly distributed along the polymer chain. 
Each reactive group can form a single irreversible bond with another reactive group in the same or in another chain.

For the non-bonded pair interaction between the polymer beads we used the following potential \cite{soddemann2001EPJE,LoVerso2015,Gnan2017}:
\begin{equation}
%\[
V_{\rm nb}(r) = \left\{
\begin{array}{lll}
V_{\rm LJ}(r)=  4\epsilon\left[\left(\frac{\sigma}{r}\right)^{12}
- \left(\frac{\sigma}{r}\right)^{6} +\frac{1}{4} \right] - \epsilon\phi &\ &\  r\leq 2^{1/6}\sigma \\
V_{\phi}(r) = \frac{1}{2} \phi\epsilon \left[\cos\left(\alpha (\frac{r}{\sigma})^2 +\beta\right)-1\right]  &\ &\  2^{1/6}\sigma < r\leq 1.5\sigma \\
 0 &\ &\ r > 1.5\sigma
\end{array}
 \right.
%\]
\label{eq:vnb}
\end{equation}
By using the parameters $\alpha = \pi(2.25 -2^{1/3})^{-1}$ and $\beta = 2\pi -2.25\alpha$,
the non-bonded potential and its first derivative are continuous both at $r=2^{1/6}\sigma$ and at the cutoff $r_{\rm c}=1.5\sigma$.
The quantities $\epsilon$, $\sigma$, and $\tau=(\sigma^2 m/\epsilon)^{1/2}$ (with $m$ the bead mass) set the energy, length, and time scales, respectively. In the following we use reduced units, $\epsilon = \sigma = m = \tau = 1$.
In the distance range $ r < 2^{1/6}\sigma$ the non-bonded interaction is a fully repulsive Lennard-Jones (LJ) potential with no local minima.
Thus, by using $\phi=0$ the interaction is already cut-off at $r=2^{1/6}\sigma$ and implicitly mimicks good solvent conditions.  
Bad solvent conditions can be implemented by switching on ($\phi > 0$) the attractive tail,
acting  on a distance corresponding to the first neighbor shell ($2^{1/6}\leq r \leq 1.5$). The solvent quality is worsened by increasing the depth $\phi$ of the attractive tail.
Besides the non-bonded interactions, permanent bonds between connected beads are implemented through a FENE (finitely extensible non-linear elastic) potential \cite{Kremer1990}:
\begin{equation}V_{\rm FENE}(r) = -\epsilon K_{\rm F}R_0^2 \ln\left[1-\left(\frac{r}{R_0\sigma}\right)^2 \right] ,
\label{eq:fene}
\end{equation}
with $K_{\rm F} = 30$ and $R_0 = 1.5$.

The synthesis of the microgel was performed in ideal good solvent ($\phi=0$).
A total of $N_{\rm cha}=36$ chains of $N=600$ beads (`monomers') were confined inside a spherical cavity of radius $R_{\rm cav} =55$.
Therefore the number density used in the synthesis was $3N_{\rm cha}N/(4 \pi R^3_{\rm cav}) \approx 0.03$. This qualitatively
corresponds to experimental concentrations of the order of $10^{-2}$ mg/mL \cite{Kremer1990,Moreno2016JPCL,gonzalezburgos2018}.
The interaction between any bead and the spherical wall was given by a repulsive LJ interaction that guaranteed the confinement
of the chains during the cross-linking process:
\begin{equation}
V_{\rm wall}(r_{\rm w}) = \left\{
\begin{array}{lll}
4\epsilon\left[\left(\frac{\sigma}{r_{\rm w}}\right)^{12}
- \left(\frac{\sigma}{r_{\rm w}}\right)^{6} +\frac{1}{4} \right] &\ &\  r_{\rm w} \leq 2^{1/6}\sigma \\
 0 &\ &\ r_{\rm w} > 2^{1/6}\sigma
\end{array}
\right. 
\end{equation}
with $r_{\rm w}$ the shortest distance from the bead to the spherical wall. 
All the chains contained a fraction of reactive groups $f= N_{\rm l}/N = 0.1$. In order to prevent trivial cross-linking events
no consecutive reactive groups were allowed in the backbone sequence.
After equilibration of the chains inside the cavity the cross-linking of the reactive groups was activated.
A permanent bond between two reactive groups was formed (imposing the permanent FENE interaction) if two conditions were fulfilled: (i) none of them was bonded to another reactive group, and (ii) they were at a mutual distance $r < 1.3$ (capture distance). A random choice was made in the case of multiple options, i.e., among the reactive groups within the same capture distance. The slow late stage of the cross-linking process ($< 10$ remaining unbonded reactive groups) was sped up as follows. We found that such groups
were homogeneously distributed inside the cavity. In such conditions we assumed similar barriers for the encountering of any pair of unreacted groups. Thus, we randomly selected a pair of the remaining unreacted groups and implemented an harmonic interaction between the constituents of the pair, in order to approach them to the capture distance and form the bond. This procedure was repeated 
for the remaining unreacted groups,
until full completion of the cross-linking.

The synthesis of the microgel was repeated for 64 different realizations of the initial conditions (same values of $N_{\rm cha}, N, f, R_{\rm cav}$ and different initial conformations of the chains). In most of the cases (53 of 64) all the polymer chains were connected, through the intermolecular cross-links, to a single network. In the rest of the cases a small disconnected cluster (formed by 1 or 2 chains) was found. Only the 53 microgels containing the whole initial set of chains were used in the analysis of the equilibrium properties and the deswelling transition.

The simulations were performed at $T = \epsilon/k_{\rm B}$ (with $k_{\rm B}$ the Boltzmann constant) under Langevin dynamics, following the integration scheme of Refs.~\cite{Izaguirre2001,Smith2009} with a time step $\Delta t = 0.005$. After completion of the cross-linking the cavity was removed, leading to the swelling of the generated microgels, which were further equilibrated.  
Finally, acquisition runs of typically $10^7$ steps were carried out for the analysis of the structural properties at $\phi=0$. In order to investigate the deswelling behavior, the equilibrated swollen microgels ($\phi =0$) were quenched (at infinite rate) to different solvent parameters  $0 < \phi \leq 1.5$. The deswelling kinetics was characterized from the quench instant ($t=0$) until equilibration. Further acquisition runs were performed to characterize the equilibrium structure at $\phi >0$.

%\textcolor{blue}{In order to compare the structural and kinetic properties of our realistic protocol with more %unrealistic realization starting from regular structures, we 
%built a  microgel  which origin from a diamond network.
%We first fixed the reactive group positions at the verteces of a diamond network. The nearest-neighbor nodes are %then connected with rods made of bead units and we proceed to the equilibration of the structure. To define this %diamond-microgel  size we select only the beads inside a sphere of radius comparable to the microgel radius.}

Not surprisingly, the cross-linking produced disordered microgel networks with topological polydispersity (see below). In order to investigate the effect of the local microstructure on the equilibrium and deswelling behavior of the microgels, we compared the results obtained by the former experimentally inspired protocol with those of a built-in regular network. We first placed the cross-linked sites in the nodes of a diamond network. Then we connected each node to its four nearest nodes through identical strings of beads. A sphere centered at the center-of-mass of the network was constructed and all the beads out of the sphere were removed, leading to a finite microgel with diamond-like connectivity, as proposed in several investigations in the literature \cite{olvera2011,winkler2014,winkler2016,Ahualli2017,Sean2018}. We tuned the length of the strings connecting the nodes and the size of the sphere
in order to obtain diamond microgels as close as possible to the disordered microgels in several properties. Namely, with the choice of 272 network nodes (cross-linked sites) and a total of $N_{\rm t} = 21692$ beads, the simulated diamond and disordered microgels ($N_{\rm t} = 21600$) had very similar values of $N_{\rm t}$, macromolecular size and fraction of elastically active cross-links (see next Section). Analogous runs to those of the disordered microgels were carried out, at the same $\phi$-values, to investigate the deswelling behavior and equilibrium structure of the diamond microgels. To improve statistics, 5 independent runs were carried out for each of the selected microgels at each $\phi$-value.

\section{Equilibrium structural characterization}\label{sec3}

\subsection{Size and shape}

We start by characterizing the global structure of the disordered microgels.
We calculate the radius of gyration $R_{\rm g} = \langle R_{\rm g}^2 \rangle$ 
and the asphericity parameter $a$ \cite{rawdon2008}.
The latter is obtained as a function of the eigenvalues of the gyration tensor, 
$\lambda_1, \lambda_2, \lambda_3$:
\begin{equation}
a = \left\langle \frac{(\lambda_2 -\lambda_1)^2 +(\lambda_3 -\lambda_1)^2 +(\lambda_3 -\lambda_2)^2}{2(\lambda_1 +\lambda_2 +\lambda_3)^2} \right\rangle .
\end{equation}
The asphericity $0 < a < 1$ quantifies deviations from spherosymmetric conformations ($a=0$).

\begin{figure}[t!]
\centering
\includegraphics[width=.5\textwidth]{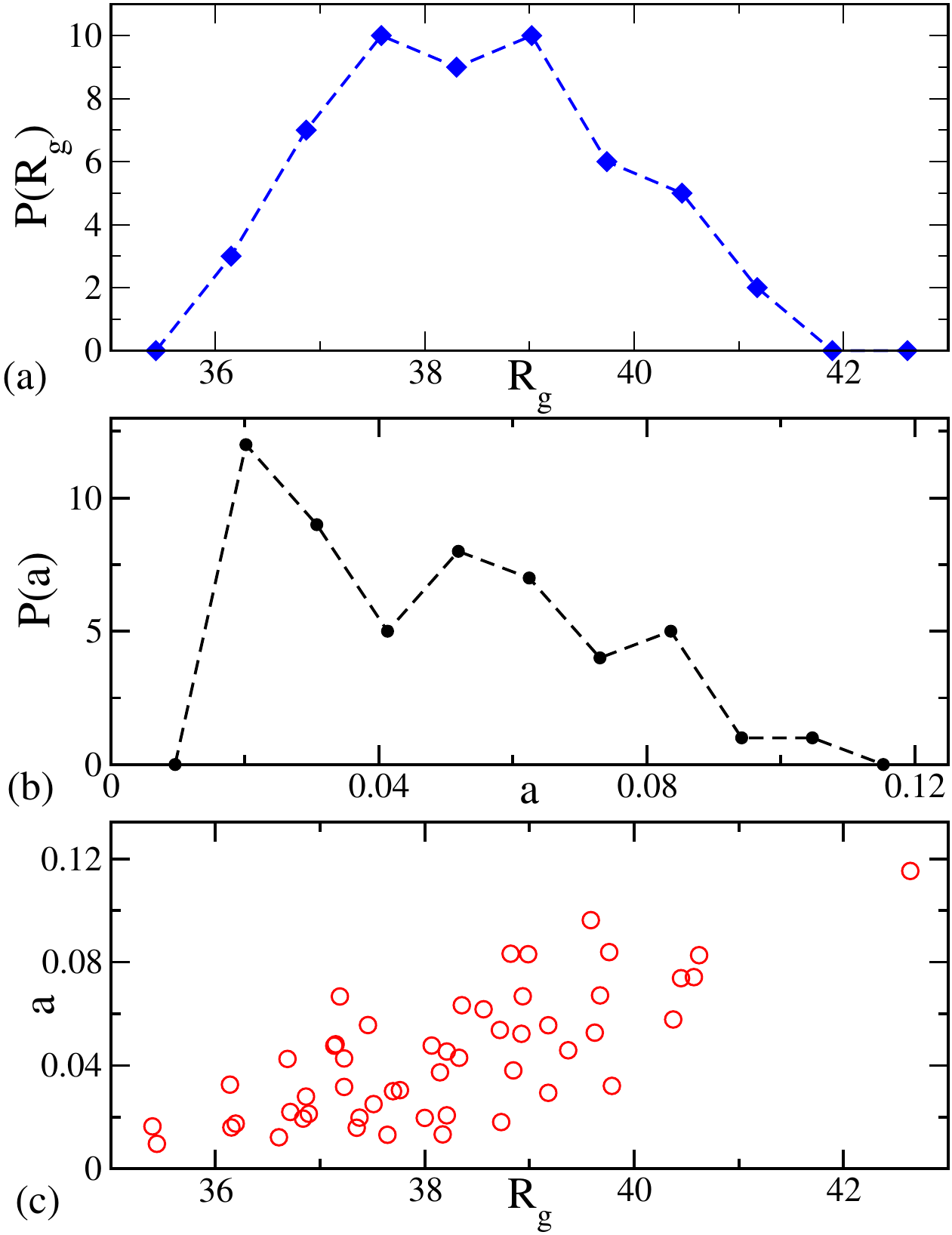}
\caption{For the disordered microgels, distribution of the time-averaged
radius of gyration (a) and asphericity (b). Panel (c) shows the correlation between asphericity and radius of gyration.}
\label{fig:rg-asph-rgasph}
\end{figure}

\ref{fig:rg-asph-rgasph} summarizes the main structural characteristics of the disordered microgels. As expected,
the cross-linking of the chains with different initial configurations and velocities produces different 
structures of the resulting bond networks, with different size and shape. Thus, for each individual microgel we average the instantaneous values of its radius of gyration and asphericity
over its simulation trajectory. We denote such time-averaged values as the $R_{\rm g}$ and $a$ of the individual microgel.
%Such time-averaged values are denoted as $\bar{R}_{\rm g}$ and $\bar{a}$.
Panels (a) and (b) of \ref{fig:rg-asph-rgasph} shows the obtained distributions of the individual 
$R_{\rm g}$ and $a$,
which account for the intrinsic polydispersity in size and shape arising from the topological polydispersity of the microgels.
The whole ensemble of microgels has an average radius of 
gyration $R_{\rm g} \approx 38.5$ with a weak size polydispersity of about 7 \% (full-width at half-maximum, panel (a)). The
microgels are also polydisperse in shape, though they are globally quasi-spherical objects (most of the distribution has $a < 0.1$,
see panel (b)). 
The analysis of the prolateness parameter \cite{rawdon2008} (not shown)
reveals that they have a slight tendency to adopt prolate over oblate conformations. 
As can be seen in panel (c), there is a correlation between microgel size and shape.
Thus,  higher $R_{\rm g}$ values correspond in general to higher asphericity values.
The diamond microgels are essentially spherical ($a < 0.01$) and have a time-averaged size of $R_{\rm g} = 37.3 $, very close to the 
mean of the distribution for the disordered microgels.

\begin{figure}[t!]
\centering
\includegraphics[width=1.\textwidth]{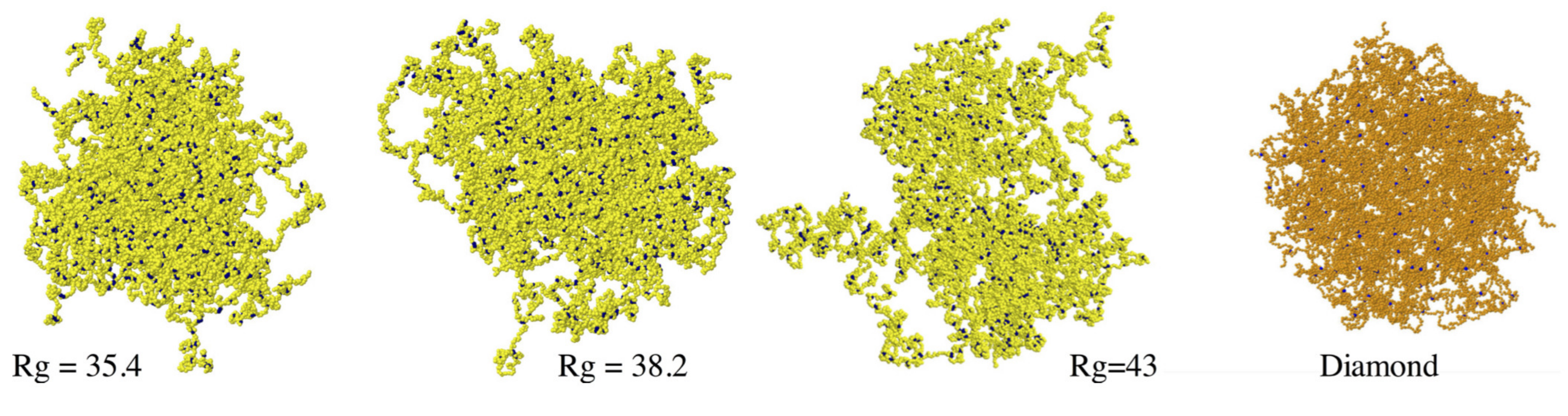}
\caption{Snapshots of typical microgel conformations, synthesized via bonding of polymer chains (yellow) and starting from a diamond-structured network (orange). Blue beads are the cross-linked sites.}

\label{fig:snap-real}
\end{figure}

\begin{figure}[t!]
\centering
\includegraphics[width=.5\textwidth]{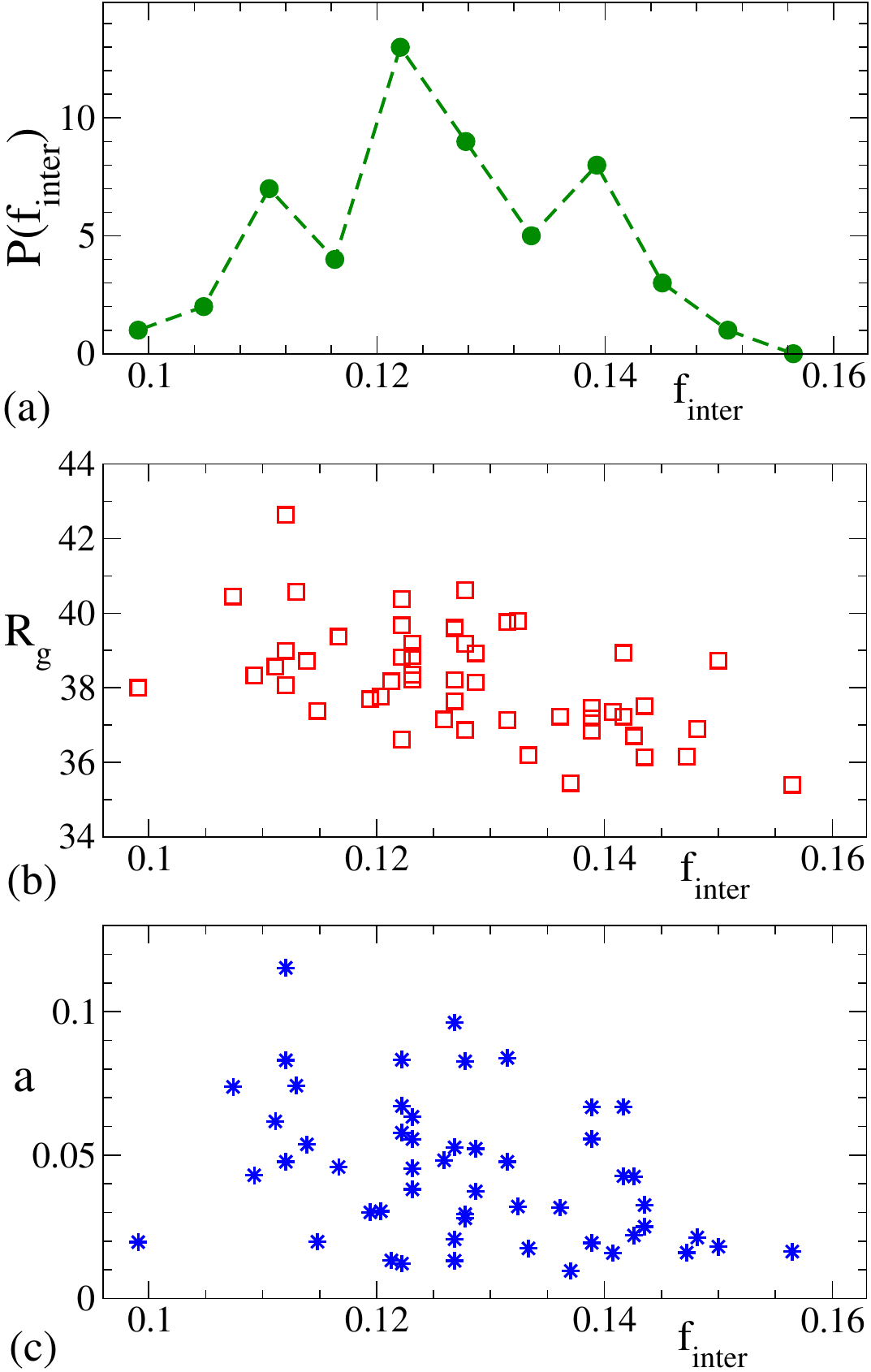}
\caption{Panel (a): distribution of the fraction of intermolecular bonds in the disordered microgels. Panels (b) and (c): correlations of the time-averaged radii of gyration and asphericities with the fractions
of intermolecular bonds.}
\label{fig:finter}
\end{figure}

In \ref{fig:snap-real} we show typical snapshots (yellow beads), corresponding to disordered microgels with radius of gyration at the
center and the two tails of  the $R_{\rm g}$ distribution of \ref{fig:rg-asph-rgasph}. In what follows we will characterize observables
for these three disordered microgels that, according to their size, we will refer to as
small ($R_{\rm g}(\phi = 0) = 35.4$), middle ($R_{\rm g}(\phi = 0) = 38.2$) and large 
($R_{\rm g}(\phi = 0) = 43.0$) disordered microgels. The respective asphericities are $a = 0.016$ , $0.021$ and  $0.11$.
In \ref{fig:snap-real} we include (orange beads) 
a typical snapshot of the diamond microgel. In all cases the cross-linked sites are depicted as blue beads.
Not surprisingly, the regular diamond microgels show the most uniform structures.
Small- and medium-size disordered microgels present more uniform and compact structures than the biggest ones, which are less spherical and characterized by more abundant and longer protrusions. 
\ref{fig:finter}a shows the distribution
of the percent of intermolecular cross-links in the disordered microgels. These are defined as bonds between pairs of reactive groups
belonging to different polymer chains. The obtained distribution reveals that only a small fraction of the whole set of cross-links 
($f_{\rm inter} \sim 0.12$) has an intermolecular origin. These cross-links are responsible for the connection of the whole
microgel network into a single cluster and therefore, are elastically active. The fraction of intermolecular cross-links in the disordered microgels is very similar to the fraccion of cross-linked sites (all them being elastically active) in the diamond microgel ($f_{\rm cl} = 272/21692 \approx 0.12$). Panels (b) and (c) of \ref{fig:finter} show the influence of $f_{\rm inter}$
in the size and shape of the disordered microgels. Through the correlation is relatively weak, increasing the number of intermolecular cross-links leads to smaller microgels. This is presumably related to a lower presence of long protrusions in the outer shell, which are expected to originate from highly intramolecularly cross-linked chains connected to the global network
just through a few intermolecular bonds. No significant correlation is found between asphericity and $f_{\rm inter}$.

%From panel (b) and (c) we can observe how this process influences the size and shape of the microgel.
%The radius of gyration decreases globally with the increasing of the inter-molecular linking. 
%The data confirm that the cross-linking between groups of different chains is crucial in order to obtain compactation and reduce %the molecule size. 
%For the low cross-linker fraction value we are considering, the asphericity is not
%remarkably influenced. 
%The inter-molecular bonding is the effective mechanism in order to avoid protrusions and empty regions. 

%\begin{figure}[t!]
%\centering
%\includegraphics[width=.51\textwidth]{plot-rg-aspher-fracbondinter-eps-converted-to.pdf}
%\caption{Correlations between time-averaged radii of gyration, asphericities, and fractions
%of inter-chain cross-links in the disordered microgels.}
%\label{fig:rg-asph-fbond} 
%\end{figure}

\begin{figure}[t!]
\centering
\includegraphics[width=.4\textwidth]{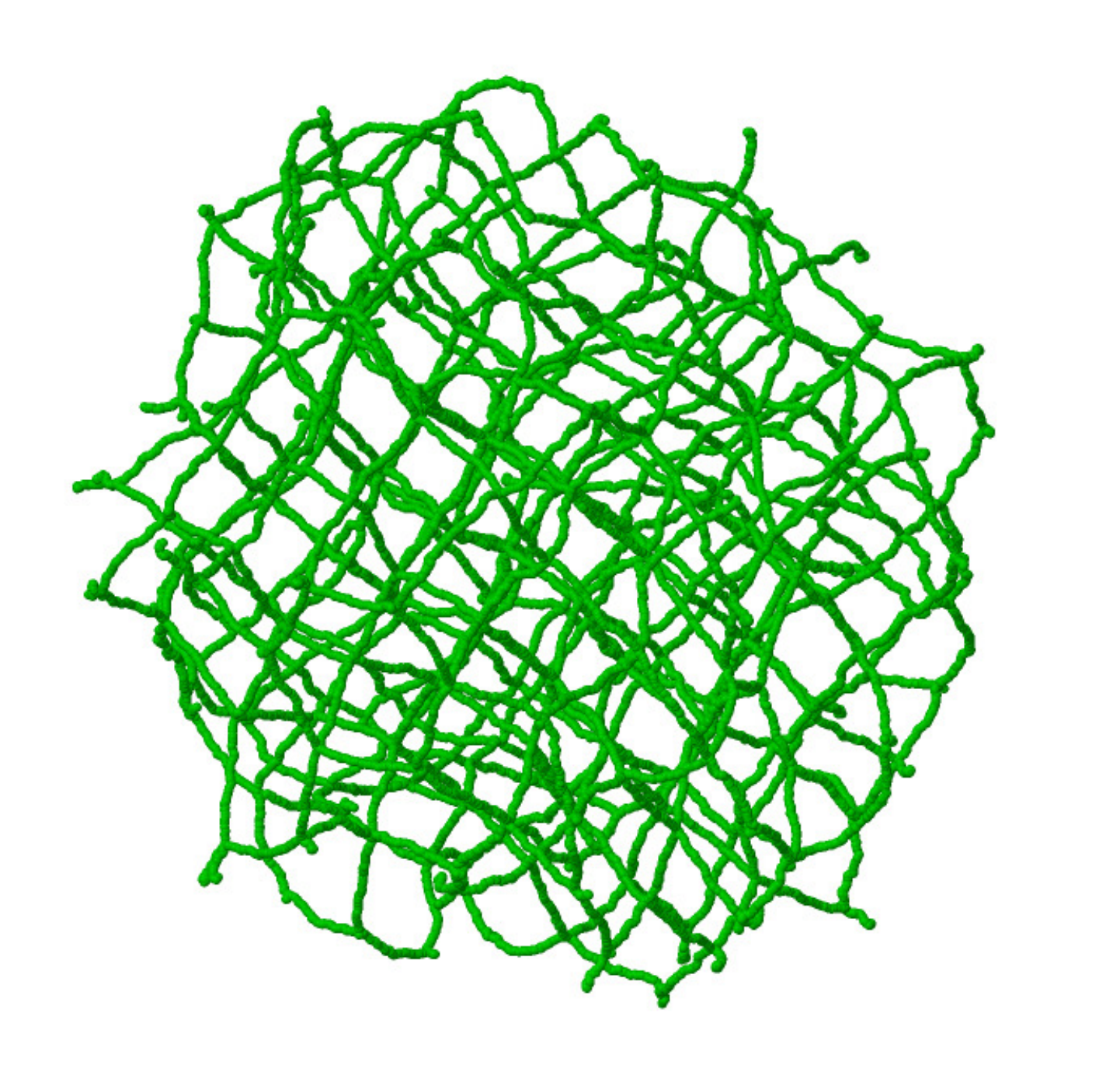}\hspace{0.3cm}
\includegraphics[width=.4\textwidth]{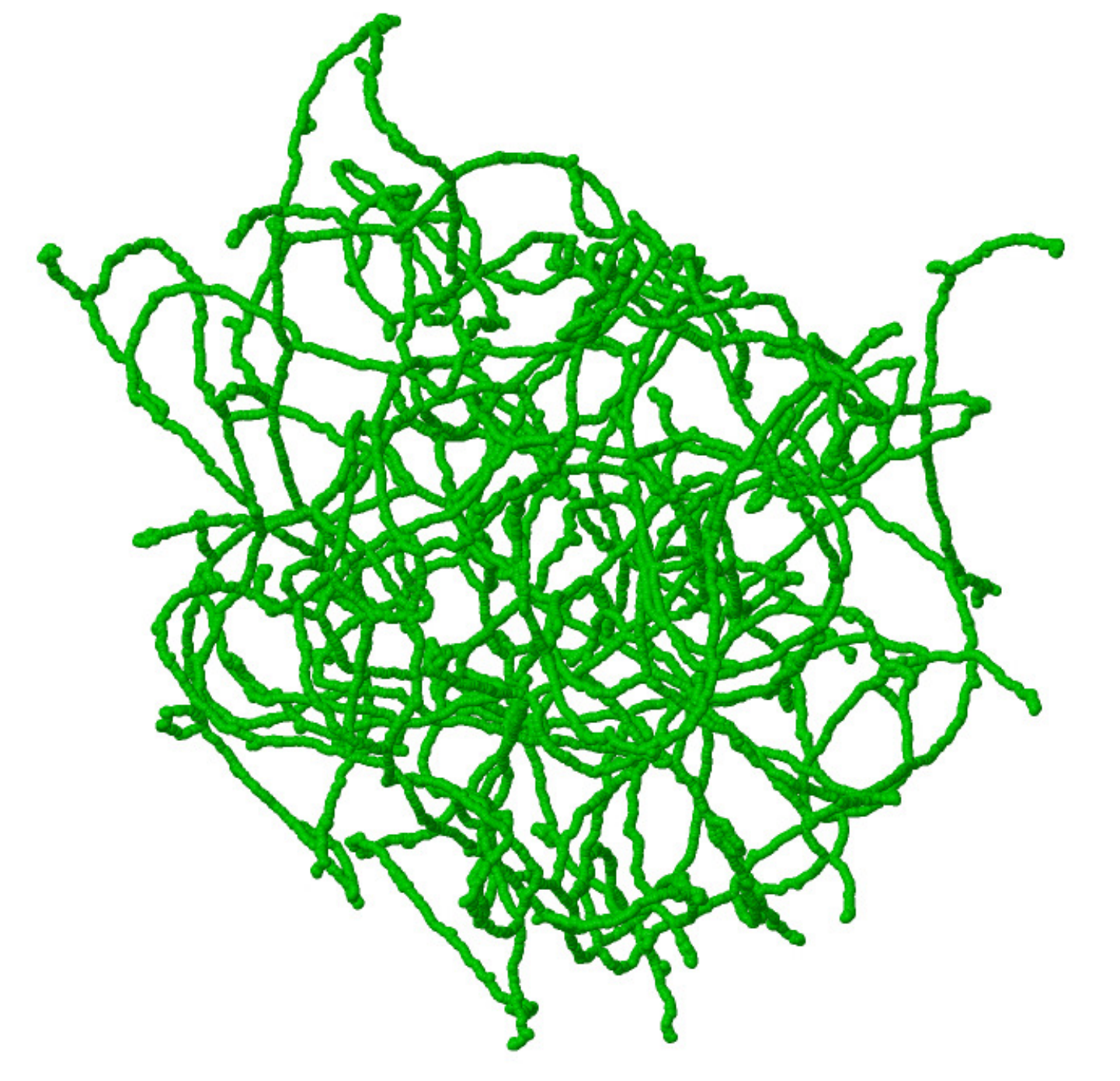}
\caption{Snapshots of typical IMPs of microgels with diamond (left) and disordered (right) bond networks.}
\label{fig:snap-imp} 
\end{figure}

To enhance the microstructural differences that may originate 
from the underlying network architectures 
we have performed an analysis based on the isoconfigurational mean path
(IMP) approach \cite{AM54}.
The concept of IMP  is based on the idea that the
tube path associated to a given macromolecular configuration can be approximated by the
coordinates of the monomers averaged over their isoconfigurational ensemble. 
In practice, for the {\it same}
initial configuration ($t=0$) of the microgel,  we generated 50 trajectories by starting with different
Maxwell-Boltzmann  realizations of the velocities, and the 50 configurations obtained at $t = 5000$ were used to get
the averaged coordinates of the monomers in the IMP. The time scale $t = 5000$ is of the order of the entanglement time \cite{Bacova2017}, so that
fast fluctuations within the tube are averaged out,
and a smooth path is generated that preserves the topological constraints.
Typical IMP configurations of the disordered and diamond microgels are
shown in \ref{fig:snap-imp}. As expected, the IMPs are smoother than the real-coordinate configurations presented in \ref{fig:snap-real}, and fast fluctuations and small-scale wiggles are averaged out.

\subsection{Static pair correlations}

\begin{figure}[t!]
\centering
\includegraphics[width=.48\textwidth]{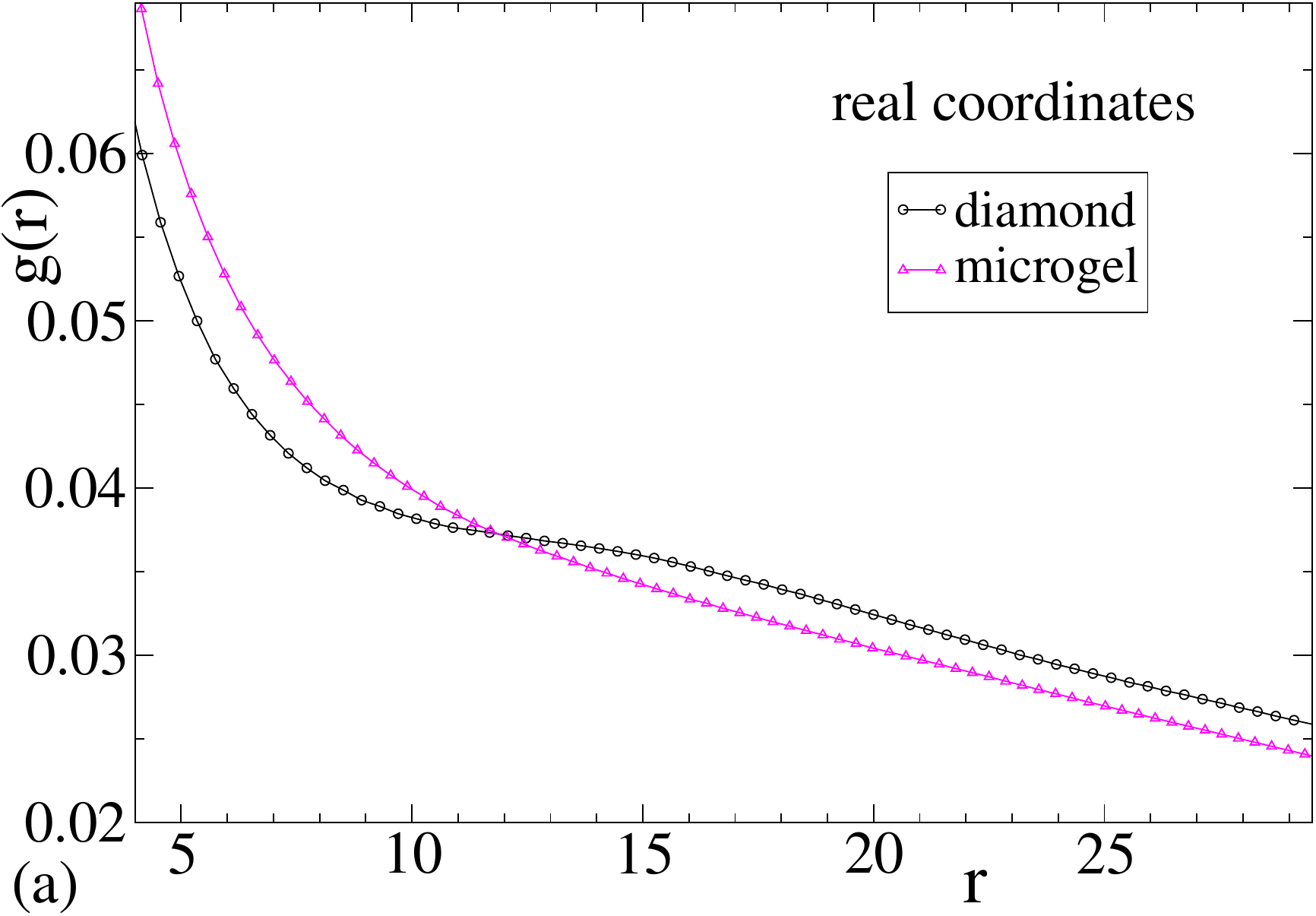}\hspace{0.3cm}
\includegraphics[width=.48\textwidth]{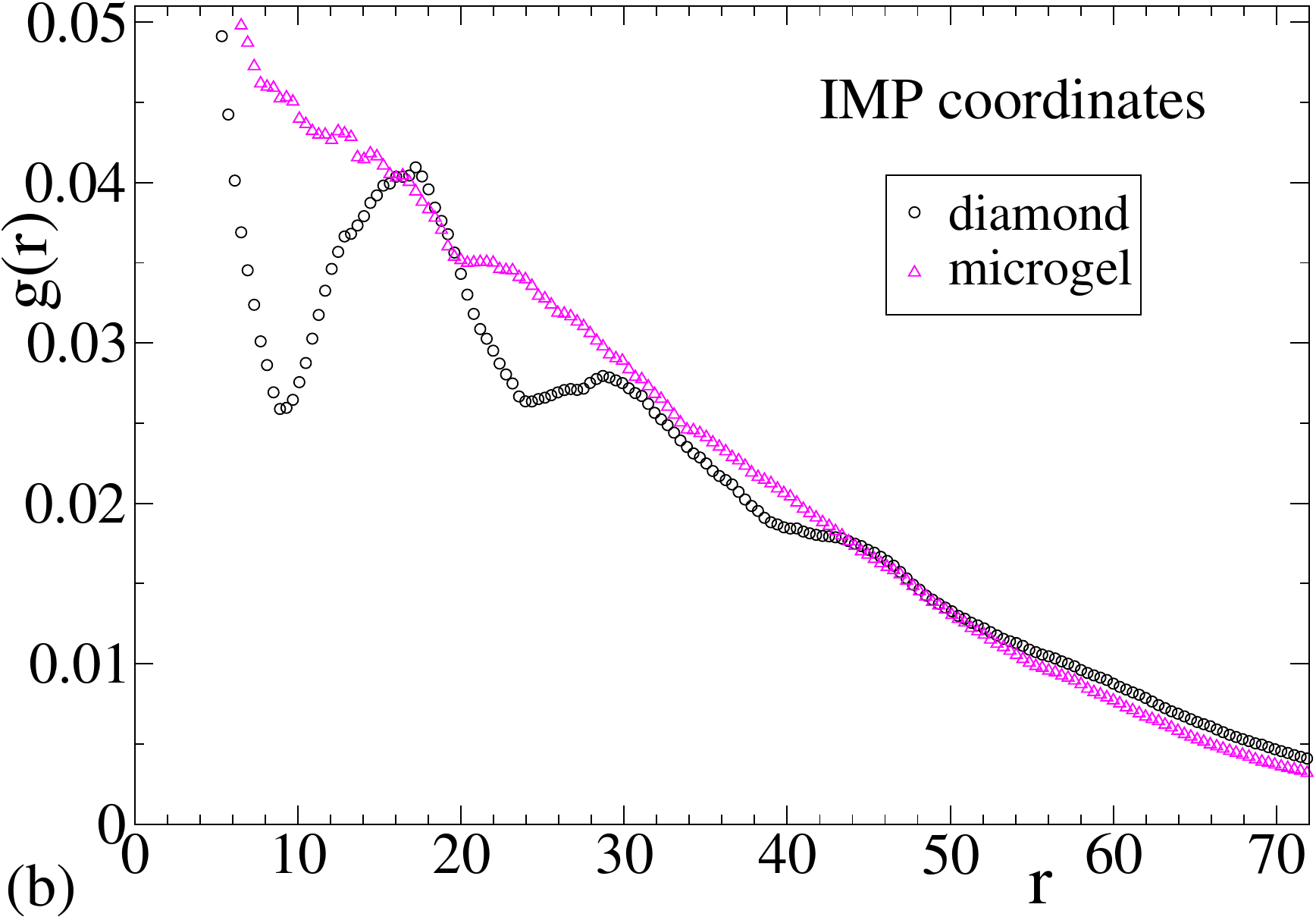}
\caption{Radial distribution function of the diamond (circles) and the middle-size disordered microgel (triangles).
Panels (a) and (b) are results obtained by using real and IMP coordinates, respectively. 
%%%The inset in (a) shows results for short distances.
}
\label{fig:gr} 
\end{figure}

\begin{figure}[t!]
\centering
\includegraphics[width=.48\textwidth]{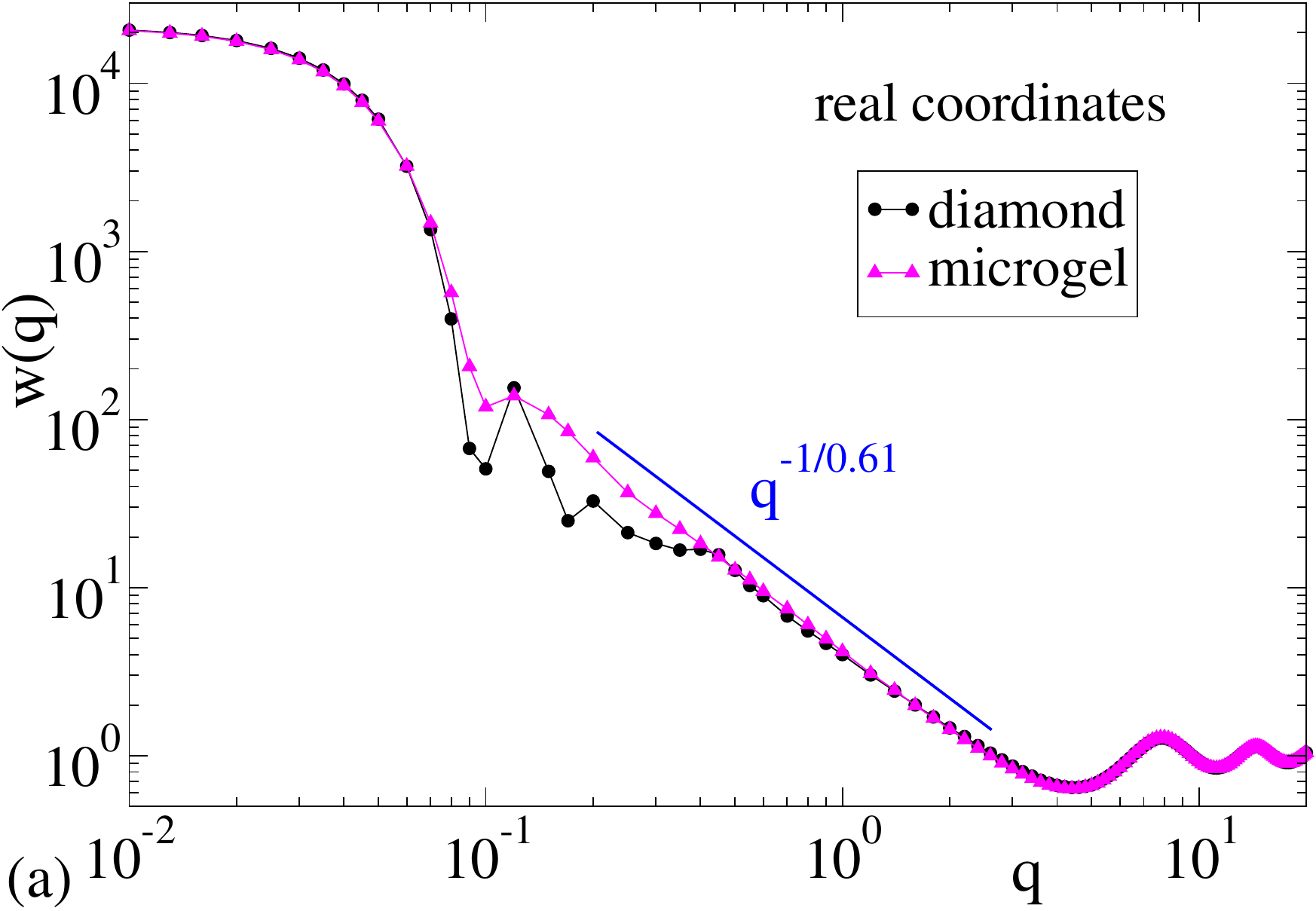}\hspace{0.3cm}
\includegraphics[width=.48\textwidth]{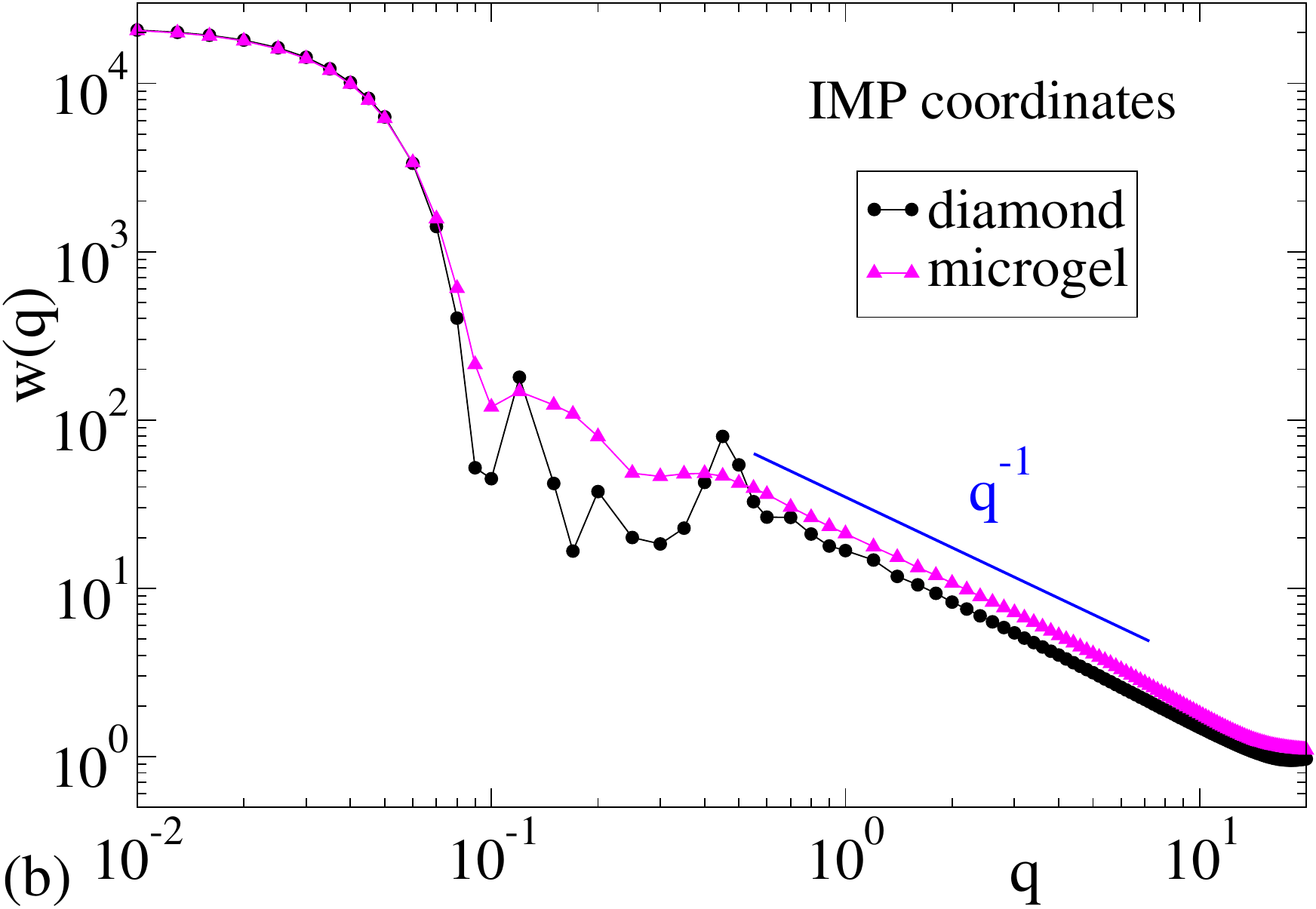}
\caption{Form factors of  the diamond (circles) and the middle-size disordered microgel (triangles).
Panels (a) and (b) are results obtained by using real and IMP coordinates, respectively. Lines indicate
power-law behavior $w(q) \sim q^{-1/\nu}$, with approximate SARW ($\nu \approx 0.6$) and rod-like ($\nu = 1$) scaling.}
\label{fig:wq} 
\end{figure}

\ref{fig:gr}a shows the radial distribution functions, for the real monomer coordinates,
of a middle-size disordered ($R_{\rm g} = 38.2$) and the diamond microgel.
This comparison does not reveal significant differences between both network topologies, except for a small bump at $r \sim 15$ in the diamond microgel.
The evident strong differences between such topologies are instead highlighted in the IMP representation, as shown in \ref{fig:gr}b.
In this case the black curve displays well-defined peaks as a
%with a periodicity of $\sim 15\sigma$,  
signature of the regular network structure in the diamond microgel.  
Similar features are found in the scattering form factor, 
$w(q) = \langle N^{-1}\sum_{j,k} \exp[i{\bf q}\cdot({\bf r}_j -{\bf r}_k)]\rangle$,
with ${\bf q}$ the wave vector and ${\bf r}_i$ the monomer positions. 
The corresponding data are represented in \ref{fig:wq} for the same middle-size disordered and diamond microgels as in \ref{fig:gr}.
%AM and FG vorrei trovare qualche plot sperimentale in letteratura per fare un paragone
% nella discussione
After the low-$q$ Guinier regime corresponding to $q R_{\rm g} < 1$ ($q \lesssim 0.02 $), the form factor of the disordered microgel
shows a sharp drop, typical of dense soft colloids with a global spherical structure (e.g., spherical brushes and high functionality stars \cite{doi:10.1021/ma00092a022,doi:10.1063/1.3494902,PhysRevLett.77.95}).
Power-law behavior $w(q) \sim q^{-1/\nu}$ is found at intermediate $q$-values, $0.1 < q < 4$ (distances $1 \lesssim d \lesssim 10$). The observed scaling exponent in such a range of distances, $\nu \approx 0.6$ (see panel (a)), reveals that the strands connecting the network nodes are essentially self-avoiding random walks (SARW) \cite{Rubinstein2003}, as expected for a swollen microgel. The counterpart of this behavior in the IMP representation (panel (b)) is rod-like scaling ($\nu \approx 1$). 
This is a direct result of the averaging of the fast fluctuations in the IMP, which
smooths the strands to quasi-linear segments (\ref{fig:snap-imp}). Finally, the peak at $q \sim 7$ (and the subsequent higher harmonics) in the real-coordinate $w(q)$ corresponds to the nearest-neighbor distance. This excluded-volume feature is absent in the IMP representation,  where the averaging procedure leads to a strong overlap of nearest-neighbor positions.
The real coordinate and IMP form factors of the diamond microgel exhibit the same former trends, though unlike in the disordered microgel, the underlying diamond structure is reflected by the presence of several peaks which are clearly highlighted in the IMP representation (black curves in \ref{fig:wq}).

\begin{figure}[t!]
\centering
\includegraphics[width=.46\textwidth]{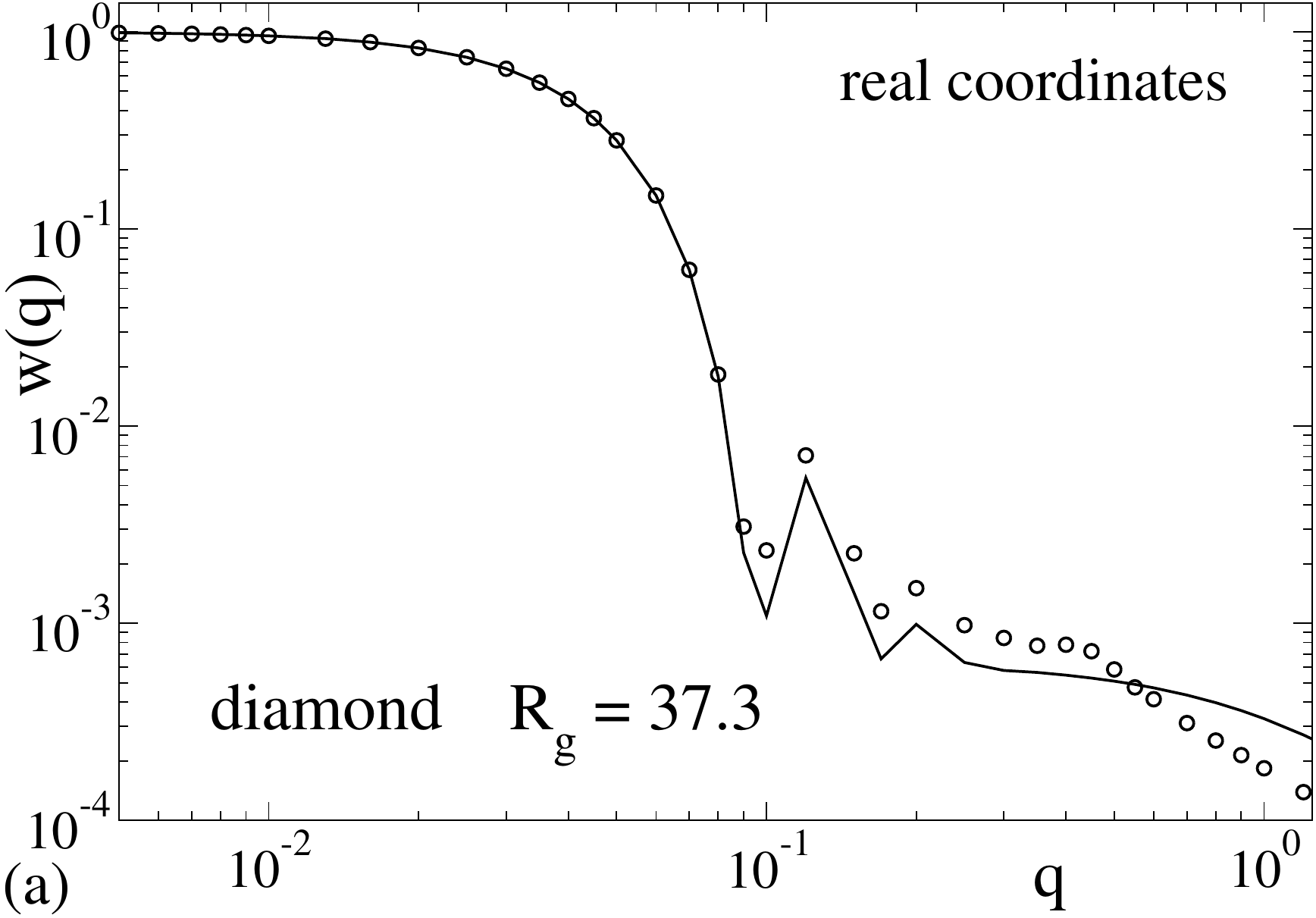}\hspace{0.3cm}
\includegraphics[width=.46\textwidth]{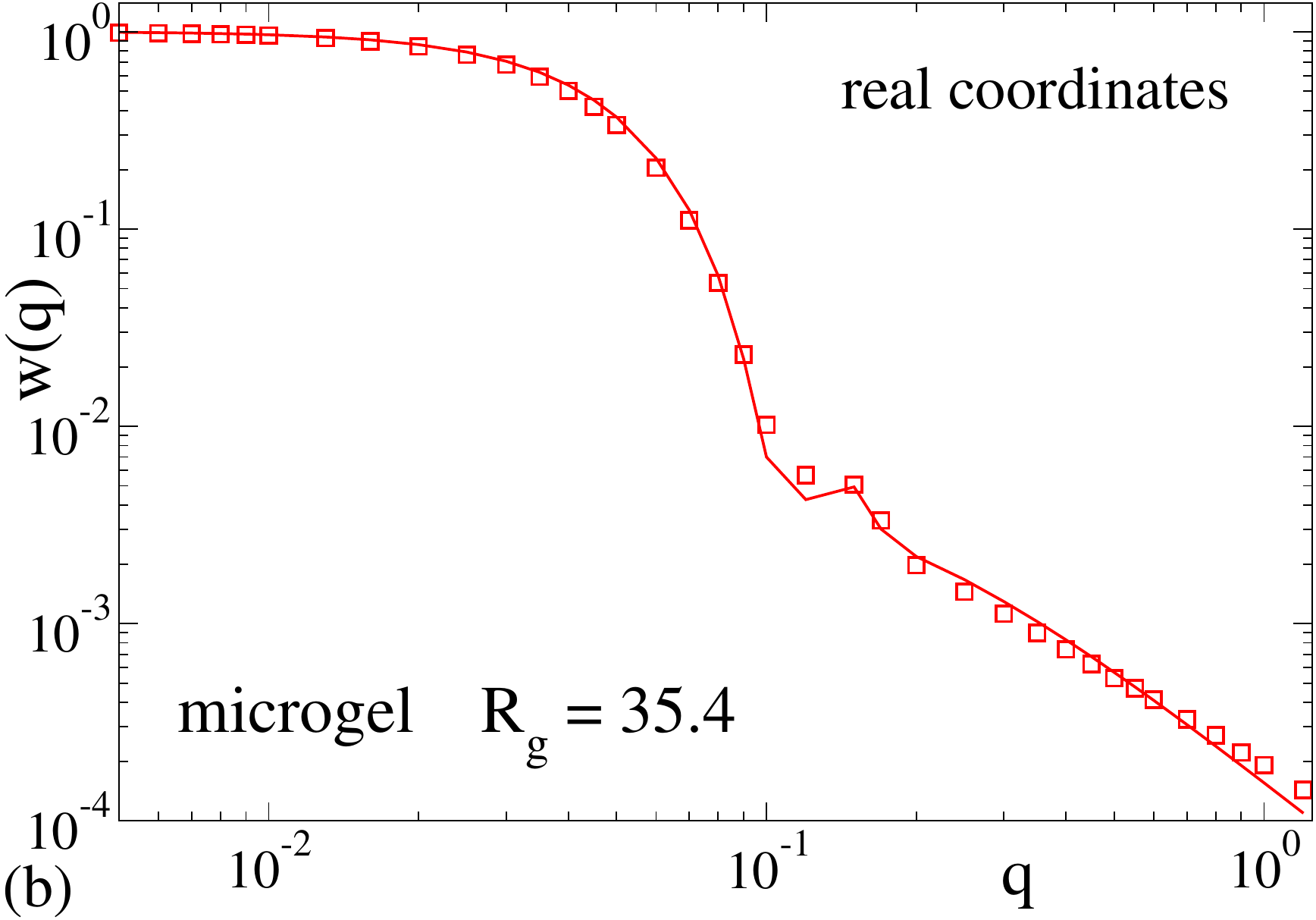}\vspace{0.1cm}
\includegraphics[width=.46\textwidth]{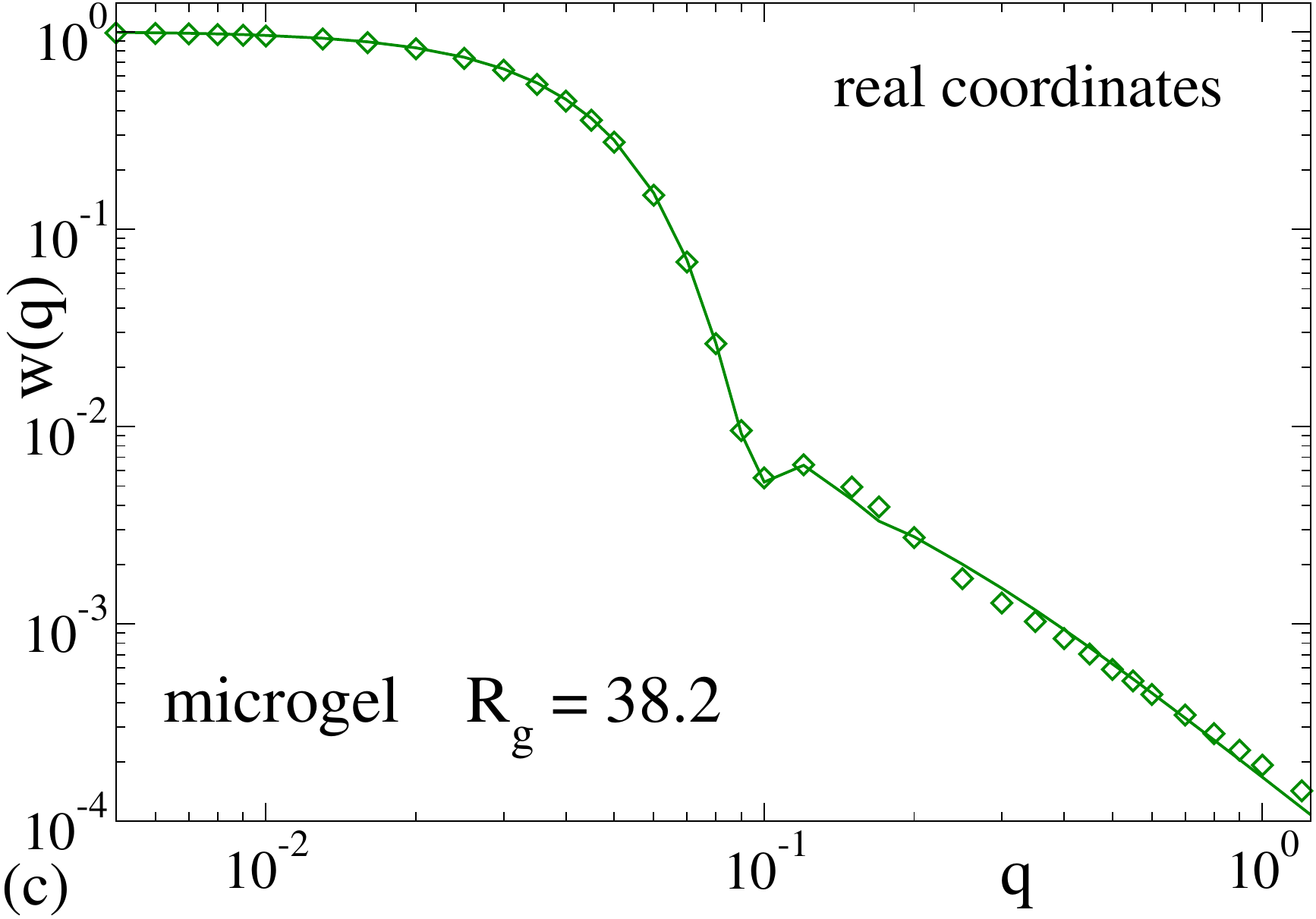}\hspace{0.3cm}
\includegraphics[width=.46\textwidth]{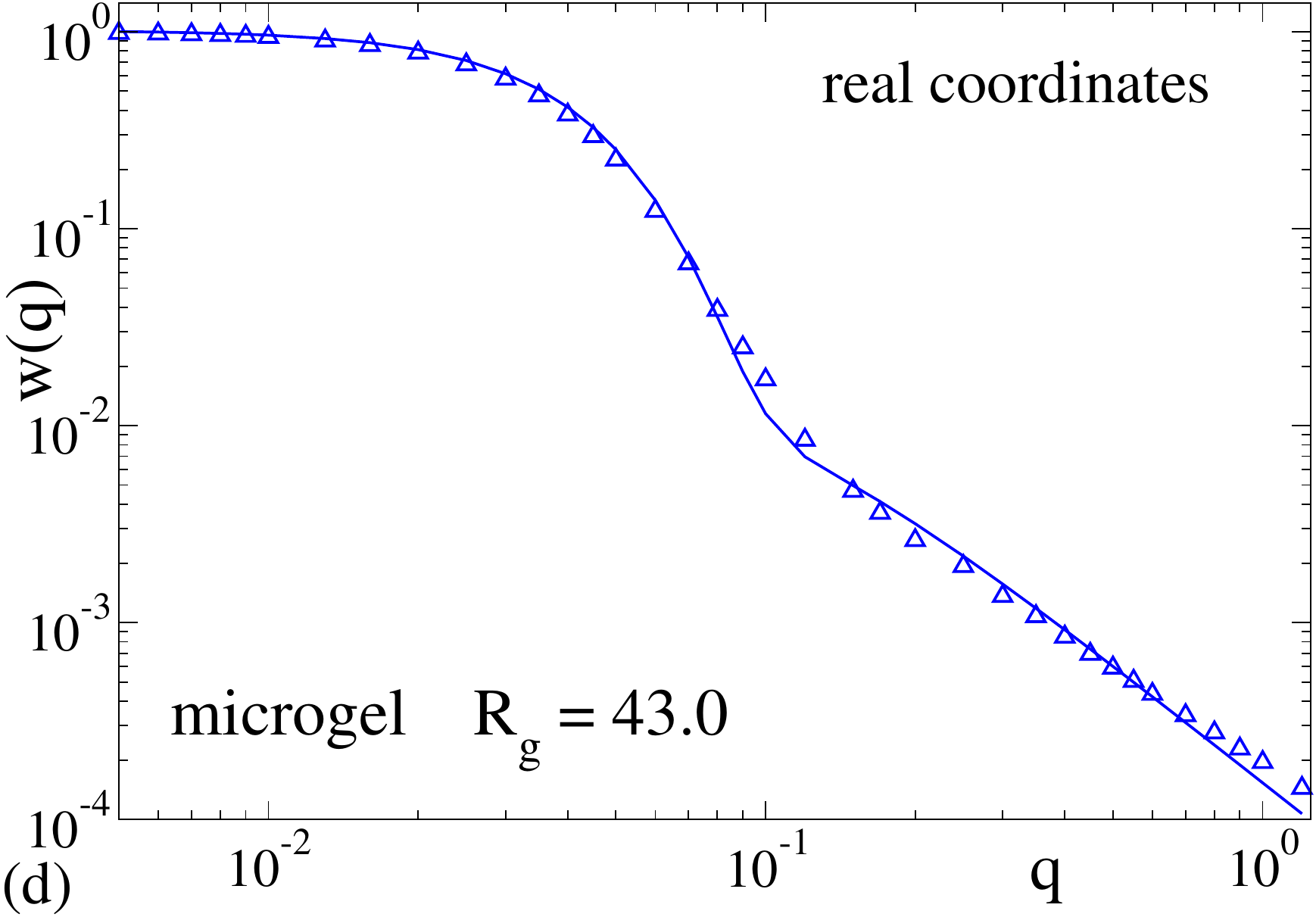}
\caption{Symbols: form factors of the diamond and three disordered microgels of small, middle and large size.
Lines: Fits to the fuzzy sphere model. The numerical values of the model parameters are given in \ref{tab:fuzzysphere}.}
\label{fig:wq-fits} 
\end{figure}

\begin{figure}[t!]
\centering
\includegraphics[width=.48\textwidth]{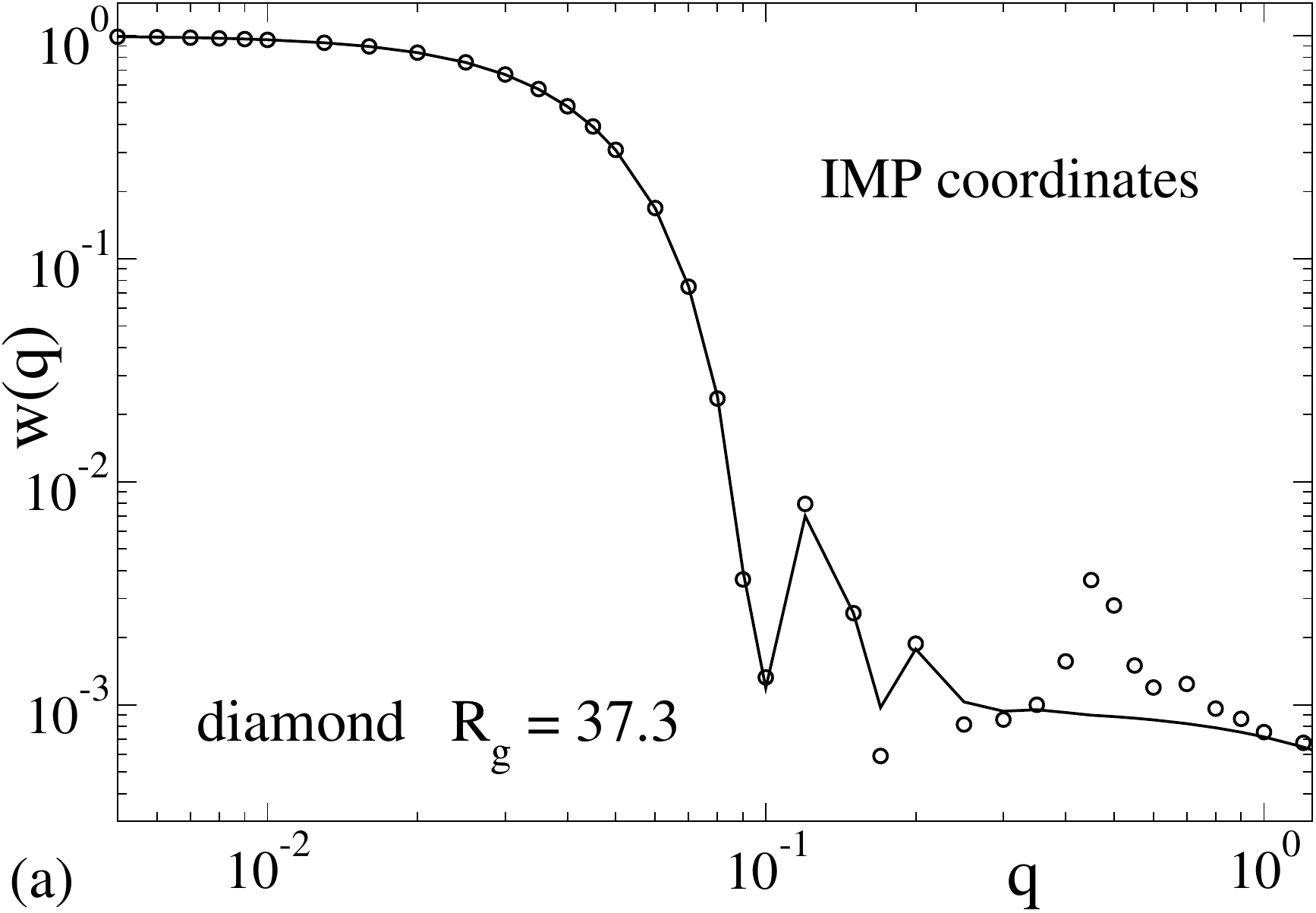}\hspace{0.3cm}
\includegraphics[width=.48\textwidth]{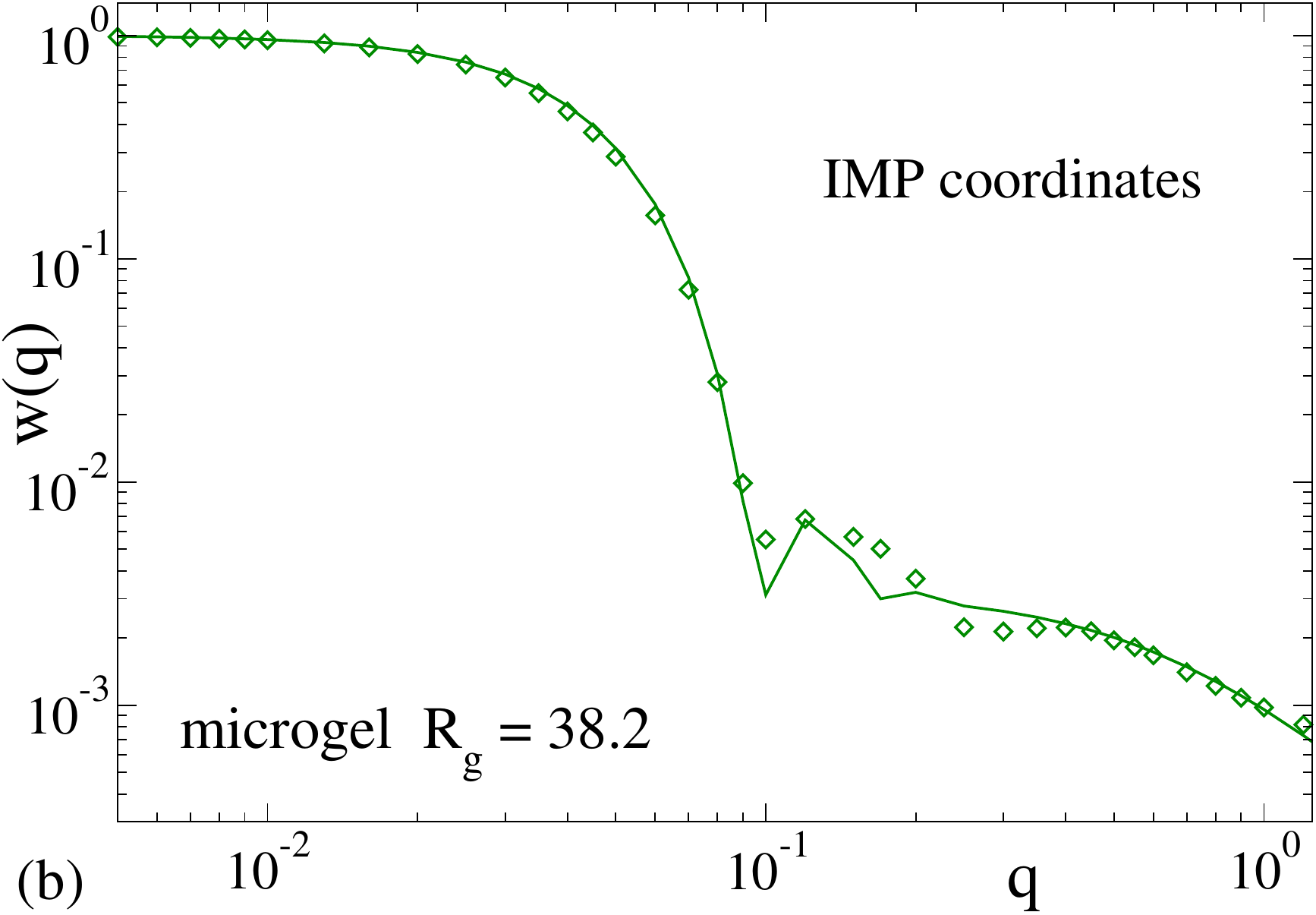}
\caption{Symbols: form factors of the IMPs of the diamond and a middle size disordered microgel.
Lines: Fits to the fuzzy sphere model. The numerical values of the model parameters are given in \ref{tab:fuzzysphere}.}
\label{fig:wq-imp-fits} 
\end{figure}

We have fitted the form factors, both for the real and IMP coordinates, to the fuzzy sphere model \cite{Stieger2004,Eckert2008}:
\begin{equation}
w(q) = \left[  \frac{3[\sin(q R') -qR'\cos(qR')]}{q^3 R'^3}\exp\left( -\frac{q^2 \sigma^2_{\rm s}}{2} \right)  \right]^2
+\frac{I(0)}{1 + q^2 \xi^2}
\label{eq:fuzzysphere}
\end{equation}
The fuzzy sphere model is generally invoked to describe the scattering form factors of microgels and to obtain
information about their internal structure. The model assumes a constant-density core of radius $R'$ with 
a shell of thickness $2\sigma_{\rm s}$.
The Lorentzian term in \ref{eq:fuzzysphere} represents the scattering (with intensity $I(0)$) from the network
fluctuations, of correlation length $\xi$. The fuzzy-sphere model was tested in the {\it in silico} microgels generated by polymerization routes in Refs.~\cite{Gnan2017,Rovigatti2018}. The form factors were consistent with the model, supporting the realistic
character of the simulated microgels. This is also the case for the synthesis route used in our simulations. 
As can be seen in \ref{fig:wq-fits}, the fuzzy-sphere model
provides a very good description of the real-coordinate form factors of the disordered microgels. Not surprisingly, the constant-density assumption for the core is an oversimplification for the regular diamond network, and the fuzzy sphere model works poorly  at intermediate distances ($0.1 < q <1$, see \ref{fig:wq-fits}a). 
This effect is even more visible in the form factor of the IMP coordinates
(\ref{fig:wq-imp-fits}a). Nevertheless, the fuzzy sphere model reasonably describes the IMP structure of the disordered microgels (\ref{fig:wq-imp-fits}b), since this structure does not present regularities and the inhomogeneities in the different
directions are strongly smoothed by the spherical average in the form factor.

The parameters of the fuzzy-sphere model obtained for the diamond and the small (I), middle (II) and large (III)
microgels are given in \ref{tab:fuzzysphere}. The smearing parameter,
correlation length and intensity of the fluctuations' contribution increase with increasing asphericity (sequence diamond-I-II-III).
For the most aspherical case (III) the inner core is not well-defined and the form factor is largely dominated
by a corona-like contribution. The IMP construction averages out fluctuations, which leads to smaller smearing parameters
and, in particular, to much smaller correlation lengths than in the respective real-coordinate configurations. 
The latter is a direct consequence of the quasi rod-like character of the strands in the IMP structure.
\begin{table}
  \caption{Parameters of the Fuzzy Sphere Model for the Form Factor}
  \label{tab:fuzzysphere}
  \begin{tabular}{lllllllllllll}
    \hline
    System         &                                 &&&   &&    $R'$     &&$\sigma_{\rm s}$      &&$\xi$     &&$I(0)$  \\
    \hline     
    Diamond        &$R_{\rm g} = 37.3$ (real coord.) &&&   &&    46.8     &&  5.3 	          &&  0.95    && 0.0006 \\
    Disordered I   &$R_{\rm g} = 35.4$ (real coord.) &&&   &&    40.6     &&  7.2 	          &&  5.4     && 0.0046 \\
    Disordered II  &$R_{\rm g} = 38.2$ (real coord.) &&&   &&    44.5     &&  9.3	          &&  6.6     && 0.0075 \\ 
    Disordered III &$R_{\rm g} = 43.0$ (real coord.) &&&   &&    0.10     &&  23.9	          &&  10.7    && 0.0177 \\
    Diamond        &$R_{\rm g} = 37.3$ (IMP)         &&&   &&    46.3     &&  2.3 	          &&  0.08    && 0.0008 \\
    Disordered II  &$R_{\rm g} = 38.2$ (IMP)         &&&   &&    44.7     &&  6.0	          &&  1.5     && 0.0032 \\ 
    \hline
  \end{tabular}
\end{table}

\subsection{Euler characteristic}

In order to further characterize the microgel structure and highlight the differences with pre-constructed ordered networks we utilize the Euler characteristic $\chi$, namely  one of the four scalar Minkowski functionals \cite{Likos1995,Hoffmann2006} that characterize 
a given surface embedded in three dimensions.
This morphometric approach is able to take into account multi-body correlations among the constituent monomers and give details on their spatial arrangement. The coefficient itself is proportional to the integral Gaussian curvature and its value is not subject to continuous, topology-preserving deformations of the surface. In three dimensions 
$\chi = \mathcal{N}_{\rm D} + \mathcal{N}_{\rm C} - \mathcal{N}_{\rm T}$, 
where $\mathcal{N}_{\rm D}$ is the number of disconnected aggregates, 
$\mathcal{N}_{\rm C}$ the number of enclosed cavities and $\mathcal{N}_{\rm T}$ the number of perforations (tunnels) that percolate through the system.
In order to compute $\chi$ we follow the procedure described in Ref.~\cite{sanchez2013design}. The system box is discretized into a cubic lattice, of spacing $d$. Each lattice site is surrounded by a Wigner-Seitz cell, in this case an elementary cube having the size of the lattice constant. We then consider the surfaces $\mathcal{S}(R)$ 
formed by spheres with a radius $R$ located at the centers of every monomer. 
We denote each lattice site inside $\mathcal{S}(R)$ as `filled' and all others sites as `empty'. 
$R$ denotes the relevant length scale: for small values of $R$ we have a collection of disjoint spheres. While increasing $R$, some of the spheres merge, and progressively form rings and cavities. For $R$ large enough the collection of spheres will fully occupy the space containing the molecule, and the ultimate value $\chi =1$ is reached. For every chosen $\mathcal{S}(R)$ family,  the Euler characteristic is computed as explained in Ref.~\cite{Likos1995}.
Due to the discretized nature of the surface $\mathcal{S}(R)$ obtained with this procedure, we choose a lattice spacing $d$ sufficiently small to avoid artifacts arising from the discretization. 

\begin{figure}[t!]
\centering
\includegraphics[width=0.48\textwidth]{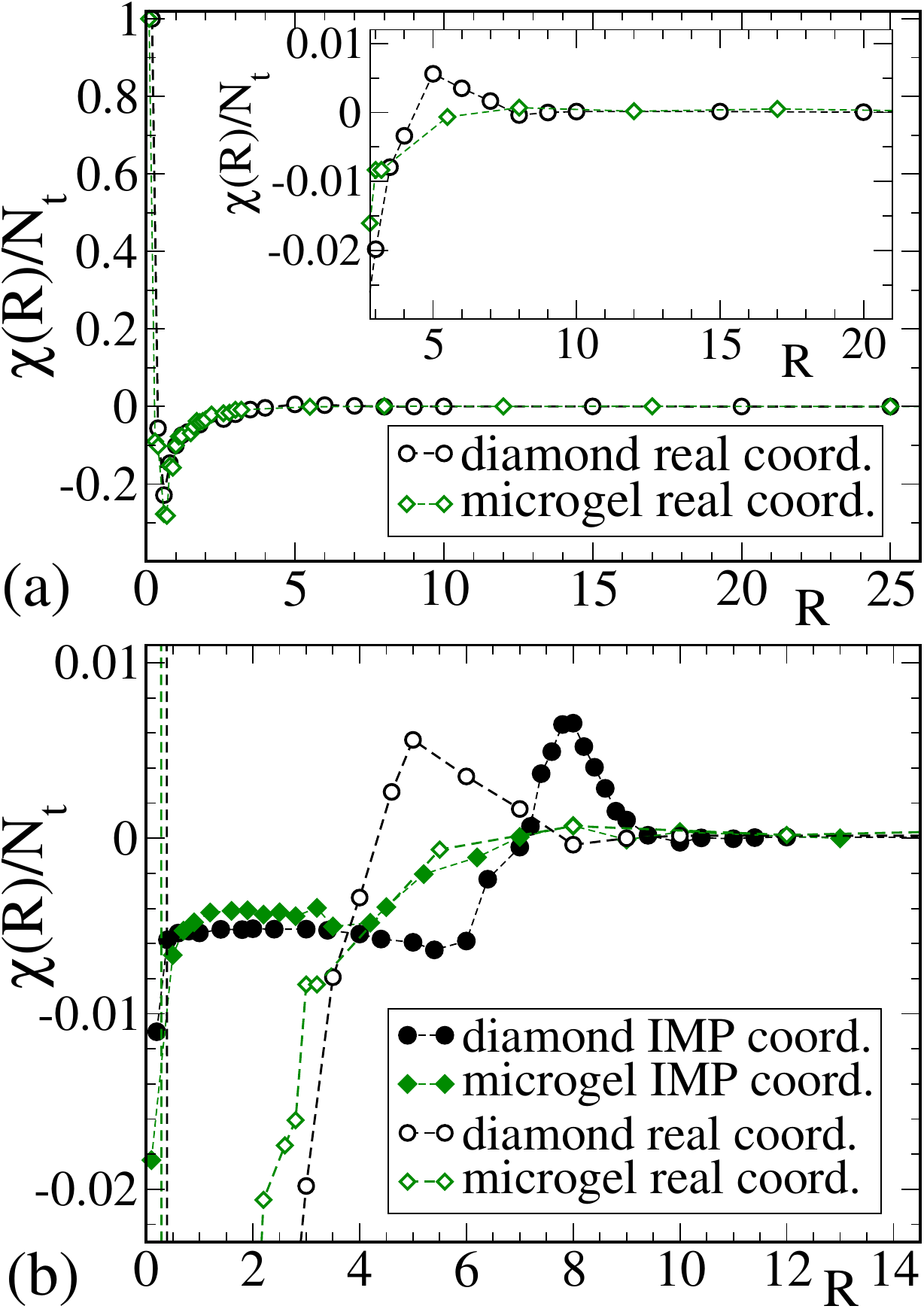}
\caption{Euler characteristic $\chi(R)$ for the diamond and for a middle-size disordered microgel. Panel (a) shows data for the real coordinates. 
The inset highlights the differences between both systems at $R \sim 5$.
Panel (b) compares the $\chi(R)$ of both systems, for the real and the IMP coordinates, in the relevant range of distances
probing the network structure.}
\label{fig:Euler}
\end{figure}

\ref{fig:Euler}a shows $\chi(R)$ for the real coordinates of the diamond (averaged over several run steps) and 
%a middle-size 
the disordered microgel. In the case of the disordered microgel, statistics have been improved by averaging over 5 different configurations and several steps each.  The data are normalized by the total number of monomers, $N_{\rm t}$, of the respective microgel. Therefore, since in the limit $R \rightarrow 0$ the surface $\mathcal{S}(R)$ is constituted by 
$N_{\rm t}$ disconnected spheres, $\chi(0)=1$. By increasing $R$ the merging of the sphere surfaces leads to the formation of rings, which contribute
negatively (as tunnels) to the Euler characteristic. This results in a minimum at a distance slightly above the monomer radius. 
As the sphere radius $R$ goes on growing up, the perforations progressively shrink and vanish, or become enclosed cavities isolated from the empty space. Thus, the positive contributions of the emerging cavities and the cancelation of the negative contributions of the tunnels lead to an increase of the Euler characteristic. For  large $R$ the whole structure merges into a single solid sphere without tunnels or cavities, 
reaching the final value $\chi = 1/N_{\rm t} \approx 4.6 \times 10^{-5}$. A closer inspection at intermediate distances 
reveals different features in the $\chi(R)$ of the diamond and the disordered microgel. Within statistics, a monotonous increase of $\chi(R)$
is observed for the disordered microgel. However, the interplay of emerging cavities and vanishing perforations
leads to a broad maximum at $R \sim 5$ for the diamond microgel (see inset in \ref{fig:Euler}a). In a network with well-defined length scales
(between nearest nodes, next-to-nearest ones, etc), the vanishment of the negative contributions from the tunnels associated to a given length scale 
(and the emergence of cavities originating from the enclosed tunnels) will lead to a steep increase of $\chi$ when $R$ probes such a length scale,
and eventually a maximum in $\chi$. 
Due to the twist of the polymer chains the maximum is not 
%Despite not being 
pronounced, 
%the maximum 
but still reflects the underlying regular diamond network. Due to the absence
of such a regular structure for the disordered microgel, no characteristic length scale can be solved by $\chi(R)$
in its real-coordinate configurations.

\ref{fig:Euler}b displays the Euler characteristics of the corresponding IMP coordinates of the diamond and disordered microgel.  
Data correspond to the average over 5 different IMP configurations. Due to the strong overlap of bonded monomers in the IMP coordinates, the merging of the fully disconnected spheres 
into rings (and the associated minimum in $\chi (R)$) occurs at $R \ll 1$.
The perforations associated to the small rings quickly vanish with growing $R$, resulting in a steep increase from the minimum. This is followed by a well-defined plateau. Because of the rod-like character of the strands in the IMP configurations, the growth of the spheres in this regime 
is much less sensitive to fluctuations and wiggles than in the real-coordinate configurations. Thus, no significant vanishment of perforations or emergence of cavities occurs by increasing $R$ over a significant range, 
and the Euler characteristic is almost unaffected up to $R \approx 6$ and $R \approx 4$ for the
diamond and microgel network, respectively. This feature is fully expected for a regular network, but is a remarkable one for the disordered one.
\ref{fig:Euler}b demonstrates that the IMP construction in combination with the Euler
characteristic is able to detect a certain underlying mesh size in the disordered microgel network.

As expected, vanishment of perforations and emergence of cavities start to arise 
at longer $R$ and, as observed before for the real coordinates, the $\chi (R)$ of the IMP configurations increases. 
In the diamond network the peak for the IMP coordinates is narrower than its counterpart 
for the real ones and is located at larger distances. This feature originates from
the averaging out of the fluctuations and wiggles in the IMP, which shifts the closure 
and vanishment of perforations to larger values of $R$.
In the disordered microgel the Euler characteristics for the real and IMP coordinates 
show negligible differences for $R > 4$. No further characteristic length scales of the disordered network are probed by increasing
the sphere radius $R$, and $\chi(R)$ grows smoothly until its final value $\chi = 1/N_{\rm t}$.

\section{Deswelling behavior}\label{sec4}

\subsection{Volume phase transition}

\begin{figure}[t!]
\centering
\includegraphics[width=.56\textwidth]{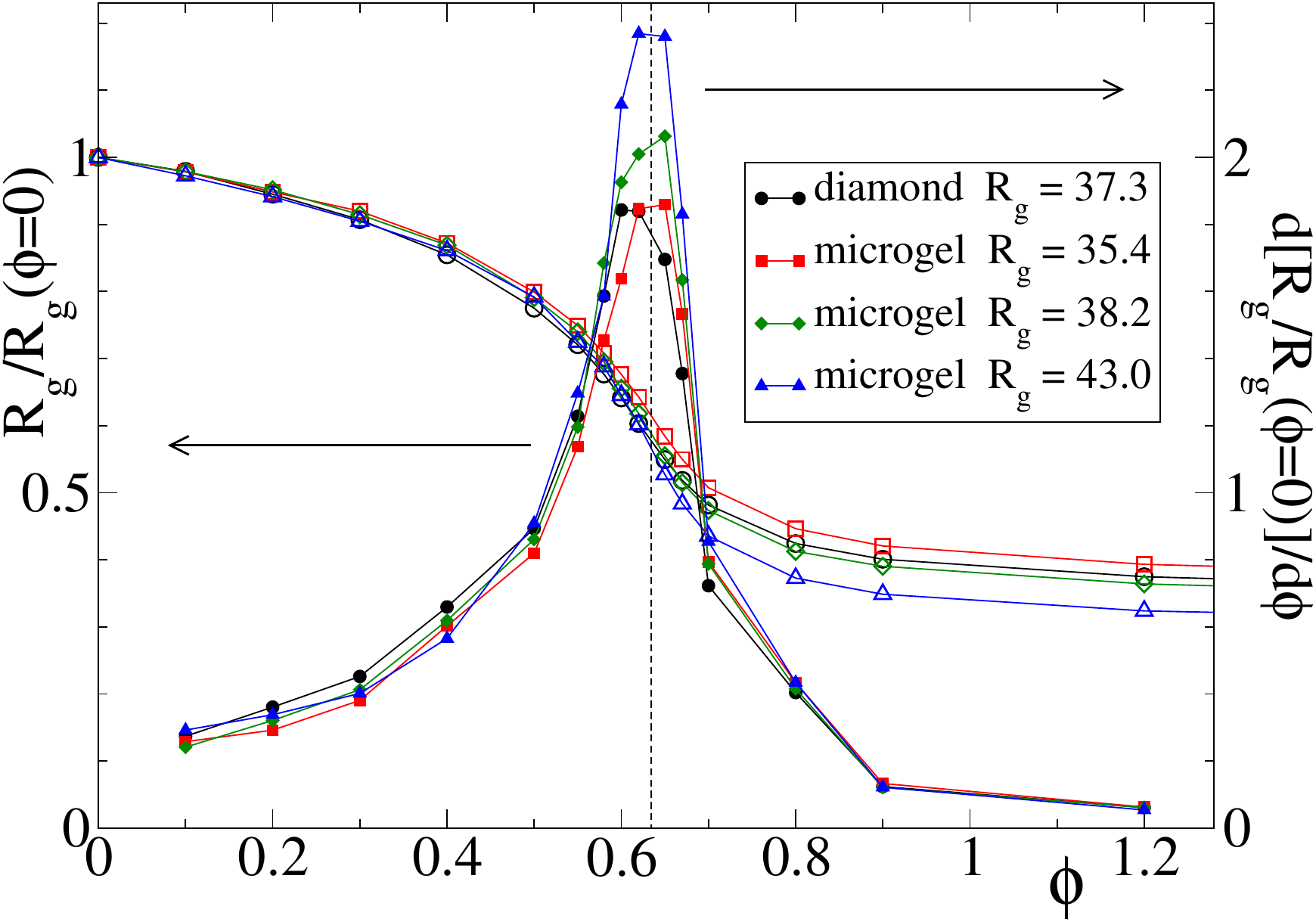}
\caption{Open symbols: Radius of gyration of the diamond and the small, middle and large disordered microgels
vs. the solvent quality parameter $\phi$. Values are normalized by the radius at ideal good solvent ($\phi=0$). Filled symbols: Derivatives of the former data sets. The vertical dashed line indicates
the approximate position of the VPT.}
\label{fig:rgvpt} 
\end{figure}

\ref{fig:rgvpt} shows the radius of gyration $R_{\rm g}$ vs. the solvent quality parameter $\phi$ for the diamond microgel and for the three representative (small, middle and large) disordered microgels.
All data are normalized by the radius of gyration at ideal good solvent ($\phi = 0$). 
As expected, the microgel shrinks by worsening the solvent quality (increasing $\phi$). 
A steep decrease of the microgel size is observed in the range $0.5 < \phi < 0.8$, followed by an ultimate plateau that corresponds to the dense
collapsed state in bad solvent (large $\phi)$. An accurate determination
of the volume phase transition (VPT) point would require to compute effective pair interactions 
(see e.g. Ref.~\cite{Narros2013}).  Still, a reliable estimation can be obtained by analyzing
the derivatives $d[R_{\rm g}(\phi)/R_{\rm g}(0)]/d\phi$ of the normalized radii of gyration. 
As can be seen in \ref{fig:rgvpt} the derivatives have a sharp peak (maximum slope in $R_{\rm g}(\phi)$) at $\phi \approx \phi_{\rm VPT}\approx 0.63$, 
which can be identified as the VPT point of the microgel. 
No significant differences in the peak positions
are found within statistics, indicating that the  VPT point is at most very weakly dependent on the specific microstructure of the microgel,
even if our results include the limits of a regular and a highly disordered network. Our results are in agreement 
with experiments in microgels with different degree of internal inhomogeneity \cite{Habicht2014}, and extend them to include the limit of regular networks.

\begin{figure}[t!]
\centering
\includegraphics[width=.51\textwidth]{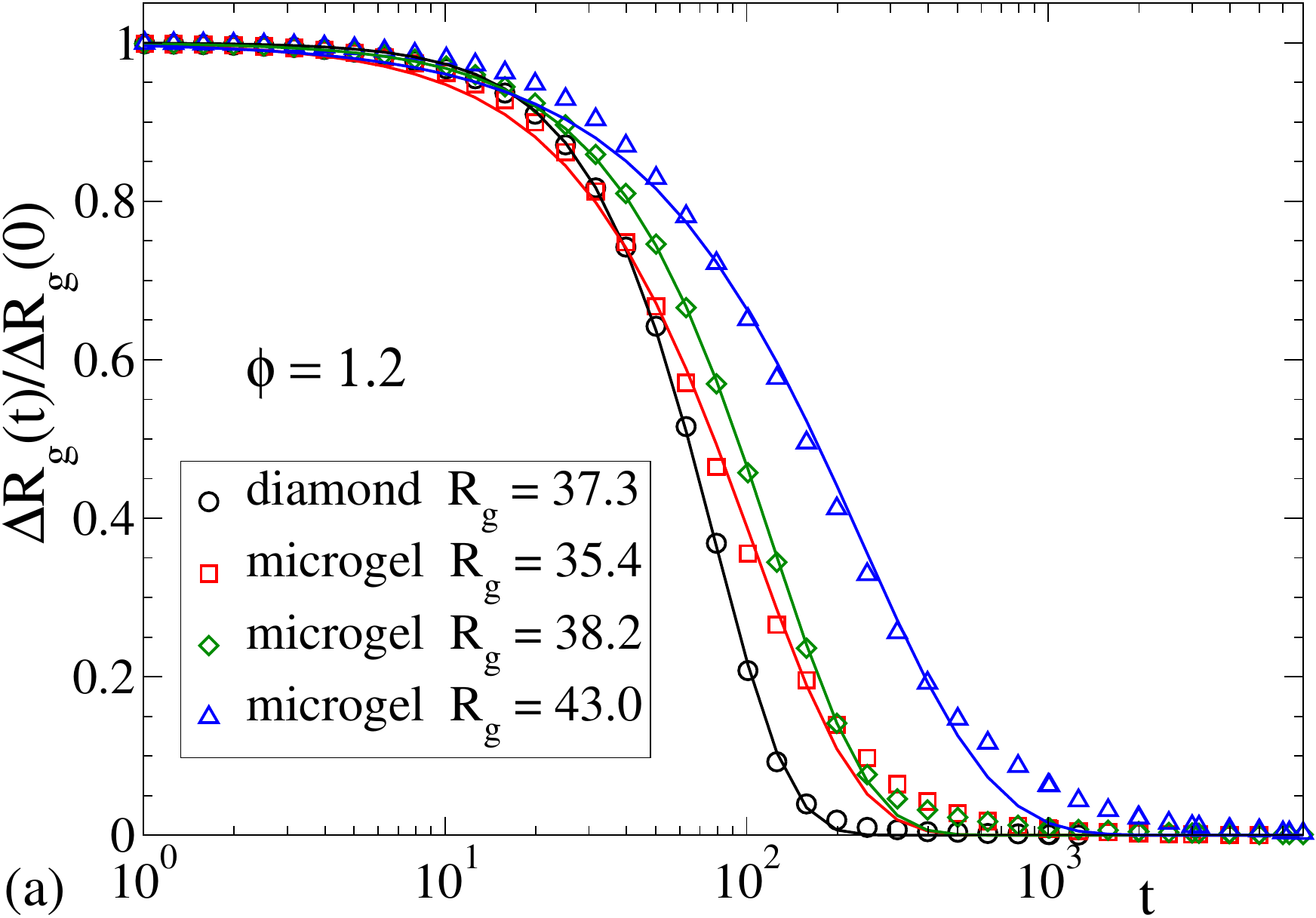}\vspace{0.3cm}
\includegraphics[width=.51\textwidth]{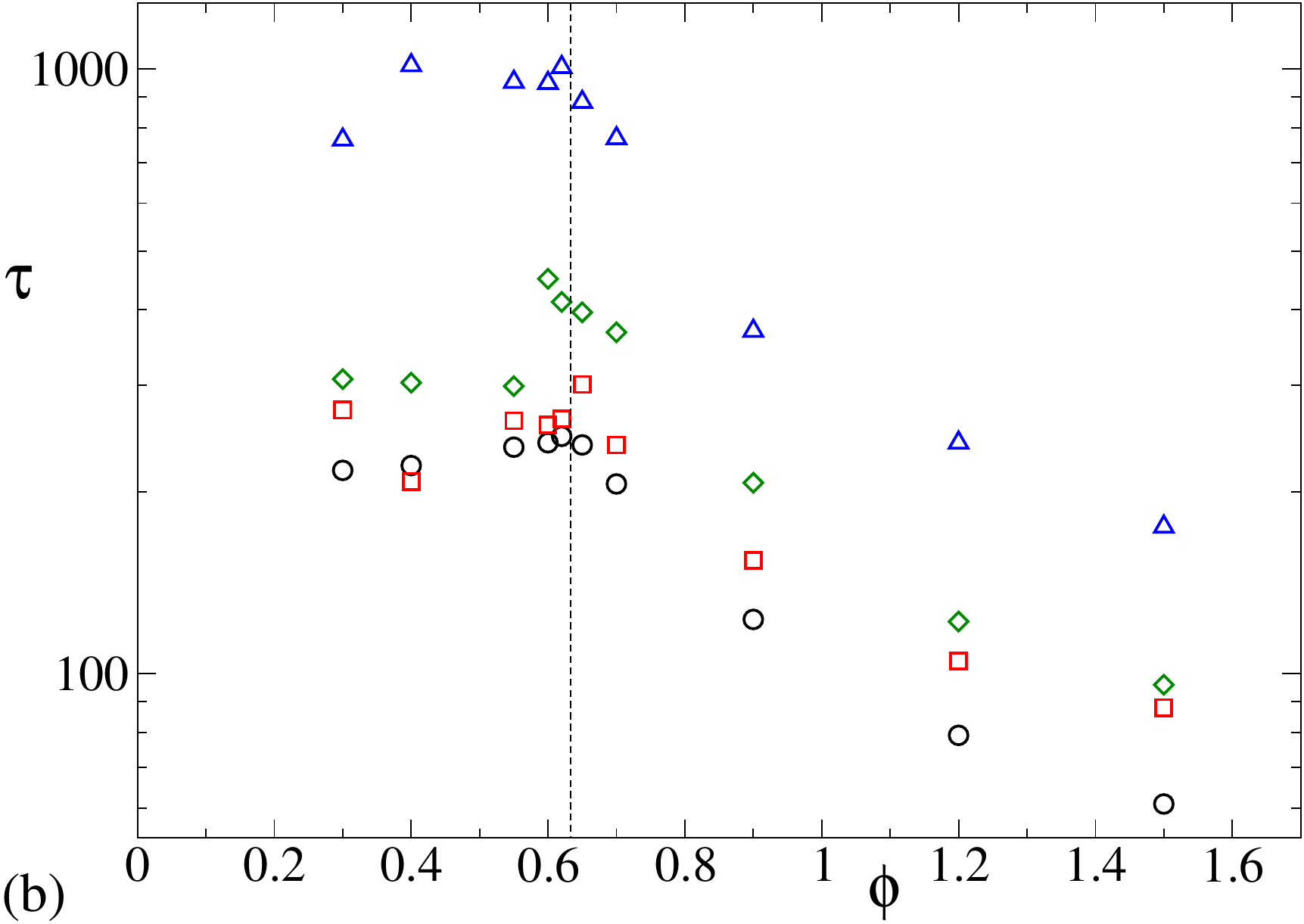}\vspace{0.3cm}
\includegraphics[width=.51\textwidth]{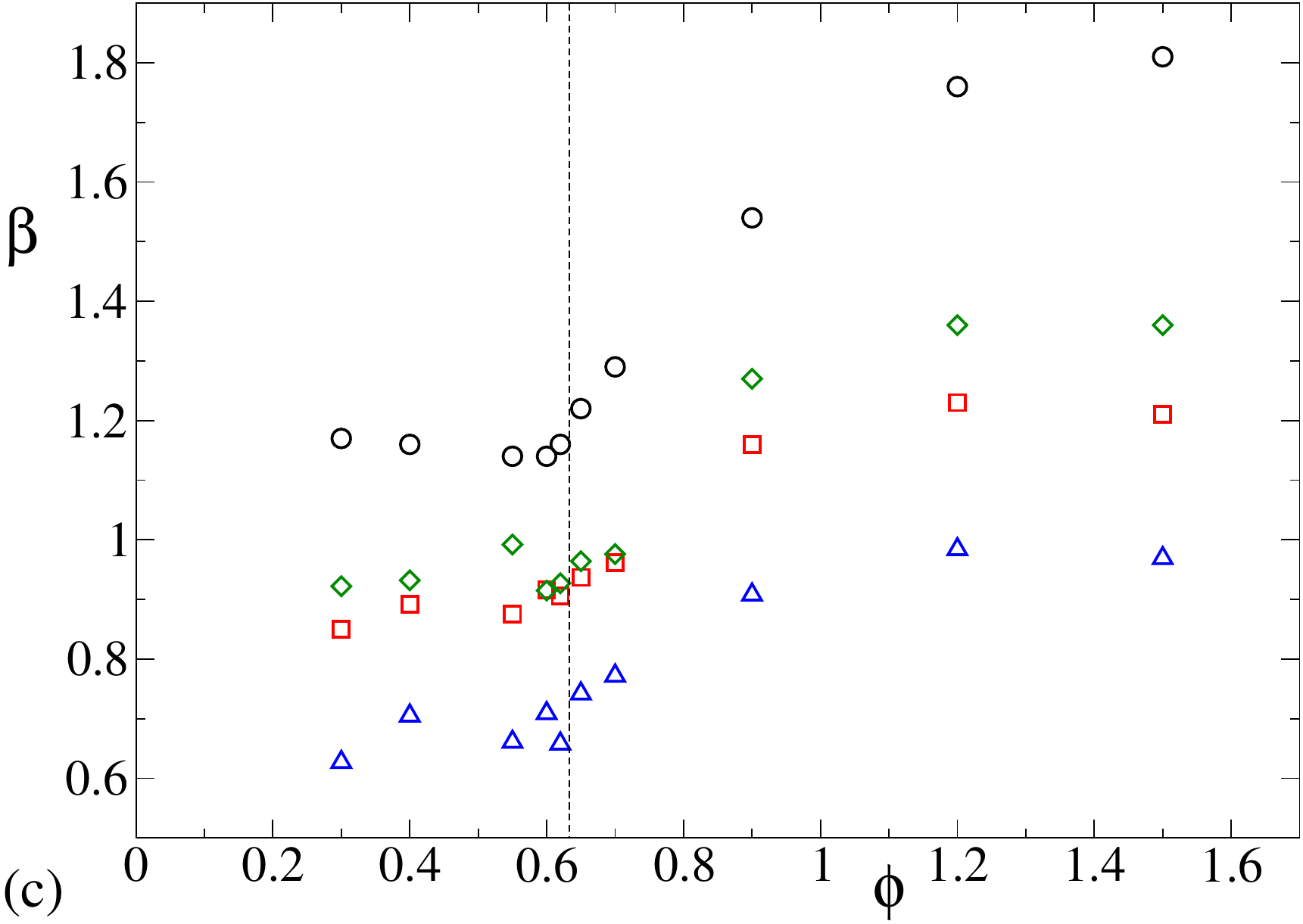}
\caption{(a): Time dependence of the ratio $\Delta R_{\rm g}(t)/\Delta R_{\rm g}(0)$, with
$\Delta R_{\rm g}(t) = R_{\rm g}(t)-R_{\rm g}(\infty)$,
during the collapse at $\phi=1.2$ of the diamond and the small, middle and large disordered microgels.
Symbols are simulation data. Lines are fits to generalized exponentials $\sim \exp[-(t/\tau)^{\beta}]$.
Panels (b) and (c) show the relaxation times $\tau$ and exponents $\beta$ obtained for
the collapse at the investigated $\phi$-values. The vertical dashed line indicates
the approximate position of the VPT point.}
\label{fig:fitskww} 
\end{figure}

\begin{figure}[t!]
\centering
\includegraphics[width=.51\textwidth]{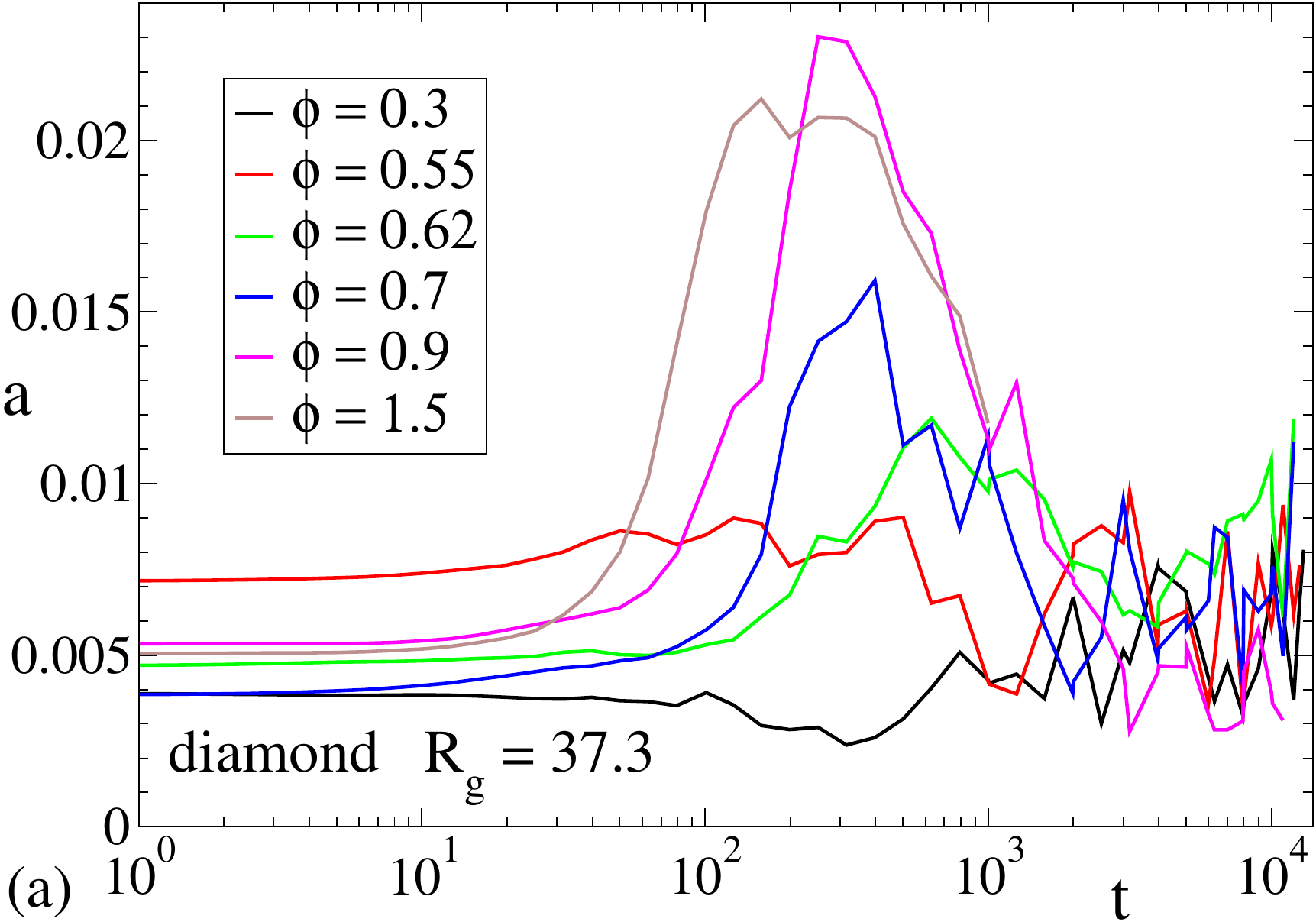}\vspace{0.3cm}
\includegraphics[width=.51\textwidth]{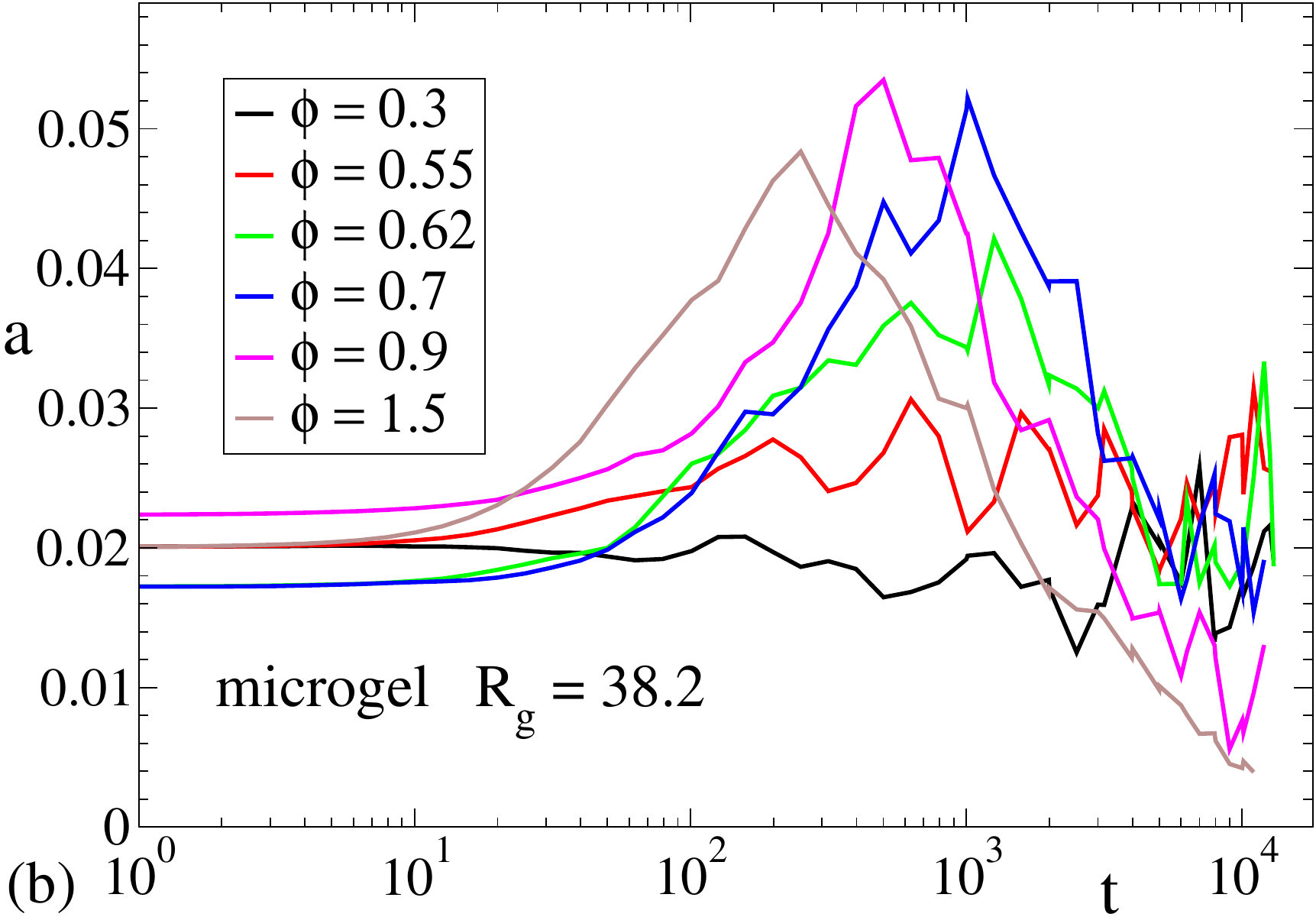}\vspace{0.3cm}
\includegraphics[width=.51\textwidth]{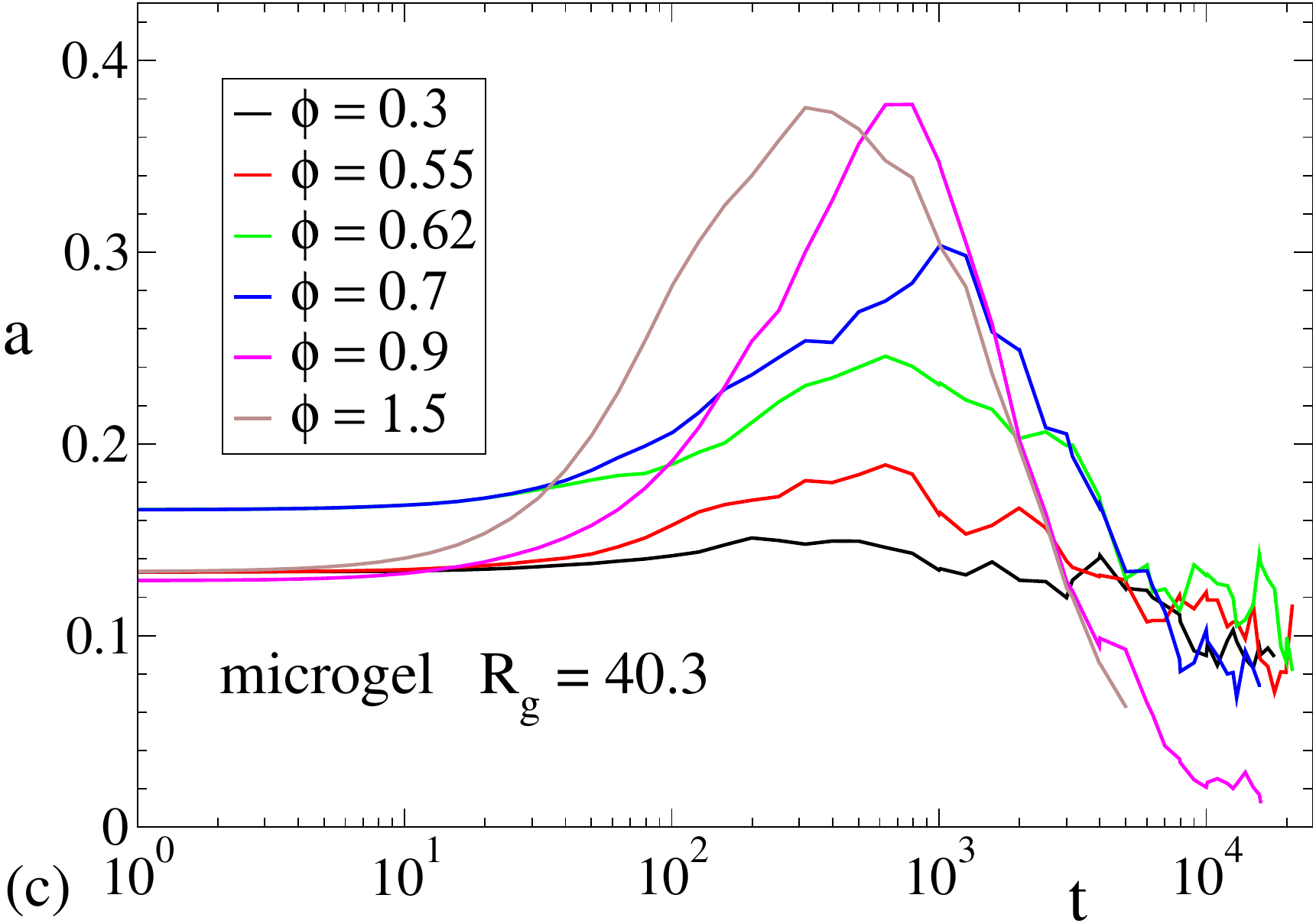}
\caption{Time dependence of the asphericity parameter during the collapse
of the diamond and two representative disordered microgels, at different $\phi$-values.}
\label{fig:asphercollapse} 
\end{figure}

Whereas no significant differences are found in the deswelling thermodynamics, the kinetics are strongly dependent on the microgel microstructure.
\ref{fig:fitskww}a shows the evolution of the ratio $\Delta R_{\rm g}(t)/\Delta R_{\rm g}(0)$ during the collapse of 
the microgel at $\phi=1.2$ (well beyond the VPT point). The quantity $\Delta R_{\rm g}(t)$ is defined as $\Delta R_{\rm g}(t) = R_{\rm g}(t)-R_{\rm g}(\infty)$,
with $R_{\rm g}(\infty)$ the final radius of the collapsed globule. By construction $\Delta R_{\rm g}(t)/\Delta R_{\rm g}(0)$
decays from 1 to 0. As can be seen in \ref{fig:fitskww}a, the time scale as well as the shape of the decay depends on the specific
microgel topology. We have fitted the simulation results to a generalized exponential $\Delta R_{\rm g}(t)/\Delta R_{\rm g}(0) \sim \exp[-(t/\tau)^{\beta}]$
(solid curves in \ref{fig:fitskww}a). The relaxation times $\tau$ and exponents $\beta$ obtained at the investigated values of $\phi$ 
are displayed in \ref{fig:fitskww}b and \ref{fig:fitskww}c, respectively. Both $\tau$ and $\beta$ are approximately $\phi$-independent 
in the swollen state ($\phi < \phi_{\rm VPT}$).
A rather different behavior is found when the microgels are driven to the collapsed state ($\phi > \phi_{\rm VPT}$). 
Deeper quenches (higher $\phi$) lead to faster (lower $\tau$)
and less stretched (higher $\beta$) kinetics of deswelling. The $\beta$-exponents seem to reach an asymptotic regime for deep quenches.
The most spherical microgel --- the diamond-like one --- shows the fastest and least stretched decay. 
The time scale and the stretching increase with the asphericity, as can be observed in the data sets of  \ref{fig:fitskww} 
for the disordered microgels (note again that $a$ increases with $R_g$ in these selected systems). It is worth stressing that it is $a$ and not $R_{\rm g}$ the key factor for the former observation. Indeed, the diamond microgel ($R_{\rm g} = 37.3$) is bigger, but {\it more spherical}, than the small disordered one ($R_{\rm g} = 35.4$), and shows a faster and less stretched decay (\ref{fig:fitskww}a).  
As can be seen in \ref{fig:fitskww}c, the relaxation to the collapsed state is super-exponential ($\beta >1$), except for the most aspherical disordered microgels (still, $\beta \rightarrow 1$ for deep quenches). Super-exponential behavior is also found in the coil-to-globule transition of a linear polymer chain \cite{Majumder2017,Christiansen2017}, with an exponent $\beta \sim 1.2$ for deep quenches. This is similar to those of the disordered microgels ($ 1 < \beta < 1.4$) and still far from that found for the diamond microgel $\beta \approx 1.8$.

A broader decay in \ref{fig:fitskww}a (lower $\beta$) suggests a more heterogeneous collapse of the different regions of the microgel.
This effect can be quantified by analyzing the evolution of the macromolecular shape at different stages of the collapse.
\ref{fig:asphercollapse} shows the time evolution of the asphericities after the quench at different values of $\phi$.
The obtained results are consistent with those of Ref.~\cite{Kamerlin2016} and extend them by investigating the effect of the specific network microstructure.
In the swollen state ($\phi < \phi_{\rm VPT}$) the initial asphericity ($t=0$) is essentially unaffected, indicating that the soft quench
just results in a weak distorsion of the conformations. For $\phi > \phi_{\rm VPT}$ the asphericity shows a peak at intermediate times followed by 
the expected decay to $a \rightarrow 0$ corresponding to the final (spherical) collapsed state. The distortion with respect to the spherical shape
at intermediate times is stronger as the intrinsic asphericity of the microgel ($a$ at $\phi=0$) increases, indicating a more heterogeneous collapse.
Still, this distorsion is relevant only for the microgels that are already significantly aspherical in the swollen state 
(see the peak in \ref{fig:asphercollapse}c at $a \sim 0.4$, starting from $a \sim 0.1$). \ref{fig:asphercollapse} also
shows that the deeper the quench, the more heterogenous (stronger deviation from sphericity) the collapse is.

\subsection{Coarsening kinetics}

\begin{figure}[t!]
\centering
\includegraphics[width=.325\textwidth]{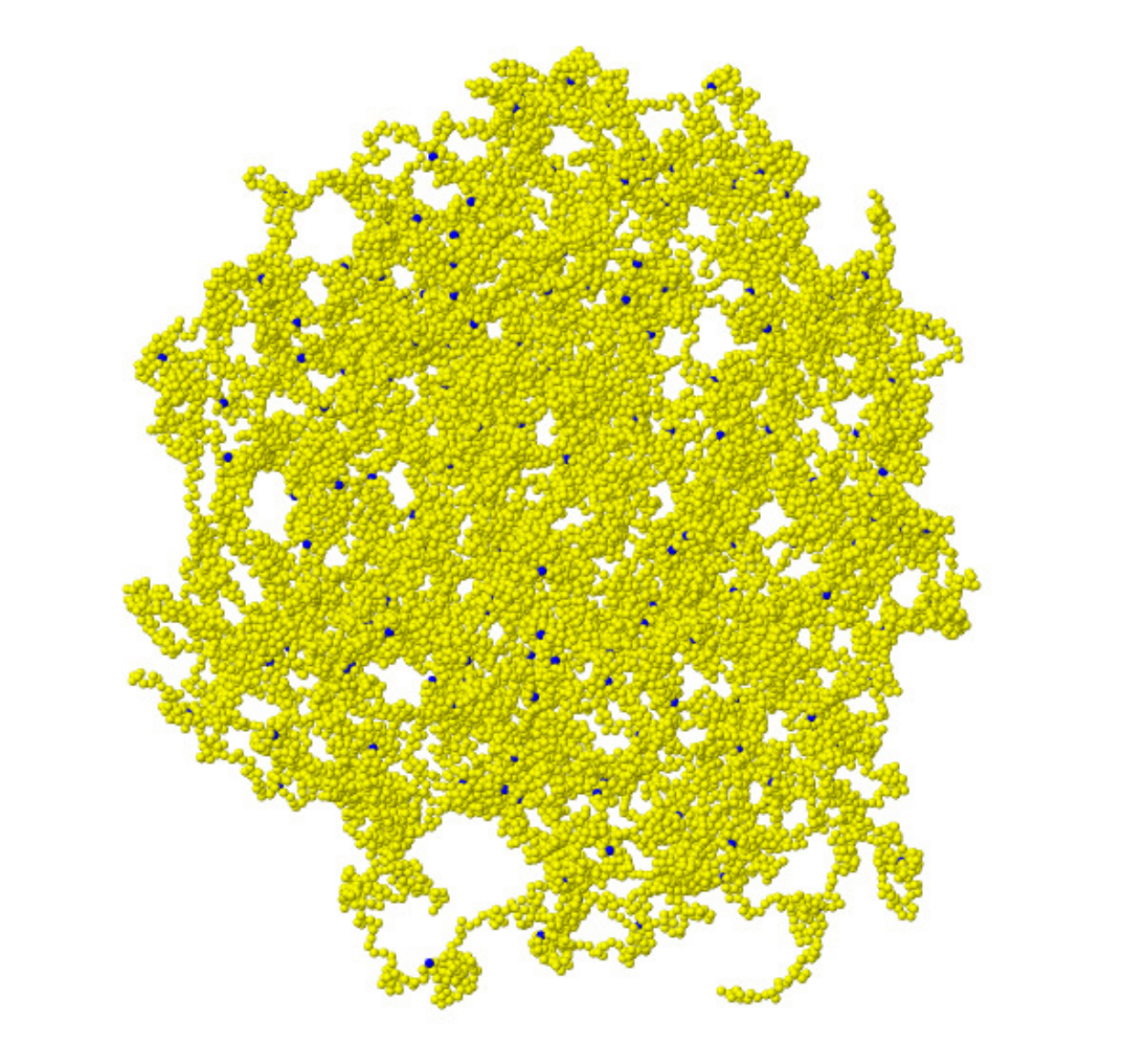}\hspace{0.01cm}
\includegraphics[width=.325\textwidth]{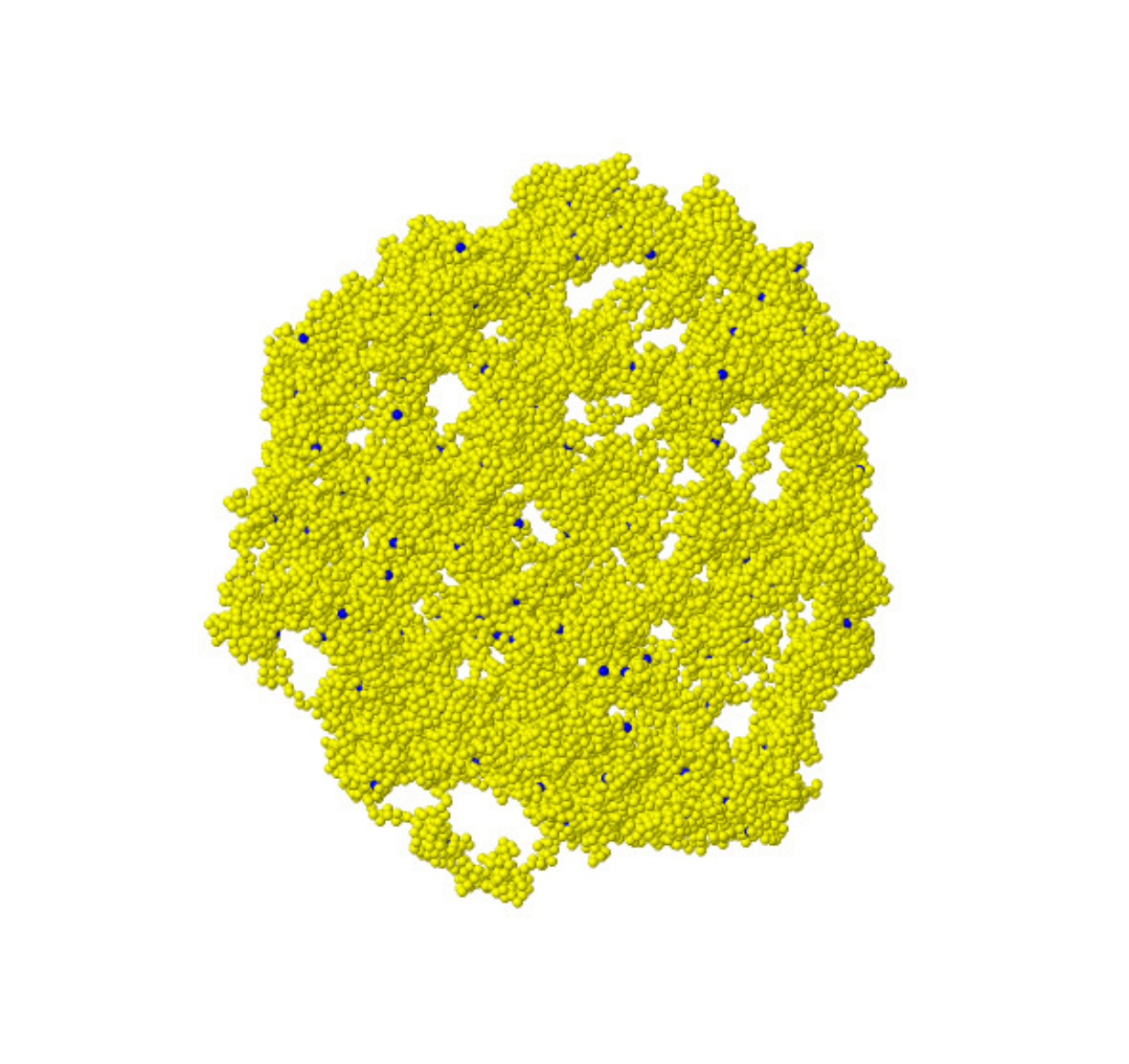}\hspace{0.01cm}
\includegraphics[width=.325\textwidth]{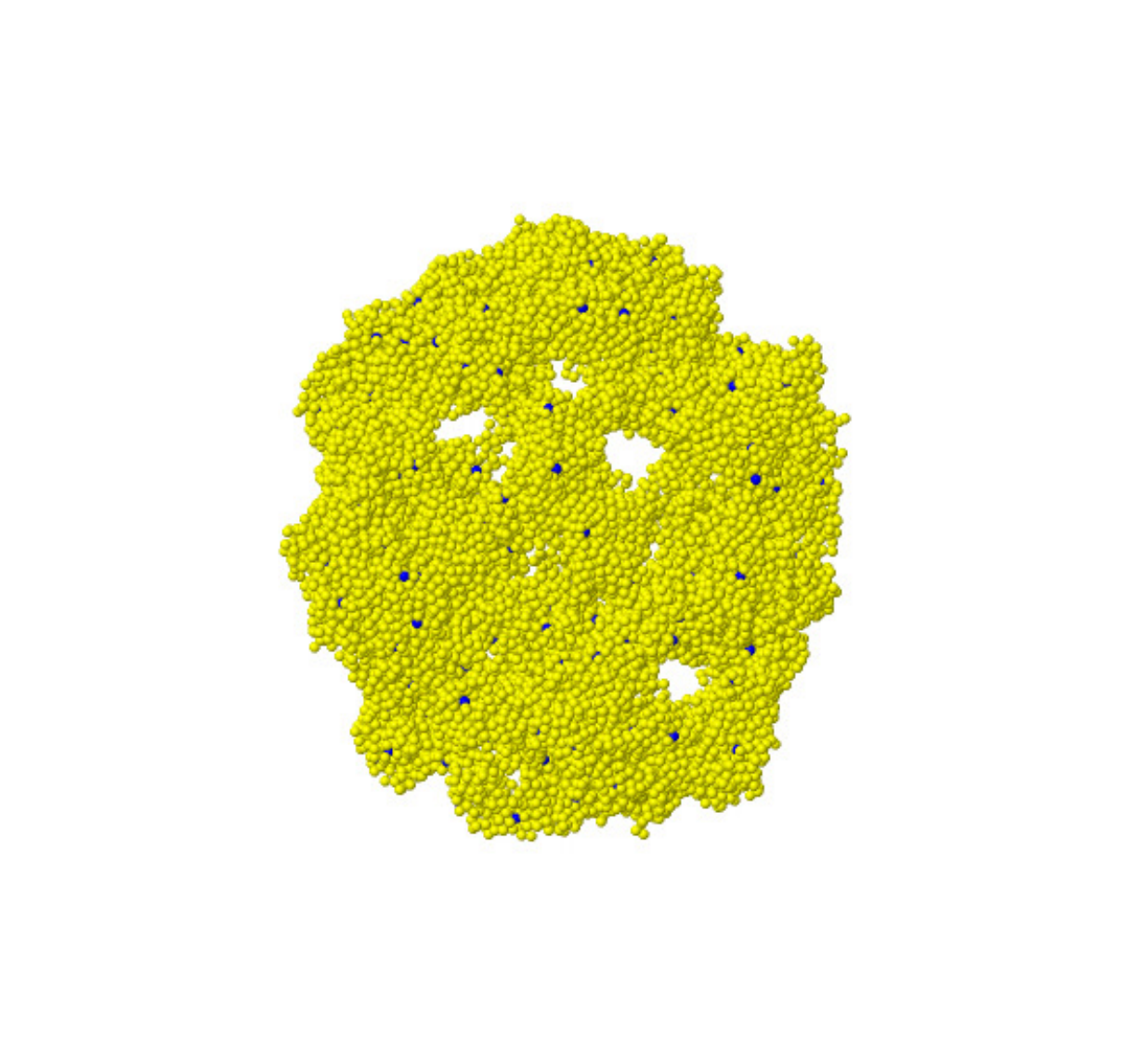}\vspace{0.01cm}
\includegraphics[width=.325\textwidth]{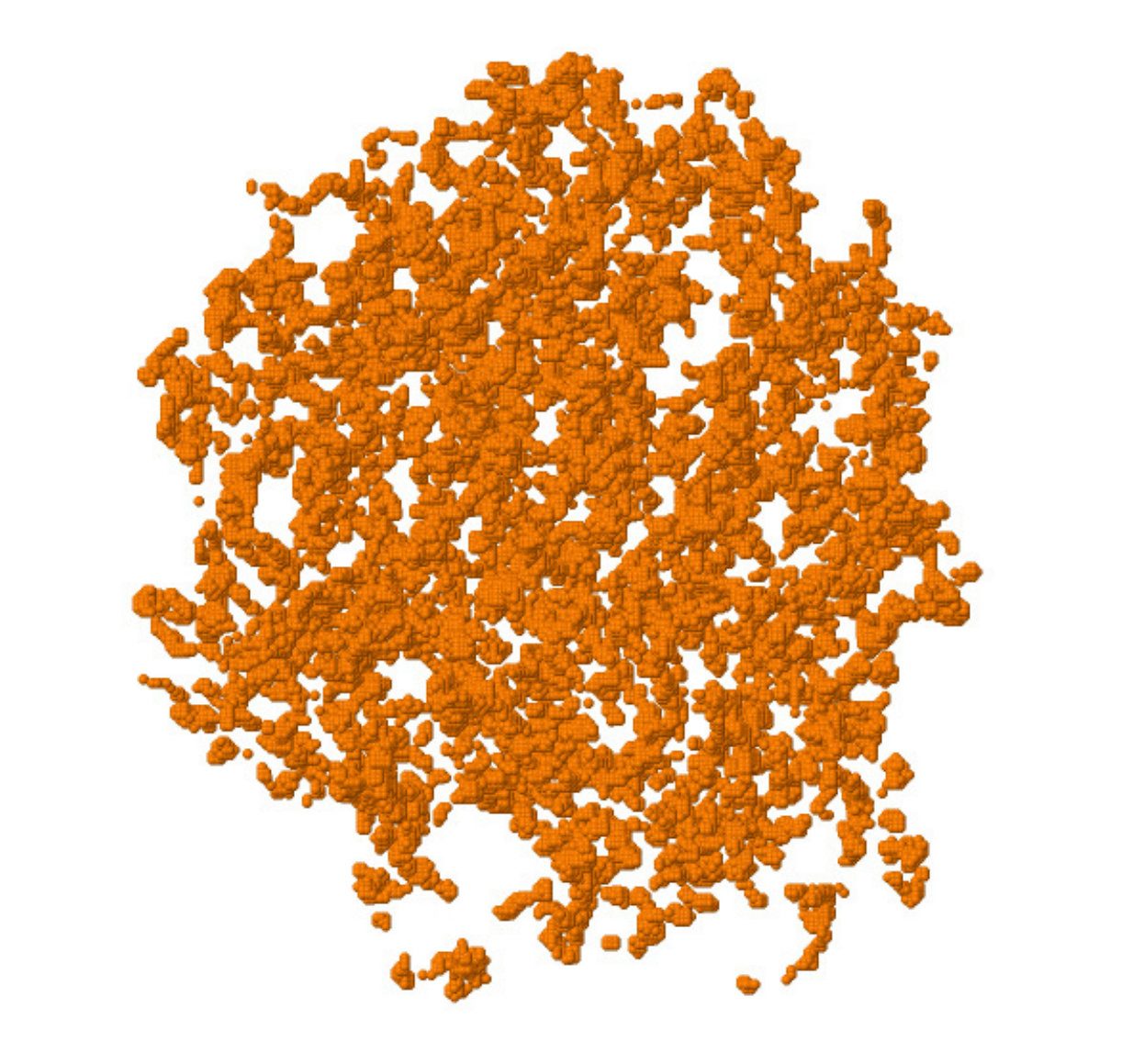}\hspace{0.01cm}
\includegraphics[width=.325\textwidth]{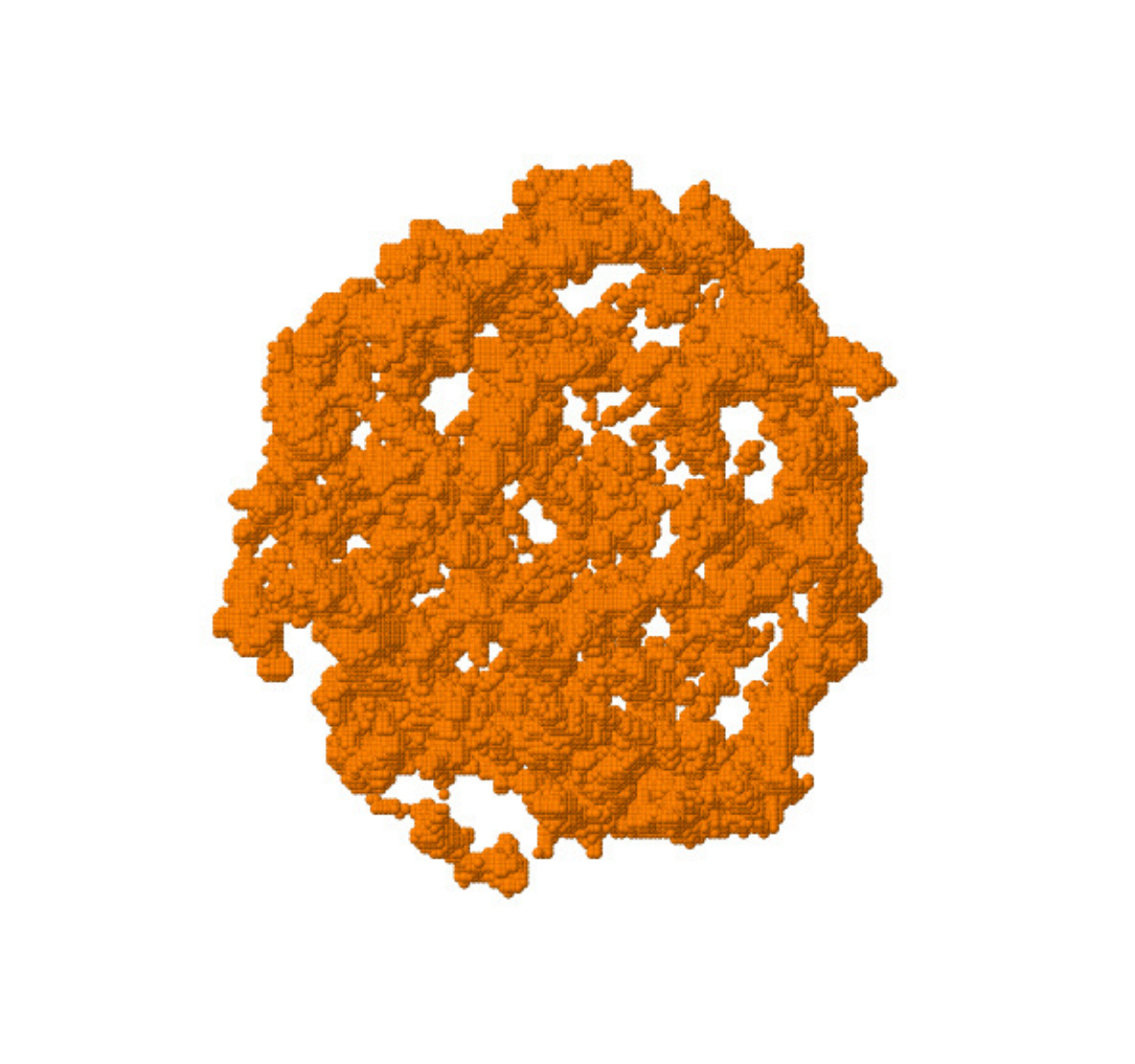}\hspace{0.01cm}
\includegraphics[width=.325\textwidth]{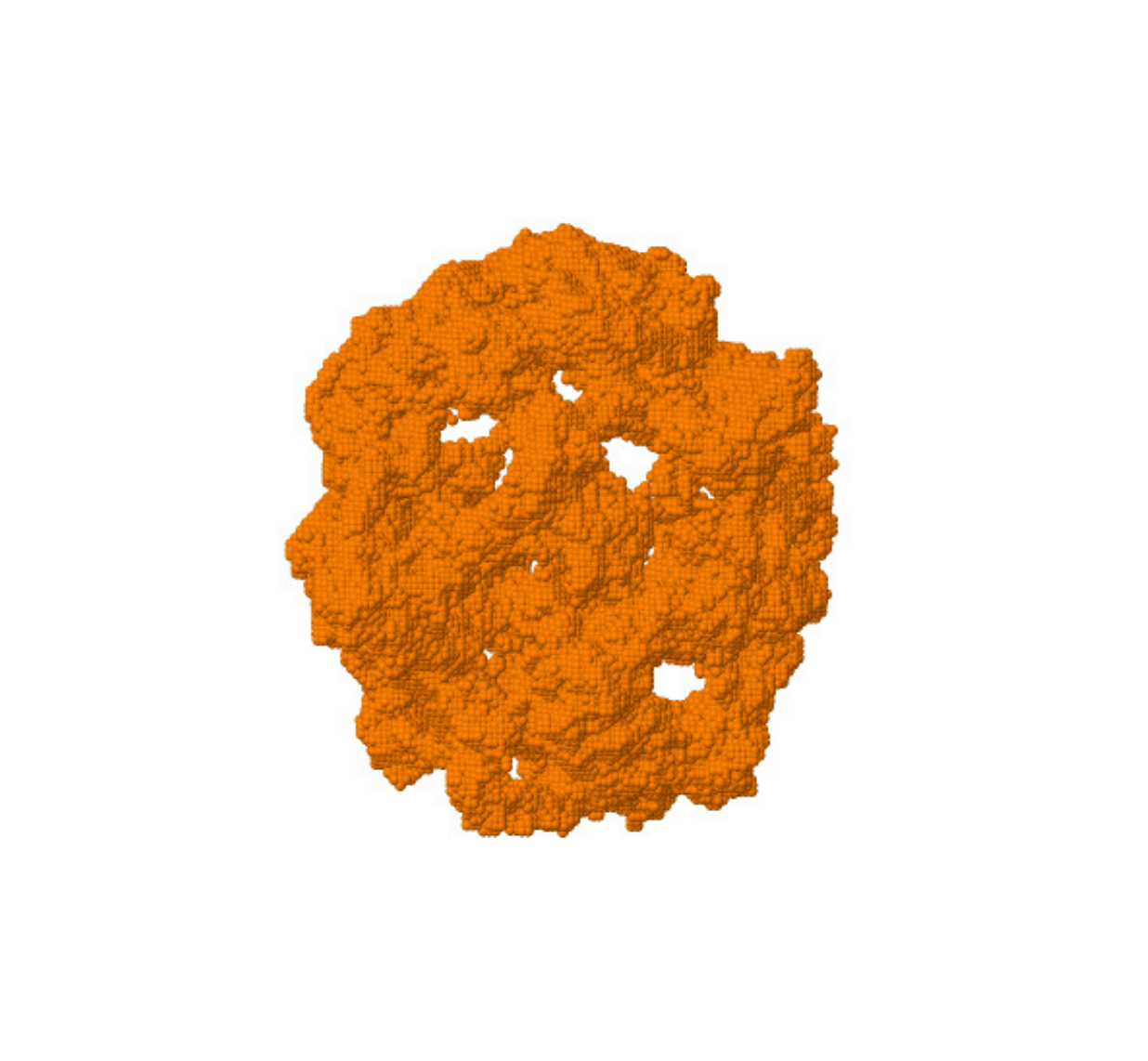}
\caption{Snapshots of real (top) and density field (bottom) coordinates of the diamond microgel
during the collapse at $\phi =1.2$. Blue beads are cross-linked sites. 
The times are, from left to right, $t= 5$, 63 and 101.}
\label{fig:snaps-field-diamond} 
\end{figure}

\begin{figure}[t!]
\centering
\includegraphics[width=.325\textwidth]{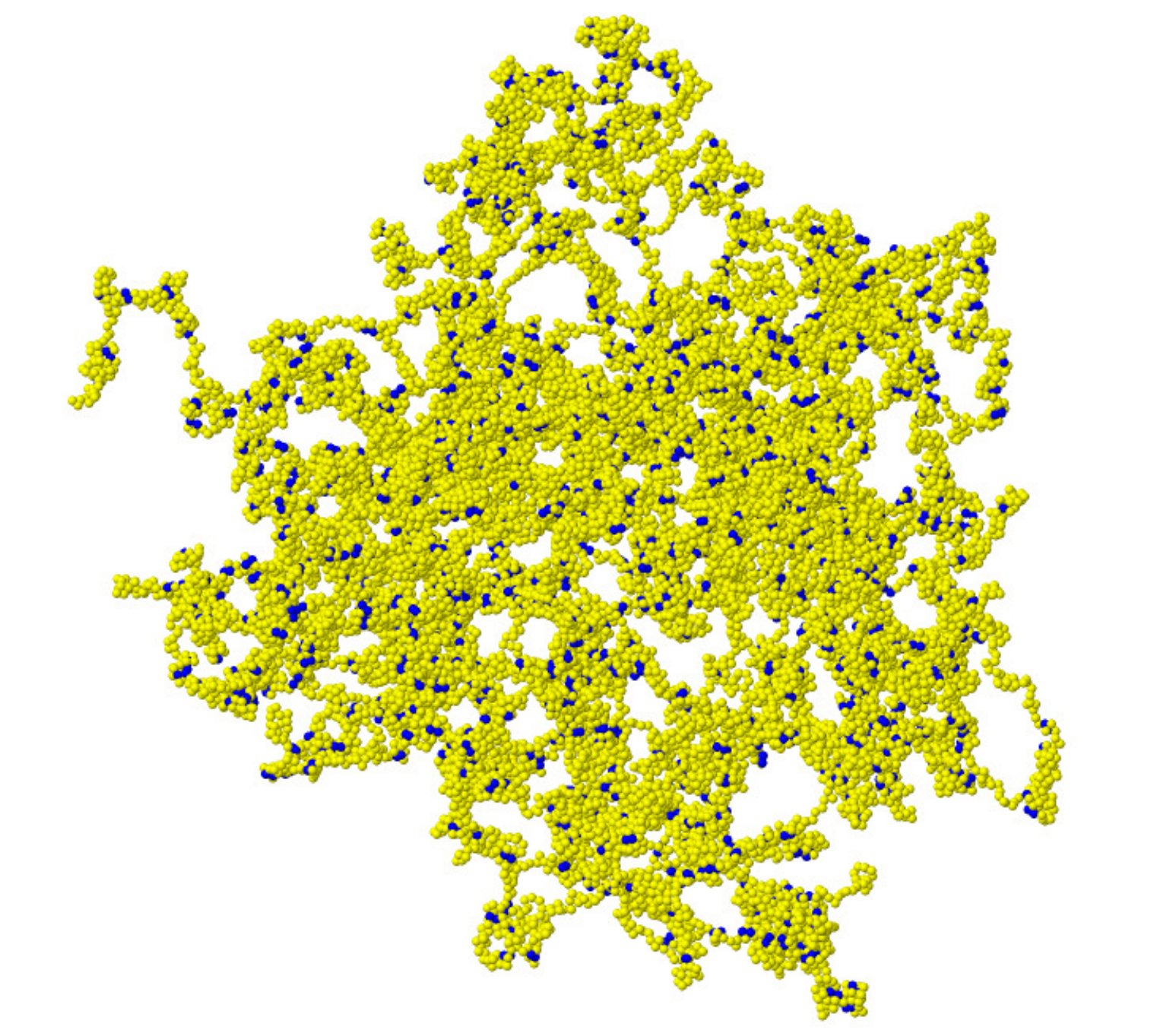}\hspace{0.01cm}
\includegraphics[width=.325\textwidth]{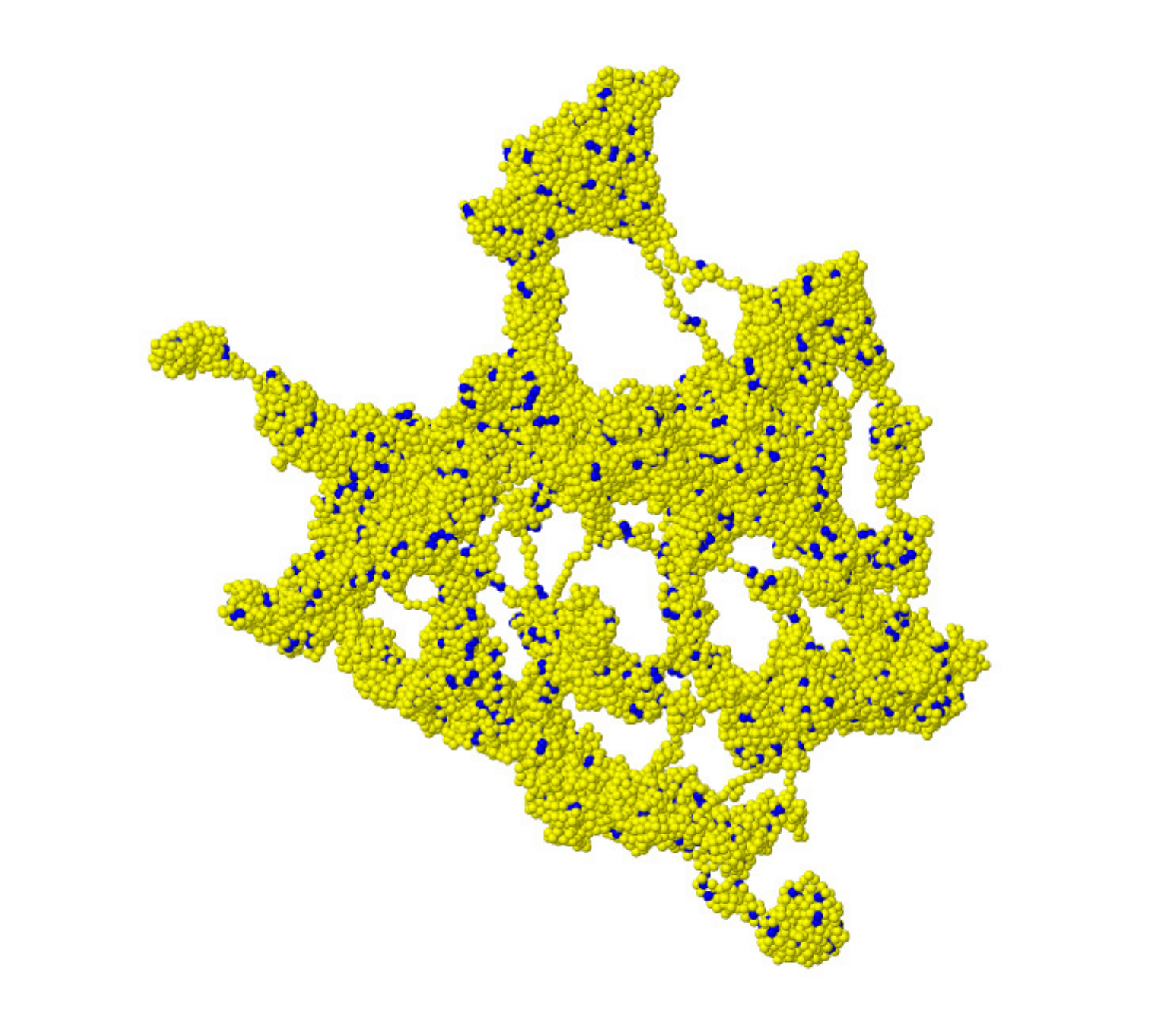}\hspace{0.01cm}
\includegraphics[width=.325\textwidth]{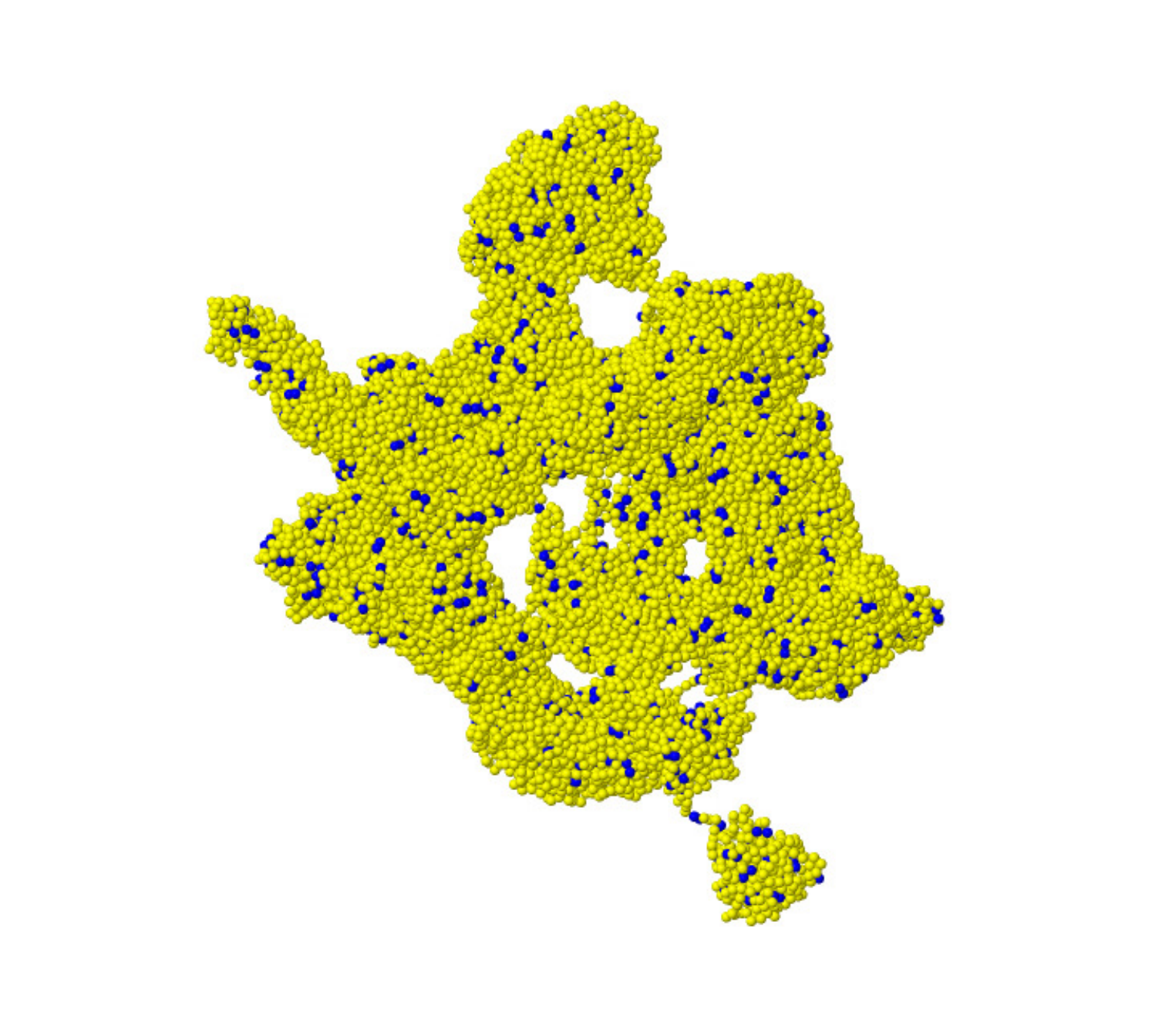}\vspace{0.01cm}
\includegraphics[width=.325\textwidth]{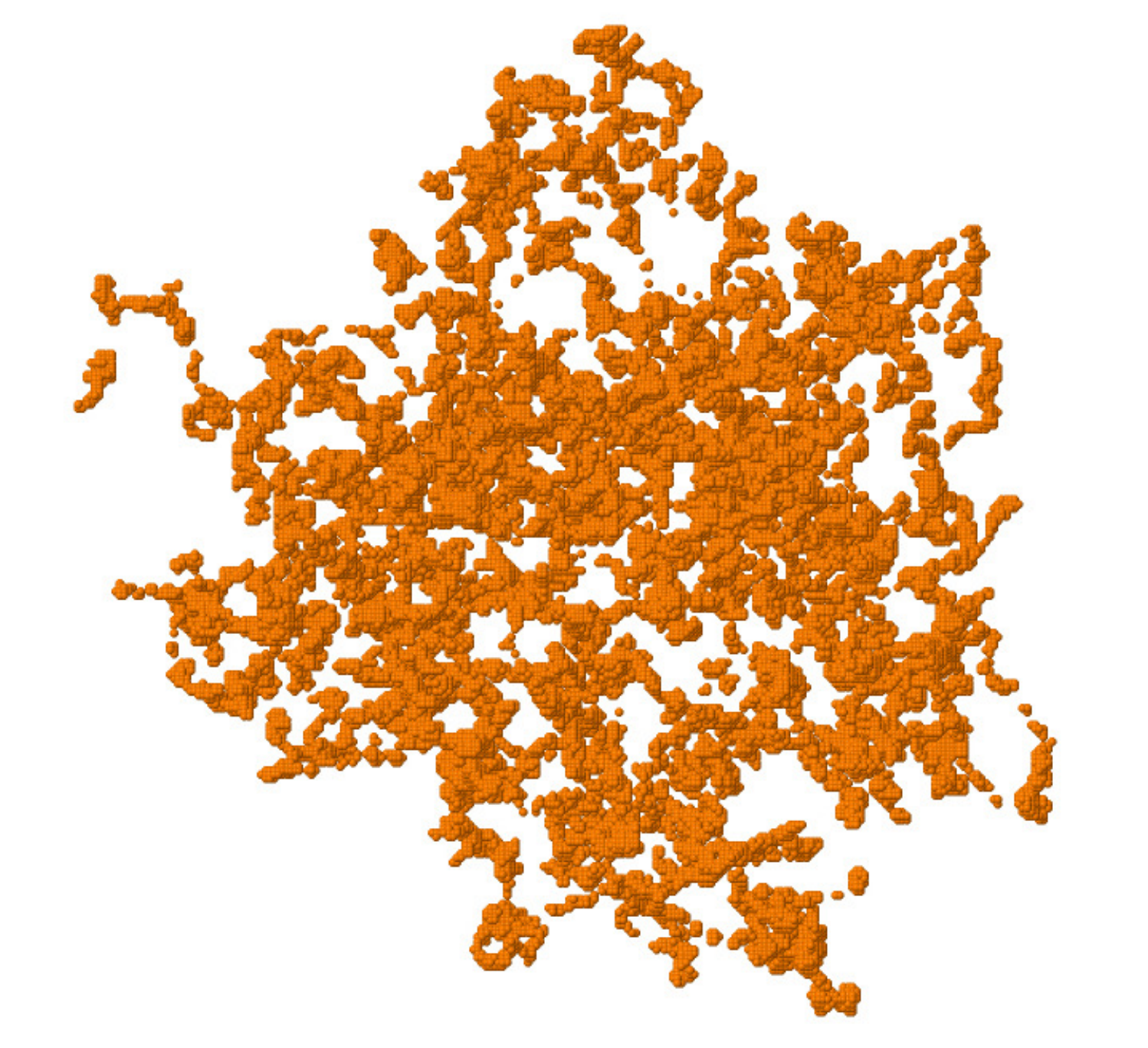}\hspace{0.01cm}
\includegraphics[width=.325\textwidth]{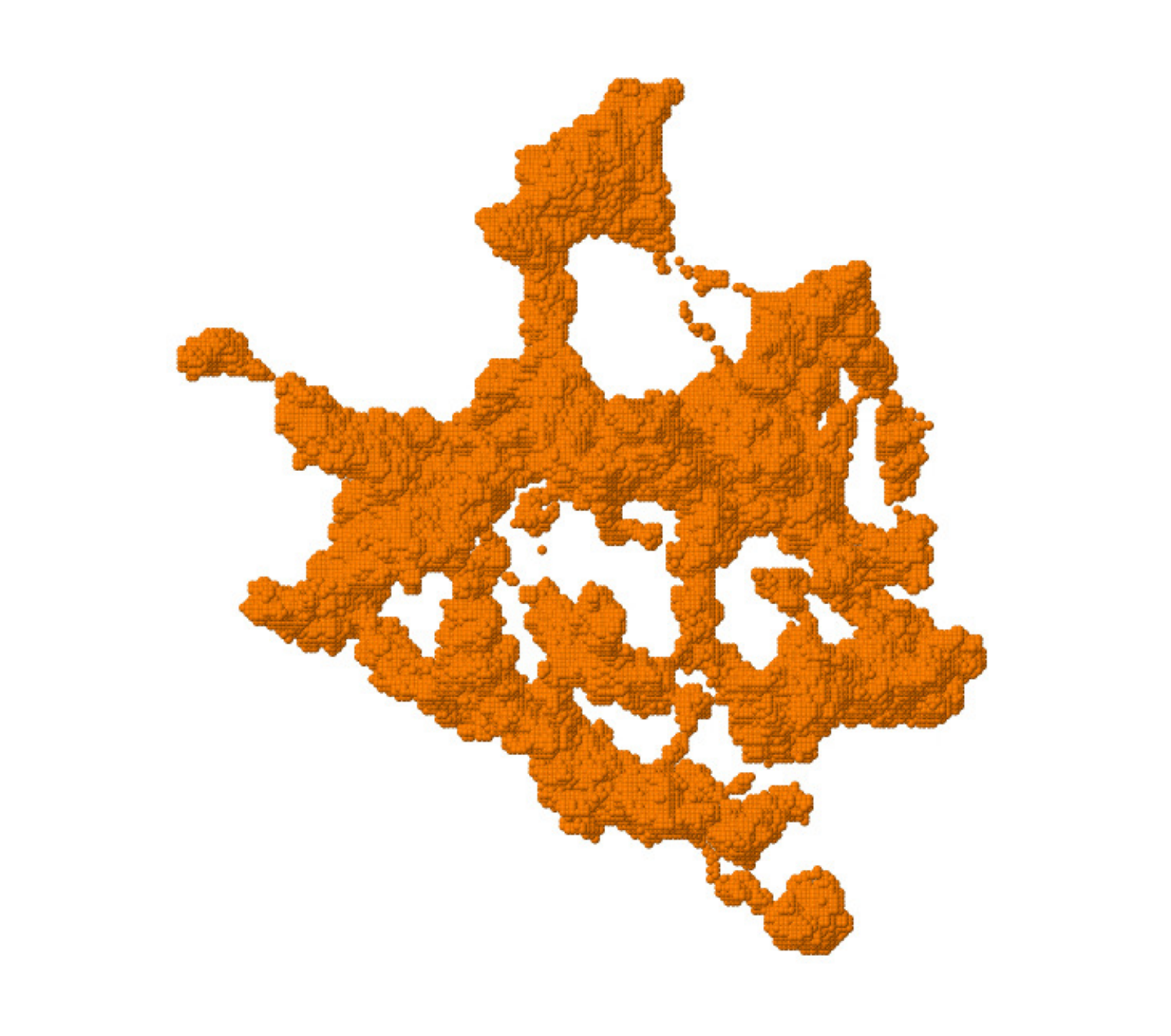}\hspace{0.01cm}
\includegraphics[width=.325\textwidth]{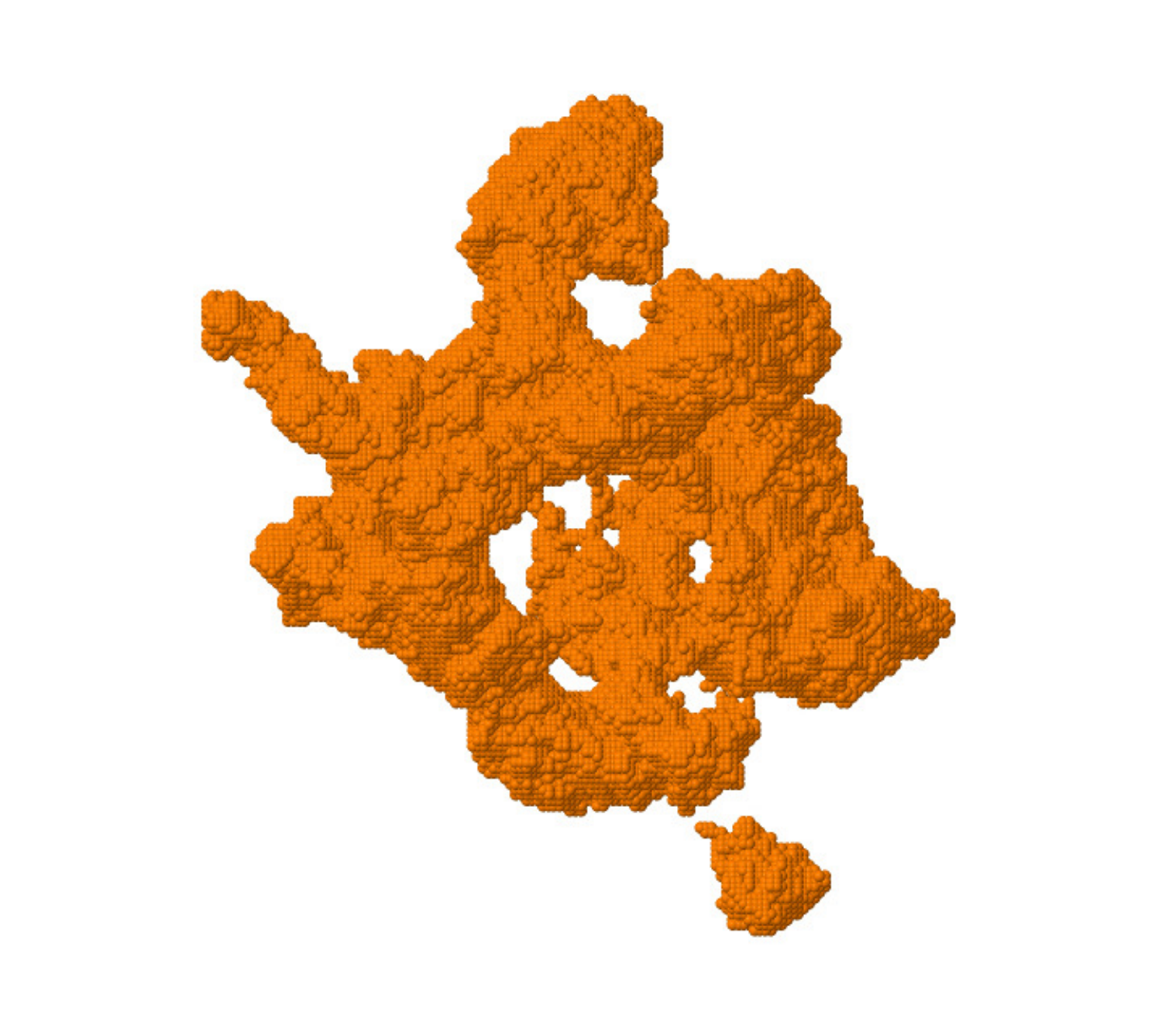}
\caption{Snapshots of real (top) and field (bottom) coordinates of a middle-size disordered microgel
($R_{\rm g} = 38.2$) during the collapse at $\phi =1.2$. Blue beads are cross-linked sites.
The times are, from left to right, $t= 5$, 79 and 158.
}
\label{fig:snaps-field-mic} 
\end{figure}

As the global conformation of the microgel collapses in bad solvent, the monomers form local globules that progressively 
merge into interconnected larger globules, and the internal holes shrink (see top panels in \ref{fig:snaps-field-diamond,fig:snaps-field-mic}).  
This {\it coarsening} process finishes in a dense fully collapsed spherical globule (not shown in \ref{fig:snaps-field-diamond,fig:snaps-field-mic}).
We characterize the coarsening kinetics of the microgel structure during the collapse by following a procedure
similar to that proposed by Testard {\it et al.}  in the context of liquid-gas phase separation \cite{Testard2011,Testard2014}.
Thus, the collapsing microgel can be seen as a coarsening `bicontinuous' structure of connected `empty' and `filled' domains, which are identified
according to their `low' or `high' local density, respectively. To facilitate the classification of the domains, the real-coordinate configuration of the microgel is substituted by a smoothed coarse-grained density field that is constructed as follows.
First we divide the space into cubic cells of side $\delta$. We define the local density at each cell
as $\rho({\bf r}) = 3n({\bf r})/(4\pi r_{\rm c}^{3})$, with $n({\bf r})$ the number of monomers at a distance 
$d \le r_{\rm c}$ from the position ${\bf r}$ of the cell center. The coarse-grained density
at ${\bf r}$ is defined as a weighted average of the local density over the surrounding cells,
\begin{equation}
\bar{\rho}({\bf r}) = \frac{1}{8} \left[ 2\rho({\bf r}) +\sum_{\bf k}\rho({\bf r} + \delta{\bf k}) \right]
\end{equation}
where the sum is performed over the six vectors ${\bf k} \in \{(\pm 1,0,0),(0,\pm 1,0),(0,0,\pm 1)\}$.
The values of the coarse-graining length scales, i.e.,  the grid size $\delta$ and the distance $r_{\rm c}$, are chosen 
in order to get a smooth density field while keeping sufficient spatial resolution in the
representation of the microgel. A good compromise can be found
by using $\delta \sim 0.5$ and $r_{\rm c} \sim 1$.  In what follows we will present results for
$\delta = 0.5$ and for two values  $r_{\rm c} = 1.0$ and $r_{\rm c} = 1.2$.
The density field can be used as a smoothed proxy of the real microgel by defining `filled' and `empty' cells
as those with $\bar{\rho}({\bf r}) > \rho_{\rm min}$ and $\bar{\rho}({\bf r}) \le \rho_{\rm min}$, respectively,
with $\rho_{\rm min}$ a threshold density.

The averaging of the local density over the surrounding cells can be seen as a generalization of the averaging 
procedure in a phase-separating Ising system \cite{Majumder2010,Majumder2011}, where `thermal noise'
is removed by replacing the spin at each lattice site by the majority spin of the surrounding sites.
In continuous systems as our microgels or those of Refs.~\cite{Testard2011,Testard2014} the use of the density field representation
smooths the interface corrugations, and fills the smallest holes in the domains of the real-coordinate structure.
This removes the shortest and narrowest protrusions that cannot be considered as real domains, and prevents that the presence of very small holes
artificially interrupts paths in real dense domains.

\ref{fig:snaps-field-diamond} shows snapshots of the real diamond microgel at different times during the collapse
(top panels), and its respective density field representations (bottom panels) for the parameters $r_{\rm c} = 1.0$
and $\rho_{\rm min} = 0.6$. The beads in top and bottom panels represent the positions of the real monomers and the
centers of the filled cells, respectively. \ref{fig:snaps-field-mic} shows the corresponding real coordinate 
and density field representations for the collapse of a middle-size disordered microgel.
The snapshots of both \ref{fig:snaps-field-diamond,fig:snaps-field-mic} correspond
to $\phi=1.2$, i.e., a bad-solvent state well beyond the VPT-point. The final state reached at late times
is, in both the real-coordinate and density field representations, a fully collapsed dense spherical object (not shown).

Once we have constructed the density fields we can characterize the size of the domains emerging during the coarsening process.
First, we define an `interface cell' as a filled cell at ${\bf r}$ with at least one adjacent empty cell (of coordinates ${\bf r'}$,
so that ${\bf r}-{\bf r'}= \delta{\bf k}$). A `chord' \cite{Testard2011,Testard2014} is defined as a straight path, parallel to one of three
axes $x,y,z$, formed just by filled cells and with two interface cells at its ends (so that an empty cell is found if the straight
path is continued by moving to the adjacent cell of any of the two ends).
The chord length is given by $L = |{\bf r}_1 - {\bf r}_2|$, with ${\bf r}_1 ,{\bf r}_2$
the centers of the two interface cells at the path ends.
To compute the distribution of chord lengths of a given configuration of a microgel at $(\phi, t)$, we sampled all the existing chords
as defined above. To improve statistics 5 random rotations of the former configuration were taken, 
and the whole procedure was repeated over the 5 independent collapse runs of the same microgel at the same $\phi$.

\begin{figure}[t!]
\centering
\includegraphics[width=.48\textwidth]{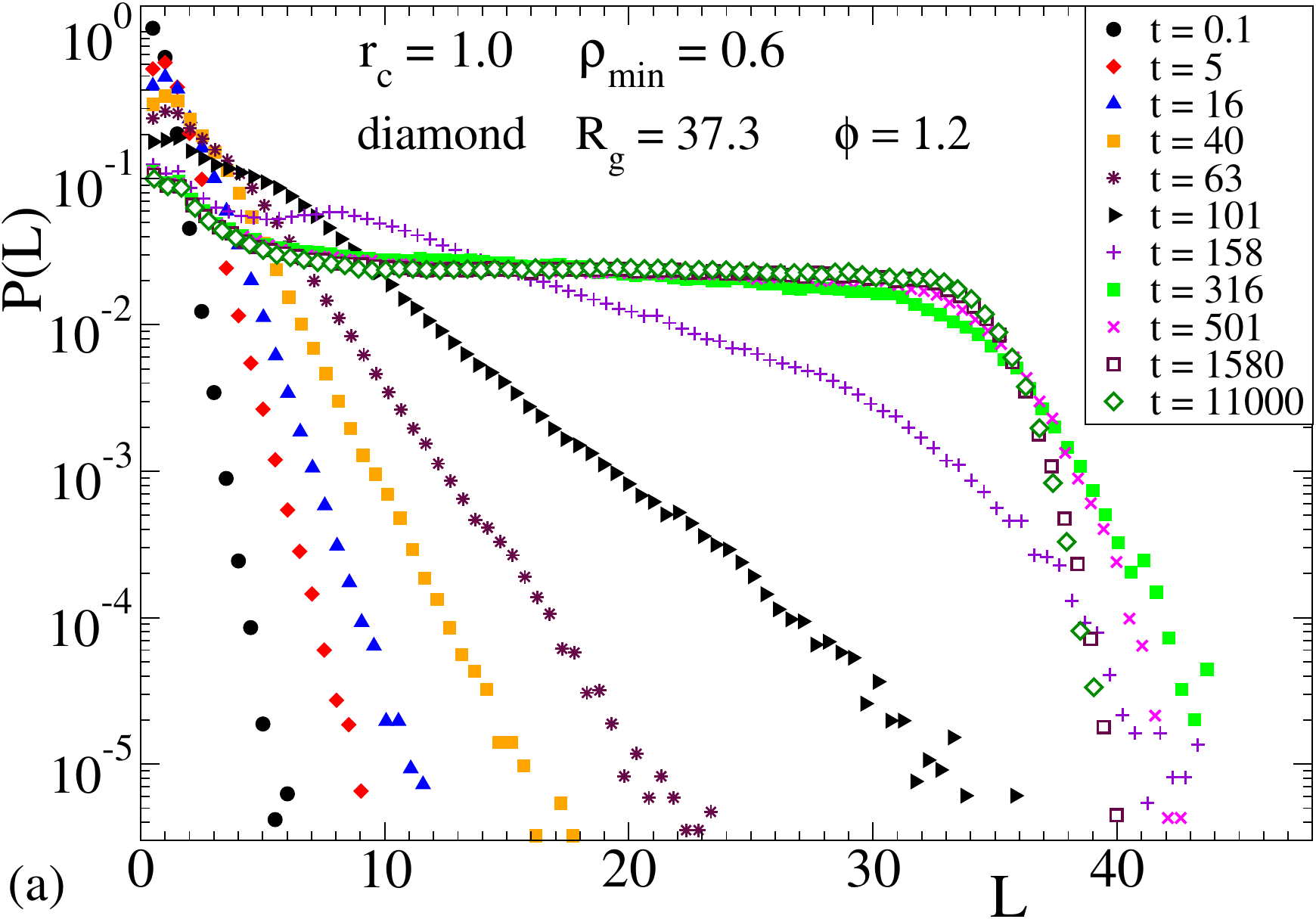}\hspace{0.3cm}
\includegraphics[width=.48\textwidth]{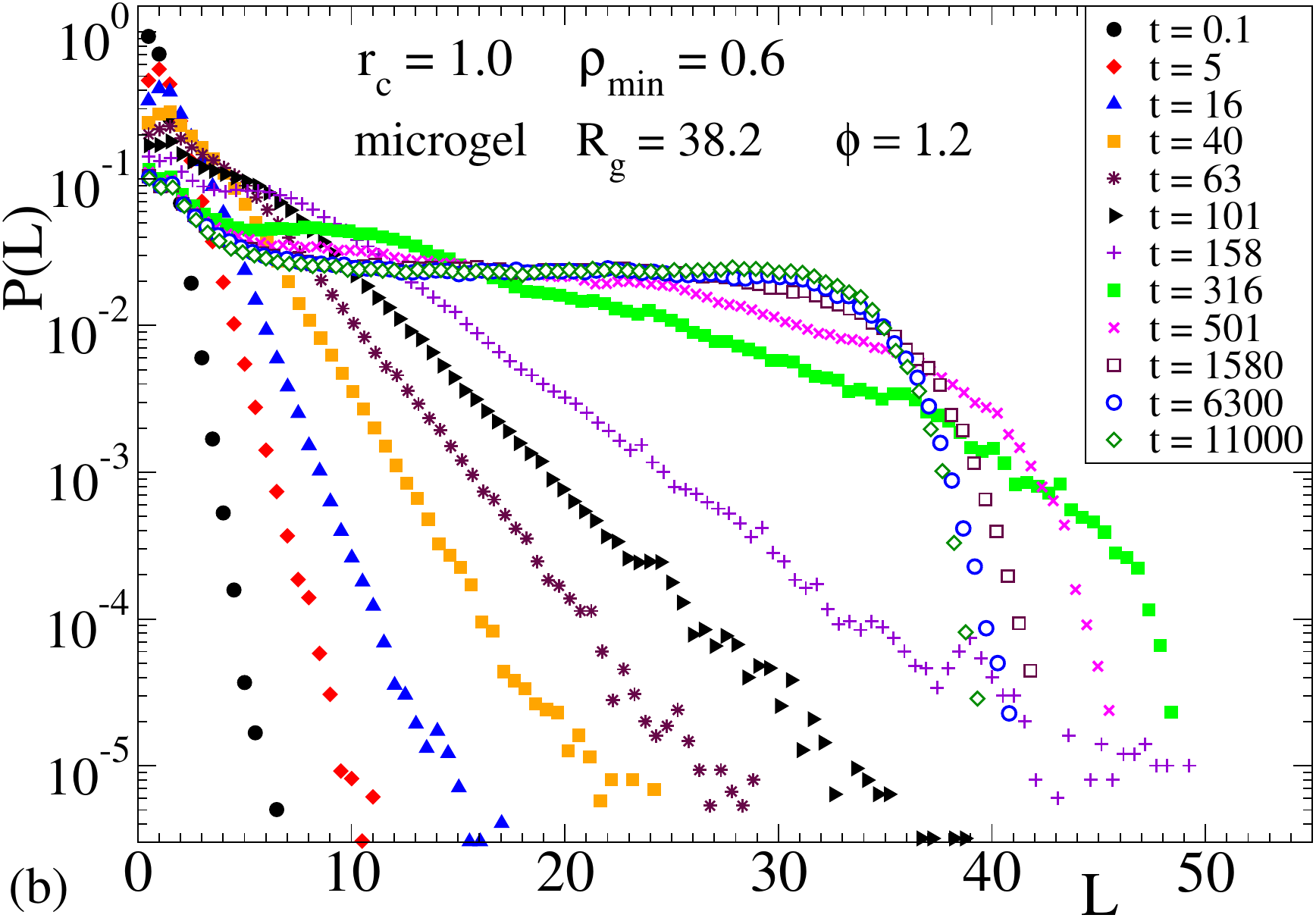}
\caption{Normalized distribution of chord lengths at $\phi=1.2$ and different times during the collapse of the
diamond (a) and a middle-size disordered microgel (b). The chord lengths are calculated for a density field
with $r_{\rm c}=1.0$ and $\rho_{\rm min}=0.6$.}
\label{fig:distlength} 
\end{figure}

\ref{fig:distlength} shows the normalized distributions of chord lengths, $P(L)$, at $\phi=1.2$ and different times during the collapse of the diamond and the middle-size disordered microgel. The displayed results  correspond to the values
$r_{\rm c} = 1.0$ and $\rho_{\rm min} = 0.6$ used for constructing the coarse-grained density field.
In agreement with observations in other coarsening systems \cite{Levitz1998,Atsuko2003,Majumder2011,MajumderEPL2011,Testard2014,Majumder2017}, 
$P(L)$ shows an exponential decay and extends over longer distances
as time increases, as a direct consequence of the growth of the filled domains during coarsening. 
As expected, this behavior saturates at long times when the microgel reaches the fully collapsed state, and a sharp
drop is found when $L$ approaches the maximum path lenght (i.e., the sphere diameter). 
The observed plateau originates from the equiprobable different straight paths that connect two points of the spherical surface, which in the fully collapsed state do not find empty cells.

\begin{figure}[t!]
\centering
\includegraphics[width=.48\textwidth]{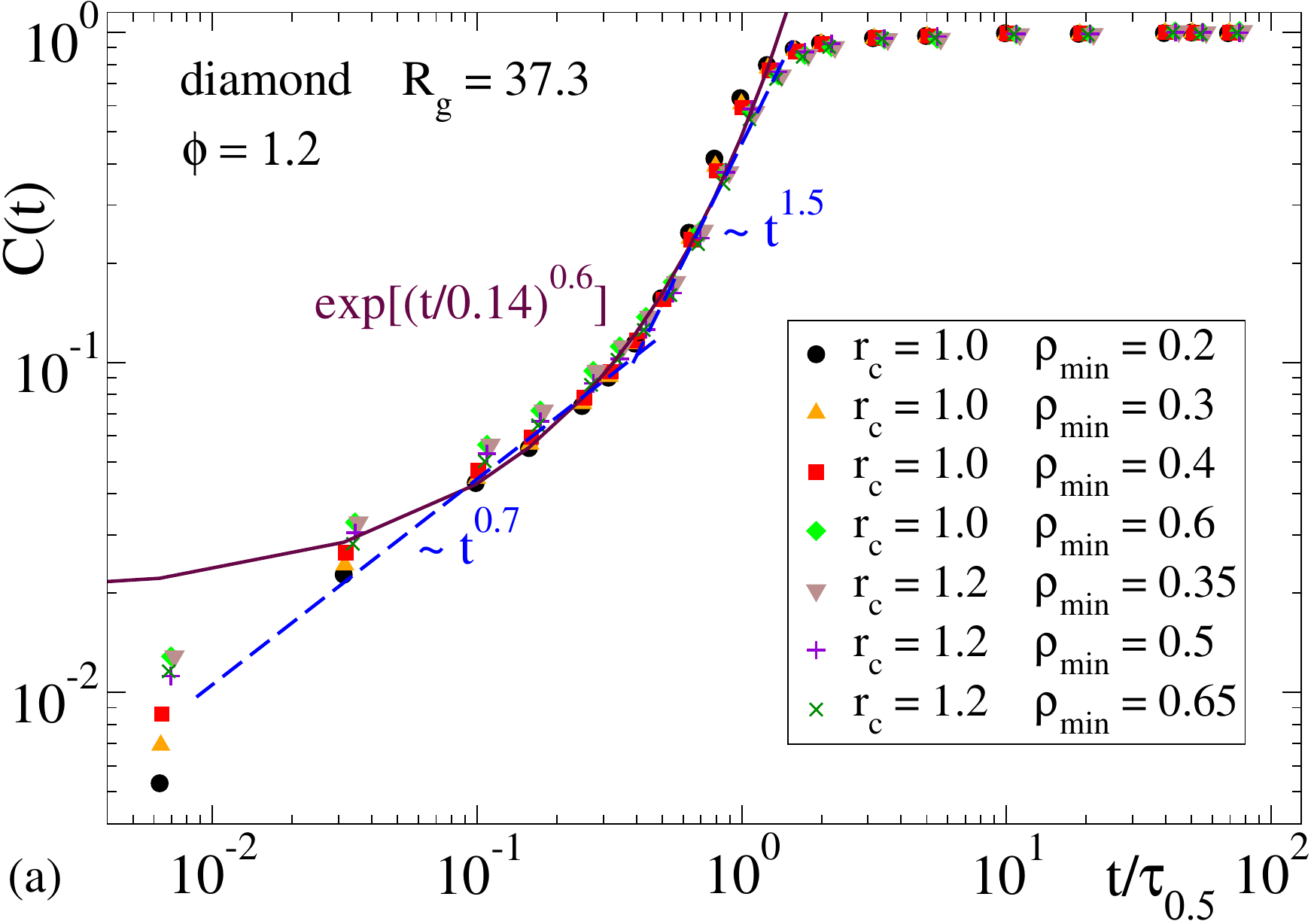}\hspace{0.3cm}
\includegraphics[width=.48\textwidth]{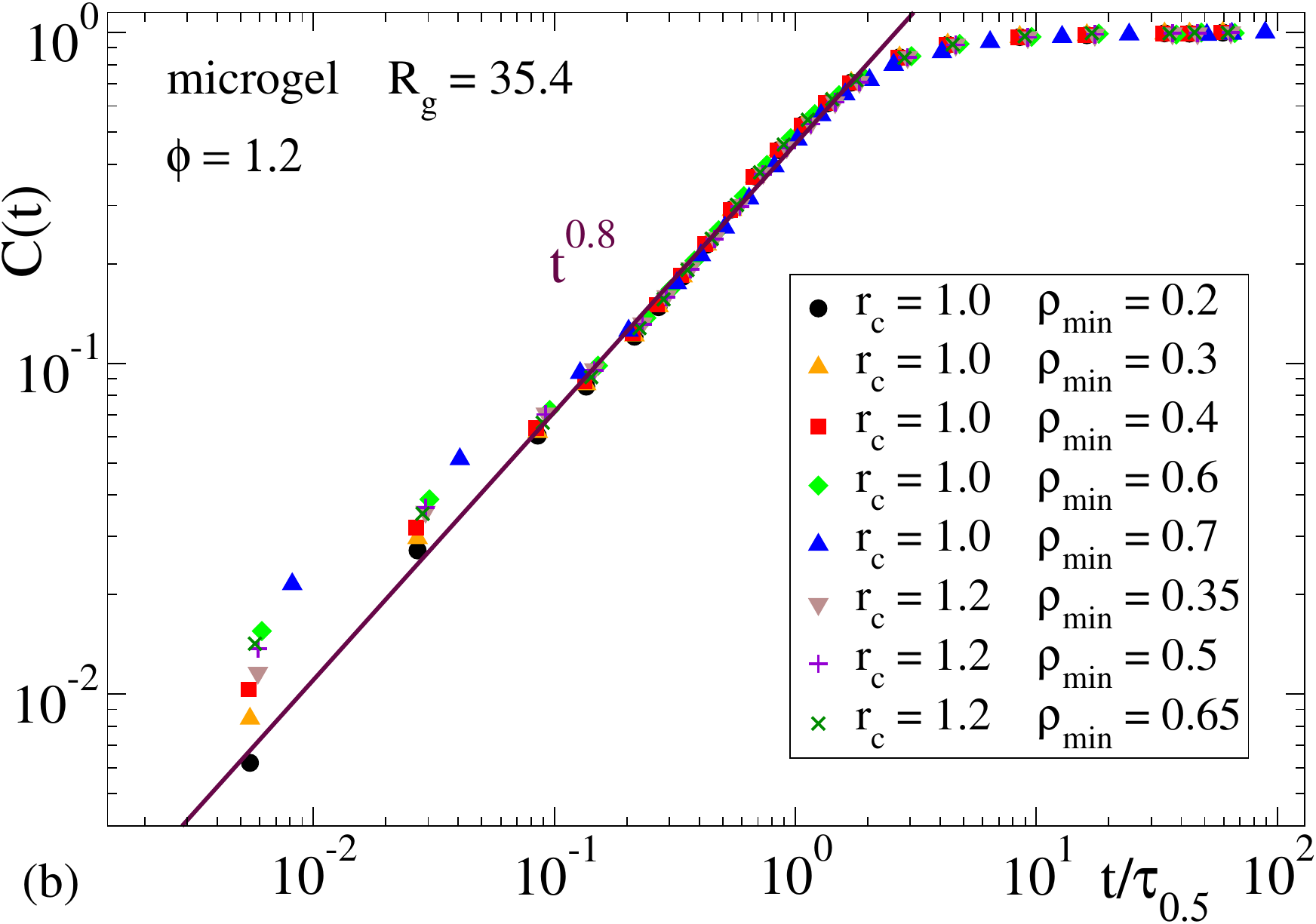}\vspace{0.3cm}
\includegraphics[width=.48\textwidth]{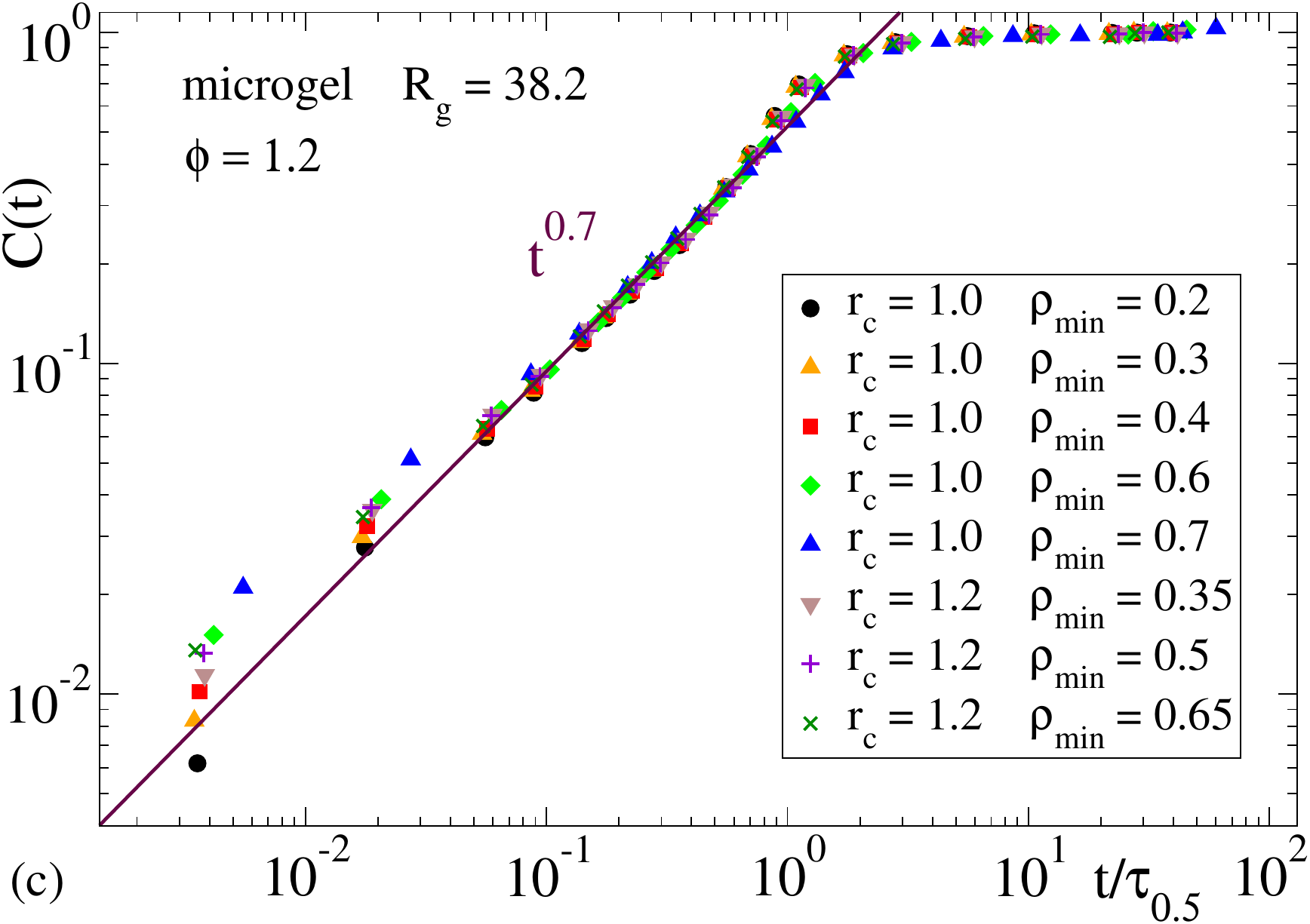}\hspace{0.3cm}
\includegraphics[width=.48\textwidth]{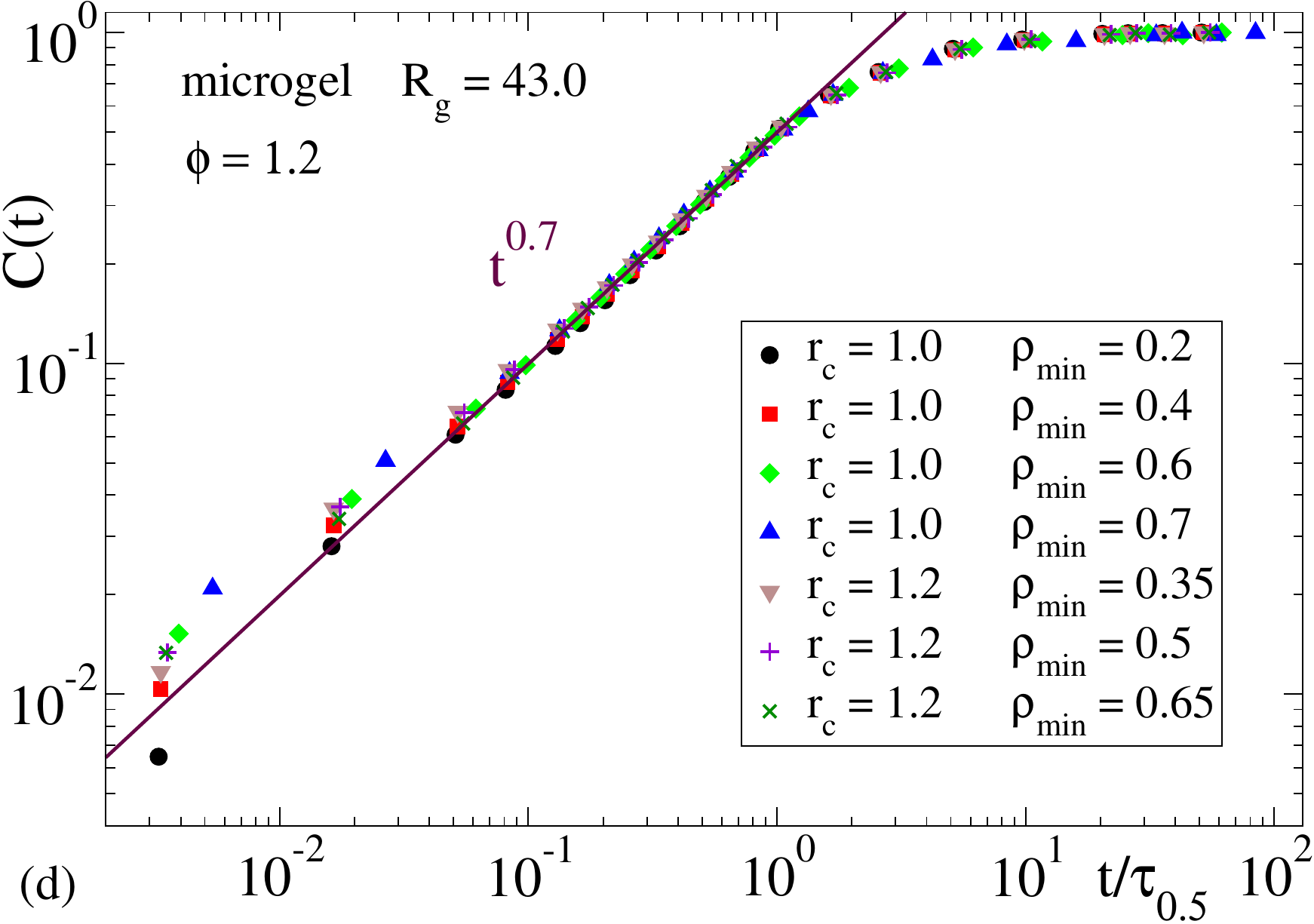}
\caption{Time dependence of the relative domain size $C(t) = (L(t)-L(0))/(L(\infty)-L(0))$
during the collapse at $\phi = 1.2$ of the diamond microgel (a) and three disordered microgels
of small (b), middle (c) and large size (d). Different data sets correspond to different values
of the parameters $r_{\rm c}$ and $\rho_{\rm min}$ used to construct the density field. The time $t$ is normalized
by $\tau_{0.5}$, defined as $C(\tau_{0.5}) = 0.5$. The curve in (a) is a fit to stretched exponential
behavior. Straight lines in all panels are fits to power-law time dependence.}
\label{fig:masterct} 
\end{figure}

\begin{figure}[t!]
\centering
\includegraphics[width=.48\textwidth]{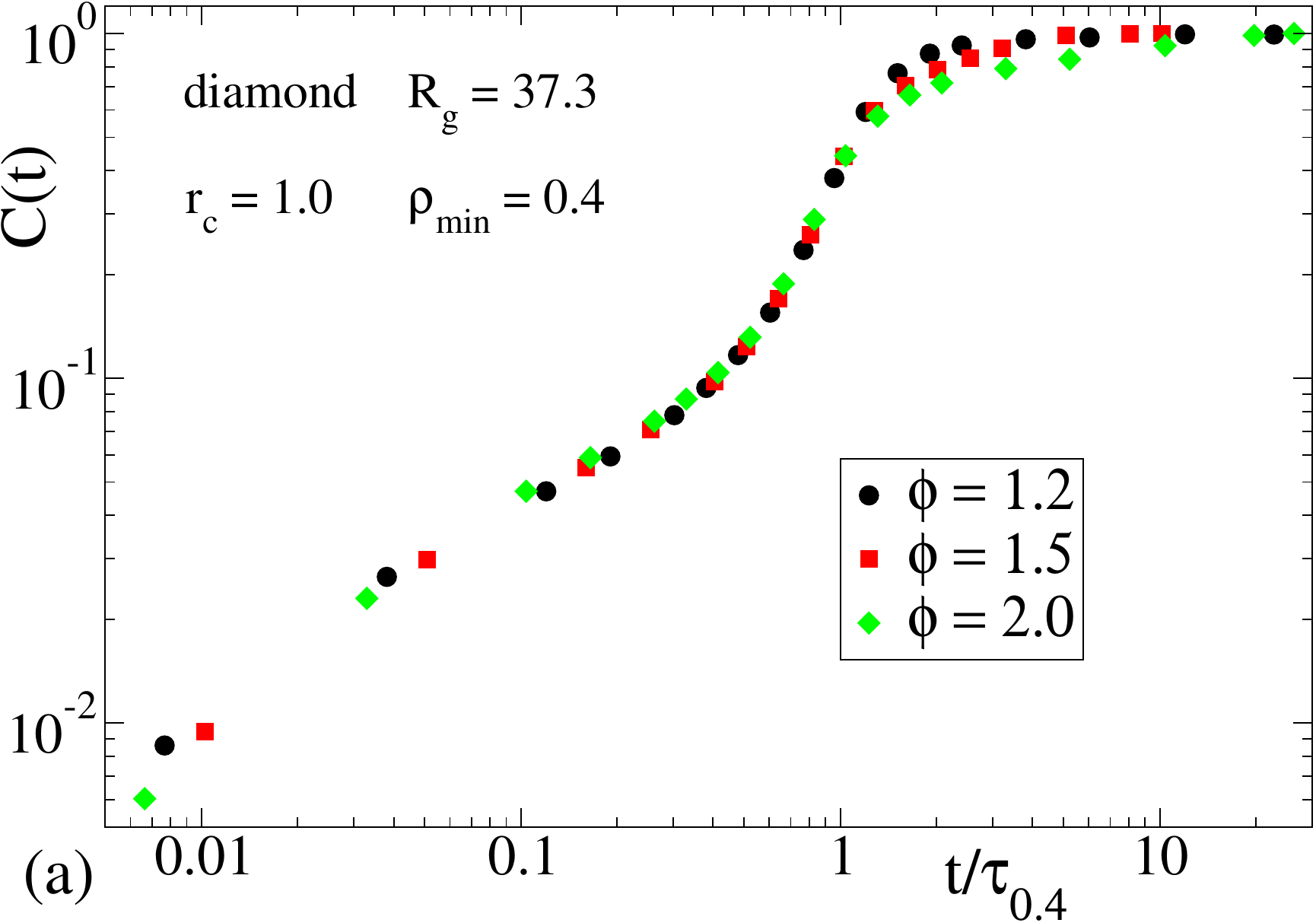}\hspace{0.3cm}
\includegraphics[width=.48\textwidth]{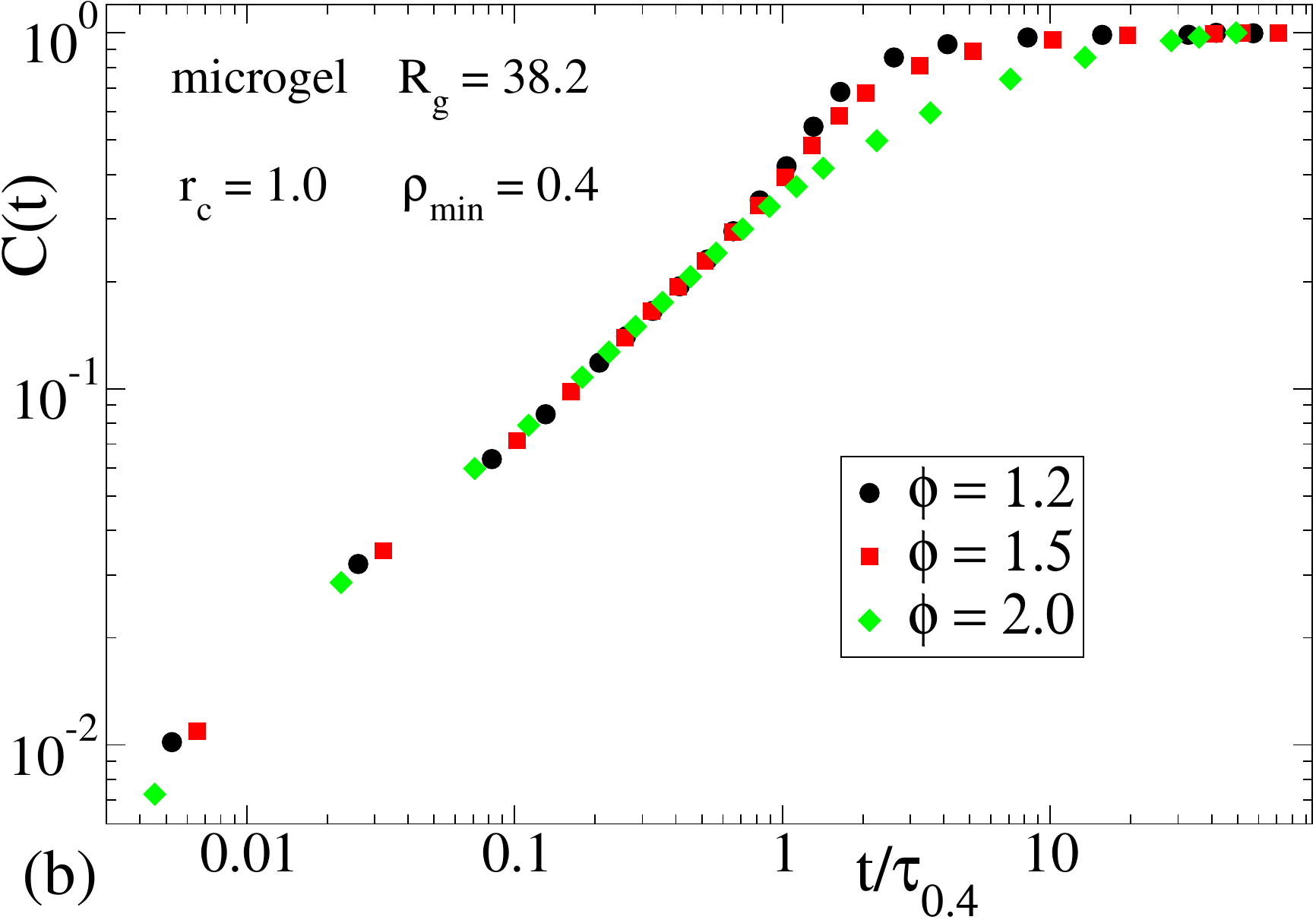}
\caption{Time dependence of the relative domain size $C(t) = (L(t)-L(0))/(L(\infty)-L(0))$
during the collapse at different $\phi$-values of the diamond (a) and a middle-size disordered microgel (b).
The chord lengths have been computed in a density field with $r_{\rm c}=1.0$ 
and $\rho_{\rm min}=0.4$. The time $t$ is normalized
by $\tau_{0.4}$, defined as $C(\tau_{0.4}) = 0.4$}
\label{fig:scalphi} 
\end{figure}

Though the same qualitative behavior of $P(L)$ is found for other choices of $r_{\rm c}$ and $\rho_{\rm min}$,
large quantitative differences are found. Thus, at fixed $r_{\rm c}$ and time, using lower values of $\rho_{\min}$ 
to define the filled cells obviously produces less empty cells and longer chords in the coarse-grained structure. 
On the other hand, using larger values of $r_{\rm c}$ at fixed time increases the local $\rho({\bf r})$, producing
more cells with coarse-grained density higher than the required $\rho_{\rm min}$, and therefore longer chords.
For the same former reasons the mean chord length, obtained as $L = \int L' P(L') dL'$, strongly depends on the specific values of the
parameters $r_{\rm c}, \rho_{\rm min}$
used for constructing the density field. Thus, to quantify the domain growth during the coarsening process
in a parameter-independent fashion we introduce the relative domain  size,
$C(t) = (L(t) -L(0))/(L(\infty)-L(0))$, which grows from 0 to 1 from the swollen ($t=0$) to the collapsed state ($t = \infty$).
\ref{fig:masterct} shows results of $C(t)$ at $\phi=1.2$ computed for different choices of $r_{\rm c}$ and $\rho_{\rm min}$,
for the diamond microgel and for the three selected disordered microgels of small, middle and large size. 
The times have been normalized by the time $\tau_{0.5}$ at which $C = 0.5$. A very good overlap of the data is found even by varying $\rho_{\rm min}$ by a factor 3.5,  demonstrating the consistency of the coarse-graining procedure. The observed master curves 
are also found for the same microgel by varying the solvent quality parameter $\phi$
(see \ref{fig:scalphi}).
Hence, in a good approximation, for each microgel there is a unique function describing the coarsening kinetics of the deswelling transition.
Interestingly, this function is almost independent on the particular topology of the disordered microgel (see \ref{fig:masterct}), and it can be described by a power-law $C(t) \sim t^x$, with $x \approx 0.7$.

The characteristic exponents for the domain growth are inherent to the underlying mechanism driving the coarsening, 
and hence they are system-dependent \cite{Wong1981,Durian1991,SaintJalmes,Barik2009,Lambert2010,Majumder2010,Testard2014,Christiansen2017,Gong2017}.
Thus, for phase-separating solid mixtures the domains are expected to grow as $t^{1/3}$, which is the result of diffusion 
driven by the chemical potential gradient \cite{Lifshitz1961,Bray2002}. For systems where hydrodynamics is relevant (as fluids and polymers), 
linear growth ($\sim t$) is predicted at long times
in the viscous hydrodynamic regime \cite{Siggia1979}. 
%A final inertial regime ($\sim t^{2/3}$) is predicted after the hydrodynamic regime,
%though this is difficult to be accessed because of the characteristic large scales for its onset \cite{Furukawa1985}.
A clear emergence of both the early diffusive ($\sim t^{1/3}$) and the late hydrodynamic ($\sim t$) regime \cite{Wong1981}
is often absent. Thus, an apparent sublinear regime $t^x$ with exponent $x \sim 0.5$ has been observed, e.g.,
in simulations of gas-liquid separation \cite{Testard2014}, which has been proposed to be an effective interpolation between the diffusive and hydrodynamic regimes. 
On the other hand a clear linear growth, $x=1$, has been found for the coil-to-globule transition in off-lattice 
linear polymer chains \cite{Majumder2015EPL,Majumder2017}.
The exponent found in our disordered microgels, $x \sim 0.7$, 
%(only fortuitously similar to that predicted for the inertial regime),
can be tentatively seen as an intermediate result between the former cases, namely faster than the $t^{0.5}$-growth for gas-liquid separation 
and slower than the linear growth for single polymer chain collapse.
Thus, the presence of thin branches connecting the nucleating centers accelerates coarsening with respect to gas-liquid separation.
However, the absence of a common backbone for `reptation' of the nucleating centers prevents them to merge coherently, leading to slower coarsening
than in the single chain case.

\begin{figure}[t!]
\centering
\includegraphics[width=.48\textwidth]{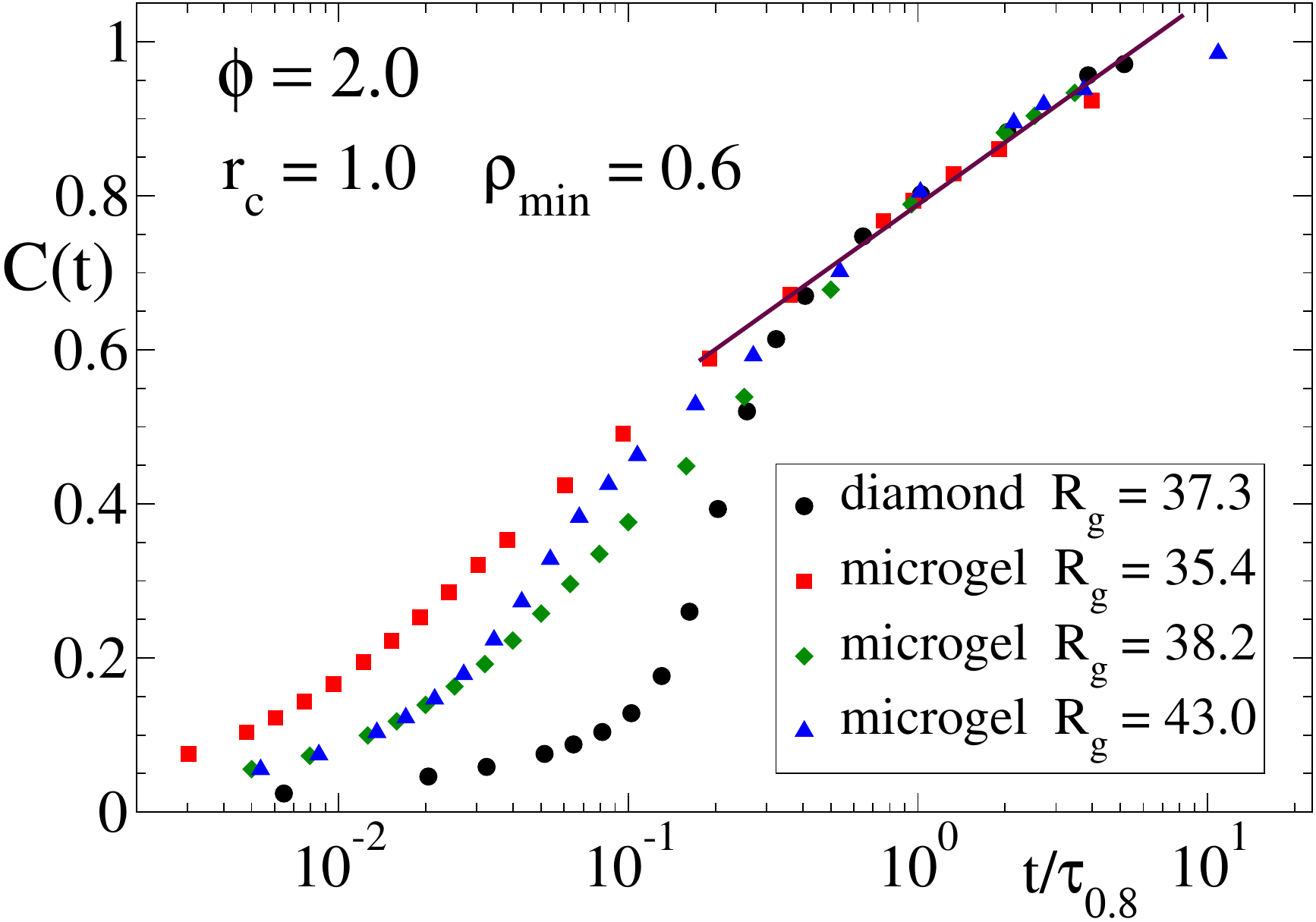}
\caption{Time dependence of the relative domain size $C(t) = (L(t)-L(0))/(L(\infty)-L(0))$
during the collapse at $\phi=2.0$ for the diamond and the small, middle and large disordered microgels.
The chord lengths have been computed in a density field with $r_{\rm c}=1.0$ 
and $\rho_{\rm min}=0.6$. The time $t$ is normalized
by $\tau_{0.8}$, defined as $C(\tau_{0.8}) = 0.8$. The data are displayed in log-lin scale to highlight
the apparent logarithmic domain growth at late times (solid line).}
\label{fig:logar} 
\end{figure}

The coarsening kinetics of the diamond microgel is rather different from that of the disordered microgels. 
Instead of a power law, a stretched exponential with exponent $\beta \approx 0.6$ provides a good description over the whole growth process 
(solid curve in \ref{fig:masterct}a).
The apparent stretched exponential growth of the domains in the diamond microgel
can be tentatively seen as the result of two processes (dashed lines in \ref{fig:masterct}a). 
Thus, a close inspection of the data shows that the initial growth  is still
roughly compatible with the power-law $C(t) \sim t^{0.7}$ observed in the whole time window for the disordered microgels.
At late times an acceleration of the deswelling kinetics is found (roughly scaling as $\sim t^{1.5}$). This unusual feature can
be understood as follows. At early times the domains 
are still small and do not feel the large-scale (regular or disordered) connectivity of the network.
However, unlike in the disordered case, the diamond network facilitates the formation of an homogeneous spatial distribution of nucleation centers,
which grow up in a similar fashion. As a consequence, all them merge together in a narrow time window, leading to the late acceleration
of the collapse kinetics.

Worsening the solvent quality (increasing the strength of the attractive tail $\phi$ in \ref{eq:vnb}) leads to 
slow monomer dynamics, and a glass transition is expected in the globular state for very large values of $\phi$.
Though the interplay of the glass transition and the domain growth is beyond the scope of this work, it is worth
noting that some effects are visible for the deepest investigated quench, at $\phi=2.0$. Indeed, a late slowing down
of $C(t)$ is observed for $\phi = 2.0$ in \ref{fig:scalphi}. To highlight this feature,
we show the function $C(t)$ in log-lin scale for the diamond and the small, middle and large disordered microgels
during their deswelling at $\phi=2.0$ (\ref{fig:logar}, data for $r_{\rm c}=1.0$ 
and $\rho_{\rm min}=0.6$). The late stage of the microgel collapse is apparently consistent with logarithmic domain growth,
as has been observed for the liquid-gas phase separation in the presence of a glass transition \cite{Testard2011,Testard2014}.

\begin{figure}[t!]
\centering
\includegraphics[width=.48\textwidth]{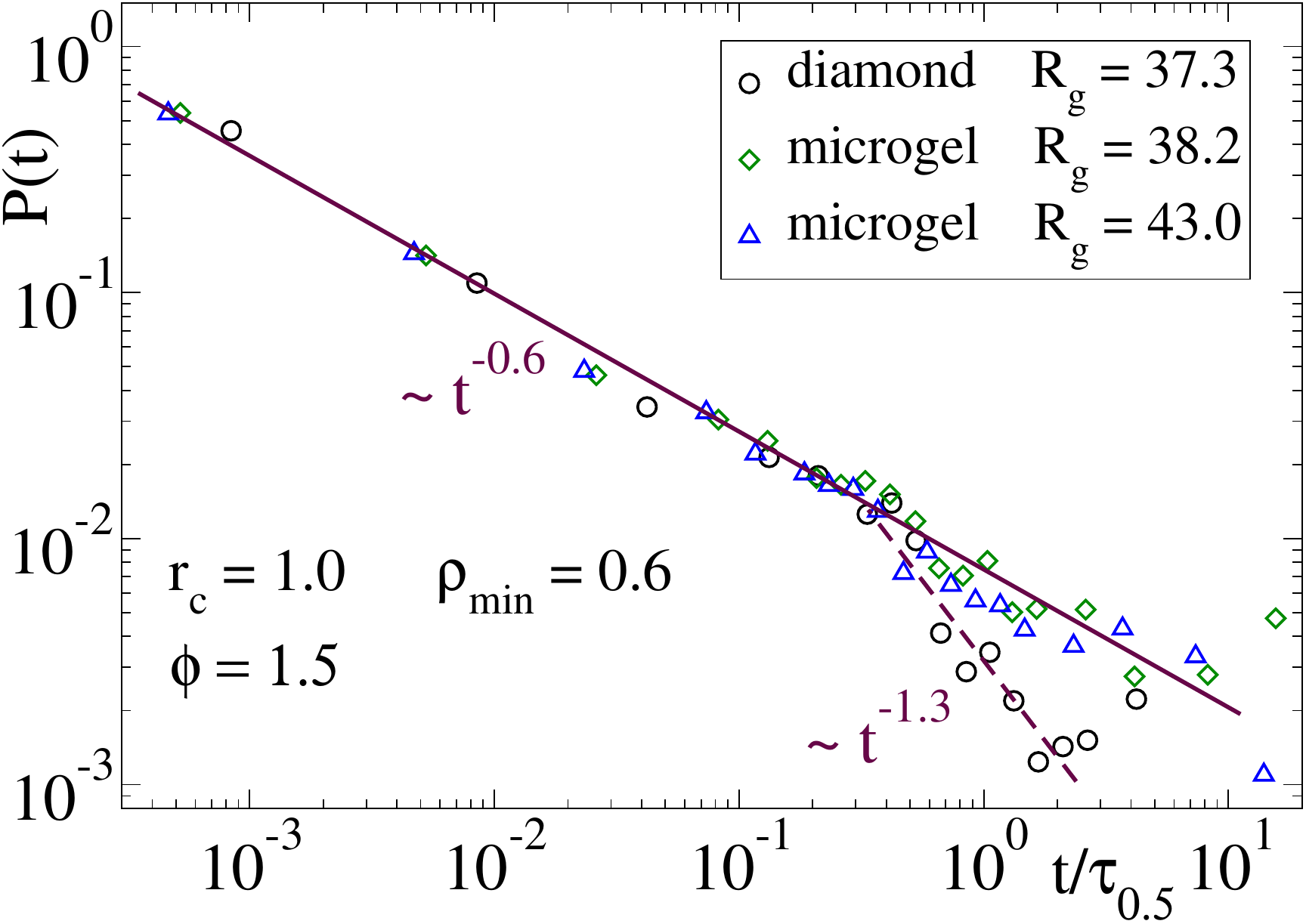}
\caption{Domain self-correlation function $P(t)$ for the diamond and two disordered microgels
during the collapse at $\phi=1.5$. The results displayed correspond to the parameters $r_{\rm c}=1.0$
and $\rho_{\rm min}=0.6$ defining the density field. In each data set the time has been rescaled by $\tau_{0.5}$,
with $\tau_{0.5}$ defined as in \ref{fig:masterct}. Lines are fits to power-law dependence.}
\label{fig:correldom} 
\end{figure}

\begin{figure}[t!]
\centering
\includegraphics[width=.48\textwidth]{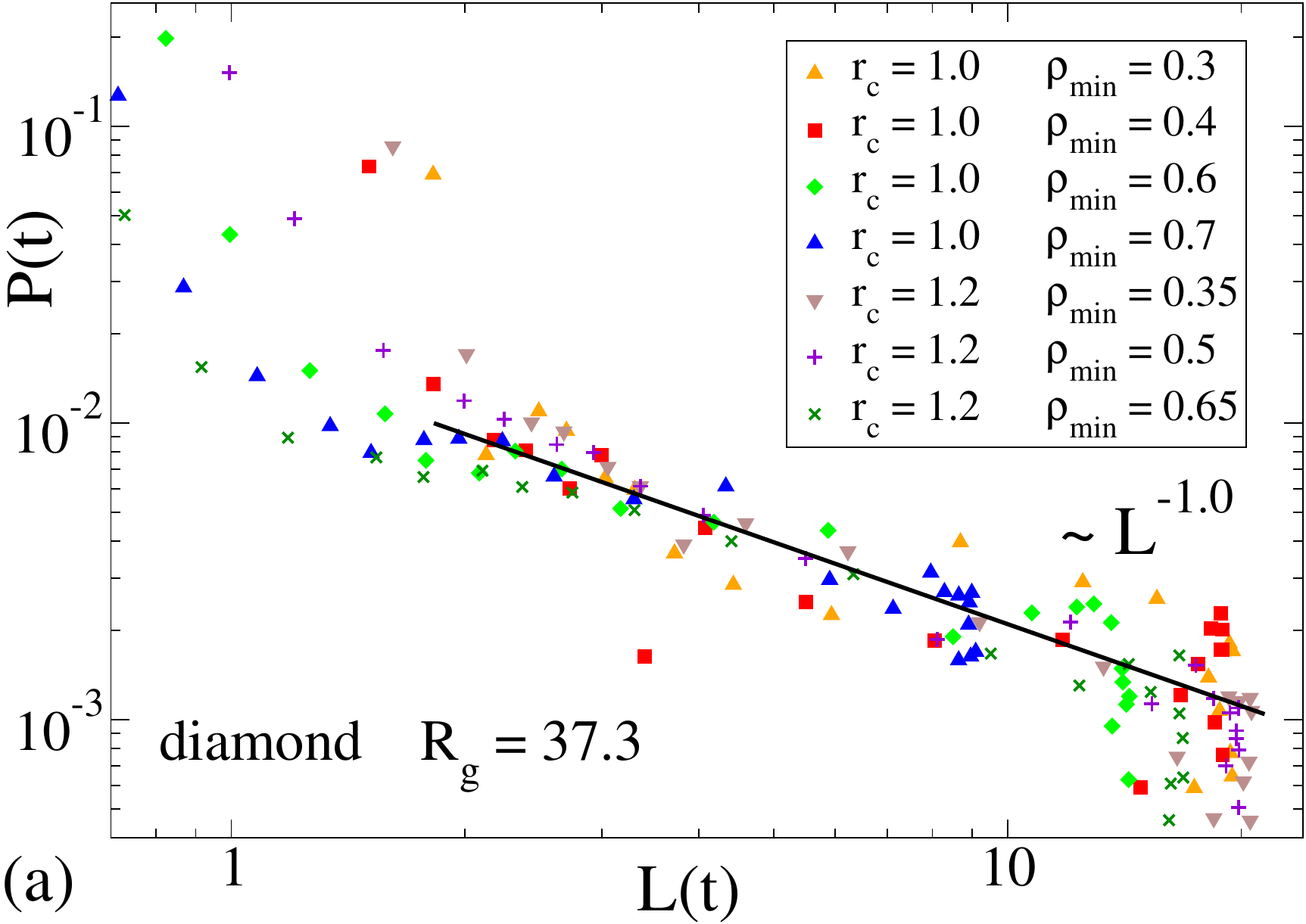}\hspace{0.3cm}
\includegraphics[width=.48\textwidth]{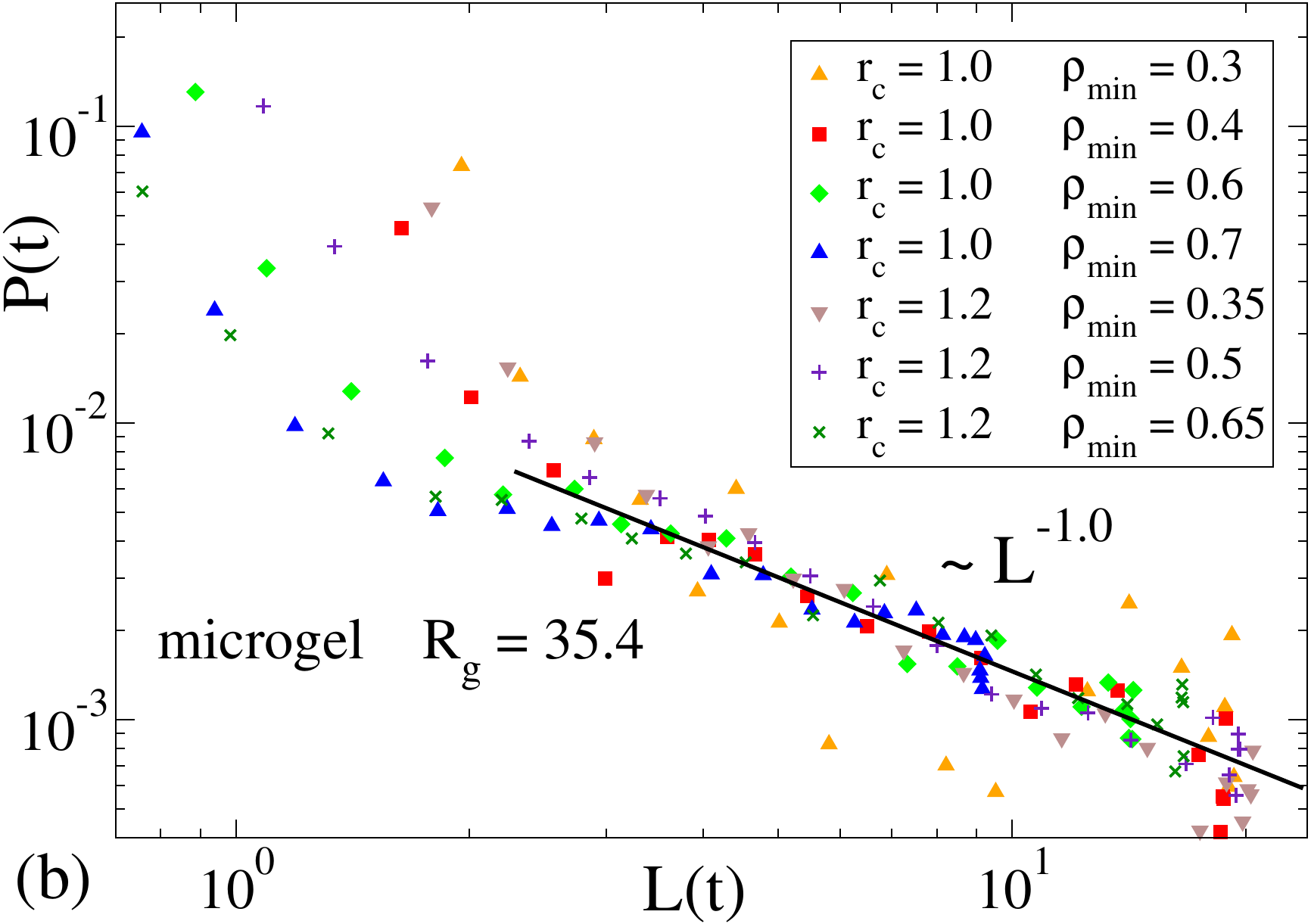}\vspace{0.3cm}
\includegraphics[width=.48\textwidth]{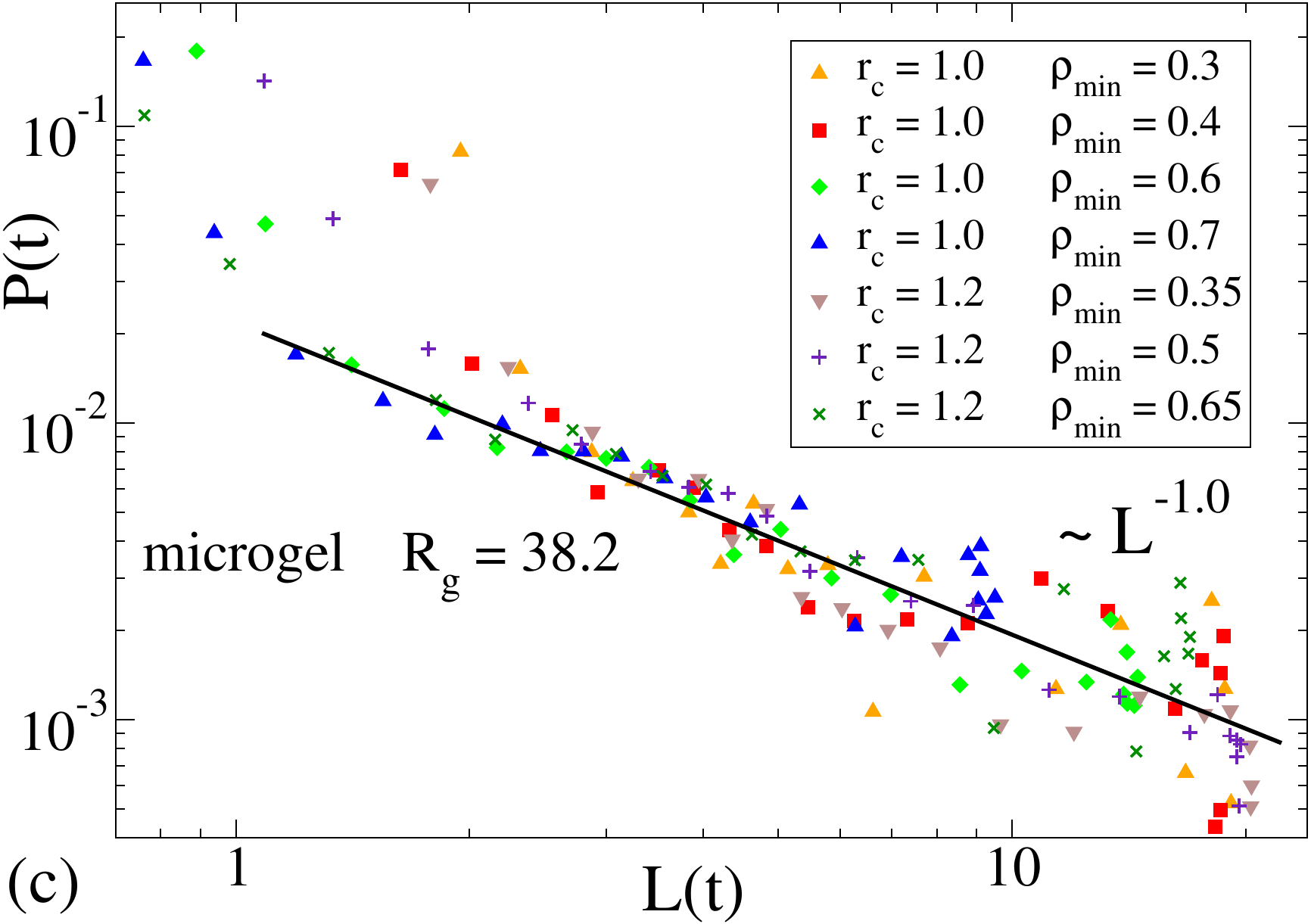}\hspace{0.3cm}
\includegraphics[width=.48\textwidth]{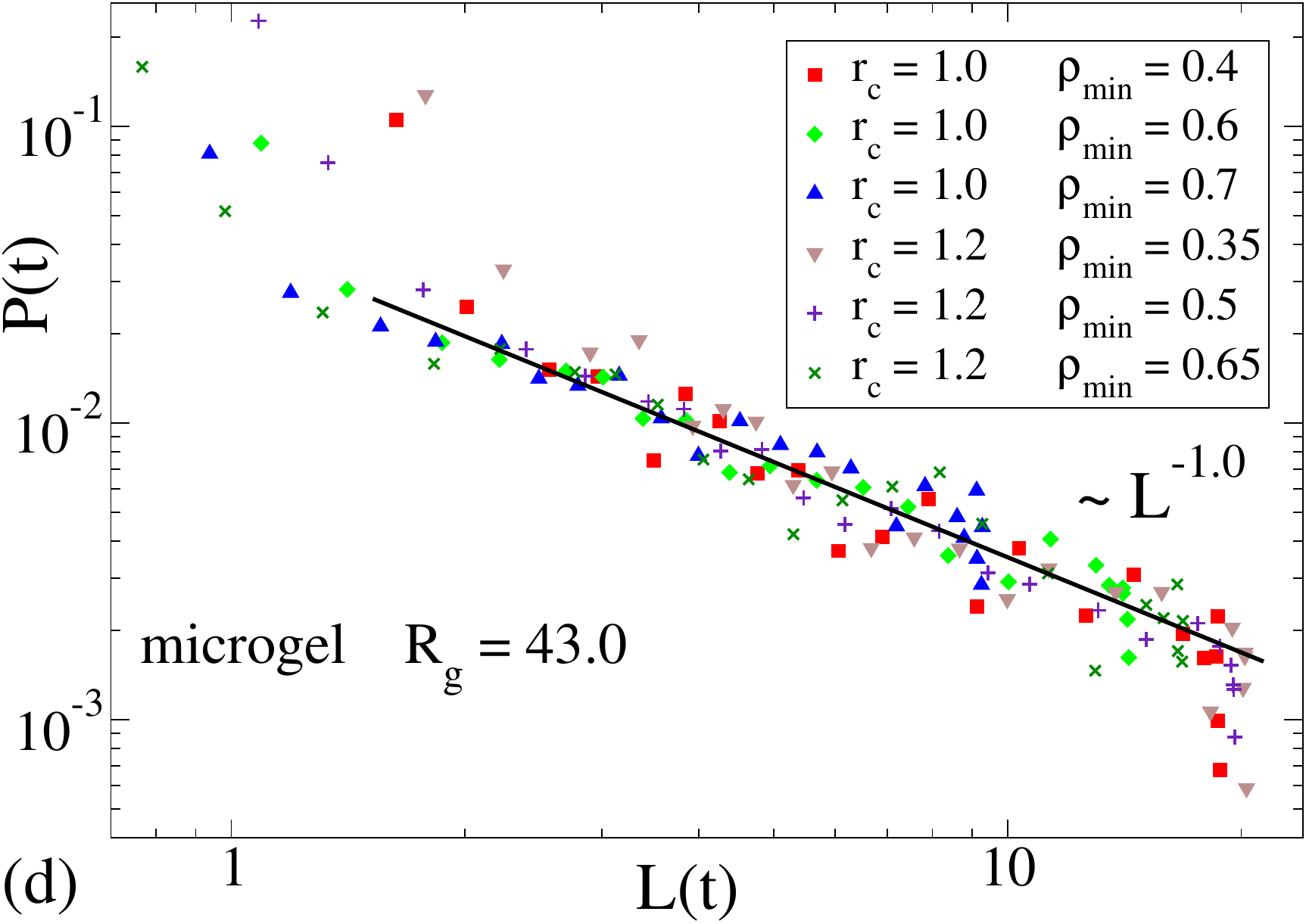}
\caption{Domain self-correlation function $P(t)$ vs. domain size at the time $t$, L(t), 
during the collapse at $\phi=1.2$. Data are shown for the diamond microgel (a) and three disordered microgels
of small (b), middle (c) and large size (d). Different data sets correspond to different values
of the parameters $r_{\rm c}$ and $\rho_{\rm min}$ used to construct the density field. The data sets in each panel 
are shifted vertically to obtain master plots. Lines are fits to power-law dependence.}
\label{fig:mastercorrel} 
\end{figure}

Inspired by the mapping to the Ising system usually performed in other systems,
we test the possibility of scaling behavior of the dynamics 
with the growing length scale of the domains \cite{Midya2015,Majumder2017}.
First, we measure the domain self-correlation function,
$P(t) = \langle S(t)S(0) \rangle -\langle S(t) \rangle \langle S(0) \rangle $,
where the variable $S$ is computed for each monomer of the microgel. $S = 1$  if the position
of the monomer in the grid used for defining the density field $\bar{\rho}$ corresponds to a filled cell.
$S = -1$ if it corresponds to an empty cell. 
\ref{fig:correldom} shows the relaxation of the domain self-correlator for the diamond and two disordered microgels 
during the collapse at $\phi = 1.5$ (with parameters $r_{\rm c}=1.0$ and $\rho_{\rm min}=0.6$ for defining the density field).
The time for each data set has been  rescaled by $\tau_{0.5}$,
with $\tau_{0.5}$ defined as in \ref{fig:masterct}.
The correlator for the disordered microgels can be described by power-law behavior $\sim t^{-0.6}$ over its whole decay.
The same power-law is found for the diamond microgel until $t/\tau_{\rm 0.5} \approx 0.4$. However, a faster relaxation,
roughly described by $\sim t^{-1.3}$, arises at late times. This crossover from
slow to fast relaxation is concomitant with the crossover (also occurring at $t/\tau_{\rm 0.5} \approx 0.4$) 
from slow to fast domain growth in the diamond microgels (\ref{fig:masterct}a).
On the other hand, the single power-law regime observed for the domain growth of the disordered microgels
has its counterpart in the single power-law decay observed in the self-correlator $P(t)$.

\begin{figure}[t!]
\centering
\includegraphics[width=.48\textwidth]{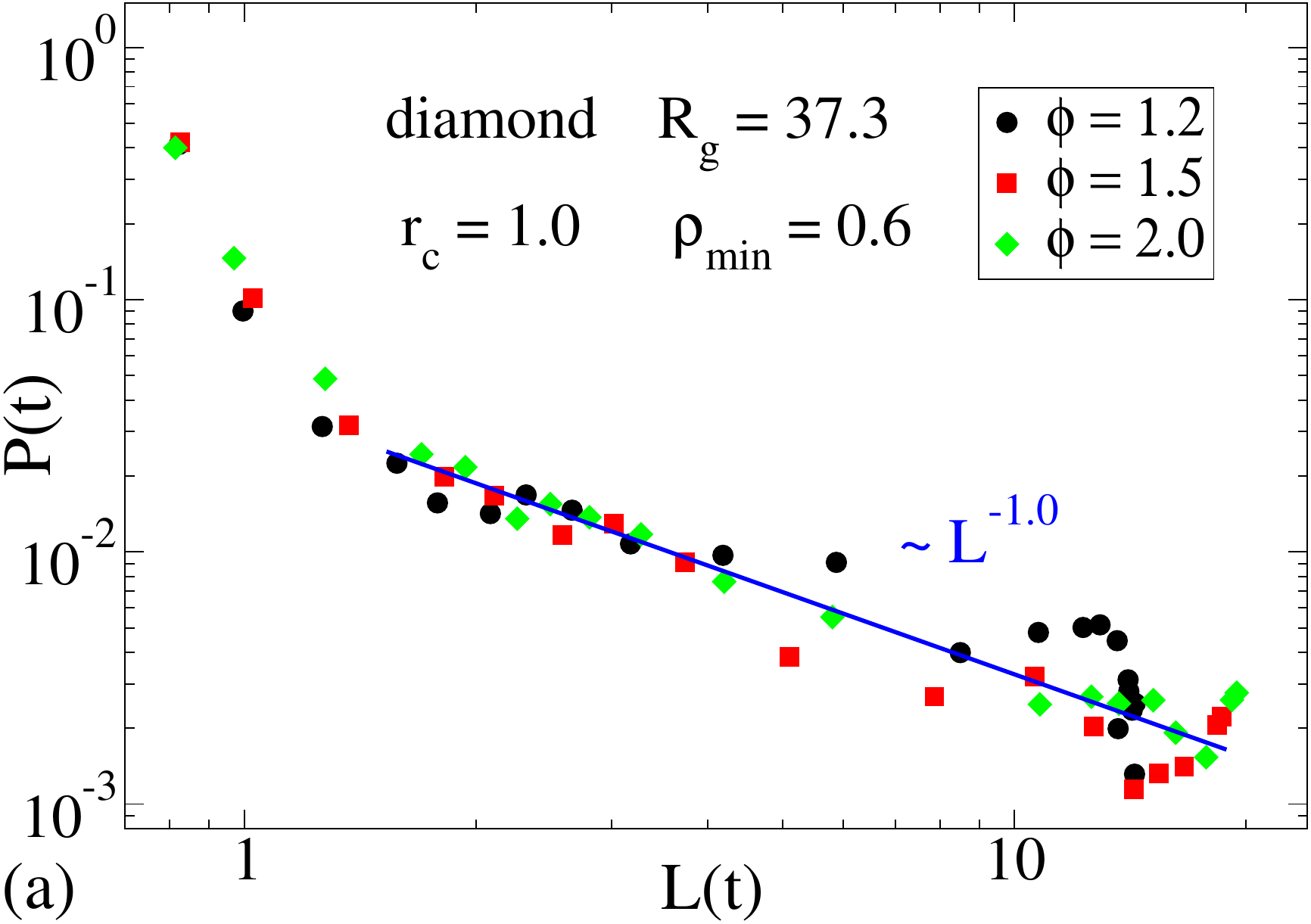}\hspace{0.3cm}
\includegraphics[width=.48\textwidth]{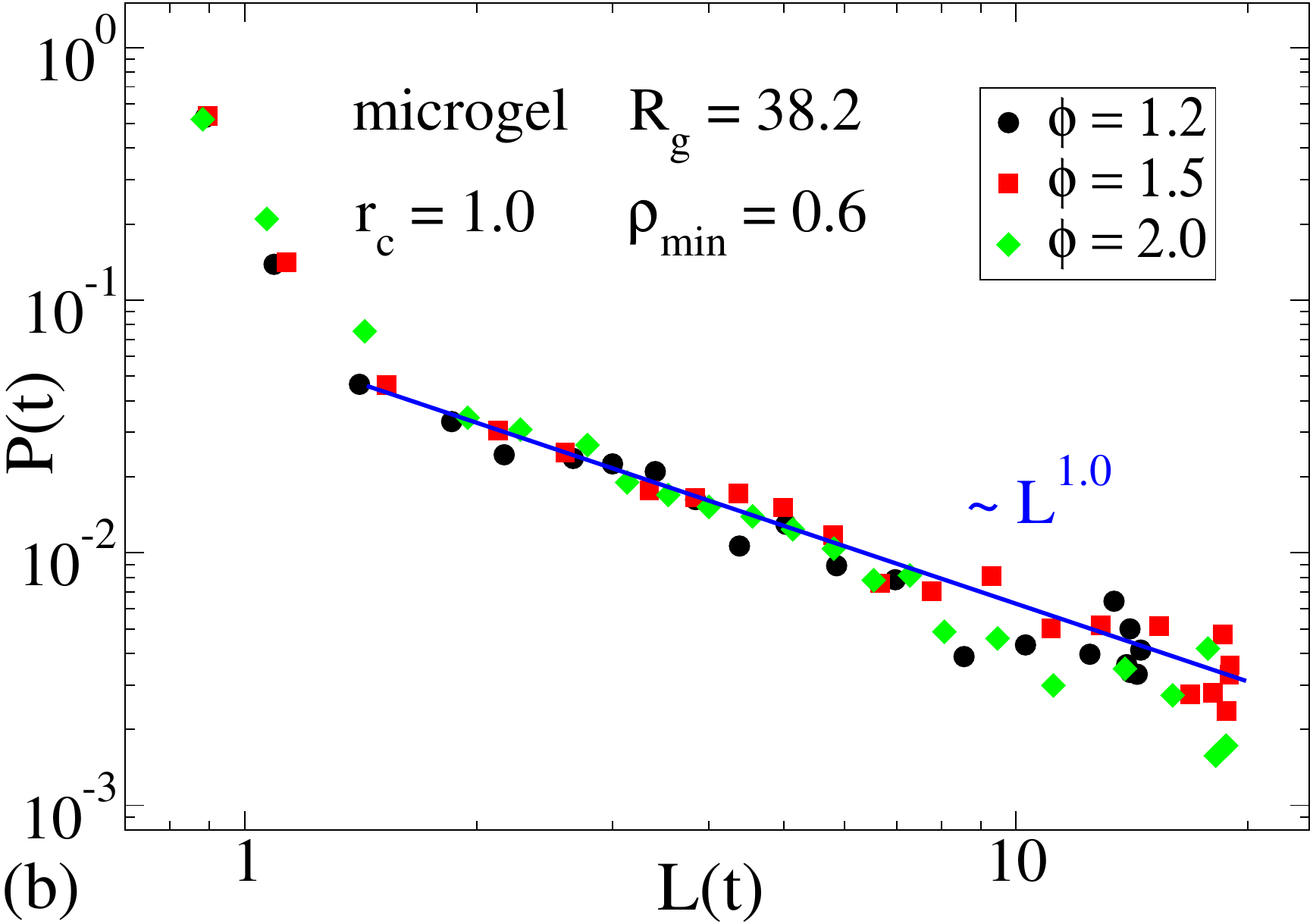}
\caption{Domain self-correlation function $P(t)$ vs. domain size at the time $t$, L(t), 
during the collapse at different $\phi$-values of the diamond (a) and a middle-size disordered microgel (b).
The chord lengths have been computed by using the parameters $r_{\rm c}=1.0$
and $\rho_{\rm min}=0.6$ for defining the density field. }
\label{fig:scalphicorrel} 
\end{figure}

From the computation of $P(t)$ and the domain size $L(t)$ a direct
relation $P(L)$ can be obtained. \ref{fig:mastercorrel} shows results for this relation in the diamond and in the small, middle
and large disordered microgel. The data correspond to the collapse at $\phi =1.2$ and, as in \ref{fig:masterct},
have been computed for a broad range of the parameters $r_{\rm c}$ and $\rho_{\rm min}$ used to define
the density field. The different data sets have been shifted vertically to obtain master plots. Though the results are not
fully conclusive because of the poor statistics, they are compatible (no systematic deviations) with a common power law in each system, irrespective
of the specific values of  $r_{\rm c}$ and $\rho_{\rm min}$. Unlike for the case of the domain growth ($C(t)$), the specific
microstructure does not seem to play a major role in $P(L)$, and both the diamond and disordered microgels exhibit scaling
$P(L) \sim L^{-1.0}$. 
The observed scaling behavior is again independent of the state point, as can be seen
in \ref{fig:scalphicorrel} for different values of $\phi$ at fixed $r_{\rm c} =1.0$ and $\rho_{\rm min}=0.6$.
Thus, the presence of two regimes in the $C(t)$ and $P(t)$ of the diamond microgels (as opposite to the single regime in
the disordered microgels) does not result in significantly different scaling of $P(L)$. Indeed, up to 
$t/\tau_{\rm 0.5} \approx 0.4$ both $C(t)$ and $P(t)$ behave in a similar fashion to their counterparts in the disordered microgels,
and the accelaration of the diamond domain growth at $t/\tau_{\rm 0.5} > 0.4$ is compensated by a similar accelaration 
(with a very similar exponent, see \ref{fig:masterct}a and \ref{fig:correldom}) of the relaxation of $P(t)$.
Interestingly, the recent analysis of the coil-to-globule transition of a linear polymer chain \cite{Majumder2017,Christiansen2017,Majumder2018}
has revealed a similar scaling ($P(L) \sim L^{-1.2}$)  to that of the microgel deswelling, 
in spite of the clear differences between the respective kinetics.

\section{Conclusions}\label{sec5}

We have presented realistic computer simulations of the synthesis of microgels through cross-linking
of functionalized polymer chains confined in a cavity. 
This model is inspired by the microfluidic fabrication of microgels 
from polymeric precursors \cite{Tumarkin2009,Weitz2010,Seiffert2010,Seiffert2011}, 
and is different from the usual polymerization routes that 
have inspired very recent simulation works \cite{Gnan2017,Rovigatti2018}. 
Most of the cross-linking events are of intramolecular character, and the global connectivity of the microgel network 
is mediated by a small fraction of inter-chain cross-links. As expected, the microgels are quite spherical and topologically polydisperse. 
Their scattering form factors can be accounted by the standard fuzzy sphere model usually invoked in real microgels.

We have investigated the effect of the network microstructure on the thermodynamics and kinetics of the deswelling transition. For this purpose, we have compared results in three topologically disordered microgels and a regular diamond-like microgel network, all them
with the same molecular weight and the same effective degree of cross-linking. 
Whereas the specific microstructure has no apparent effect
on the locus of the volume phase transition, it strongly affects the kinetics of deswelling. Following the methodology proposed in
a different context \cite{Testard2014}, we have characterized the coarsening kinetics 
through the analysis of the growing domains of a smooth density field representation of the microgel. 
This representation removes the fast interfacial fluctuations of the real-coordinate structure, and allows for a consistent
characterization of the domain growth, with no significant dependence on the specific parameters used to define the density field.
The internal structure of the microgel strongly affects the coarsening process during deswelling. Thus, the domain size grows in a stretched exponential fashion in the diamond network, whereas power-law behavior $\sim t ^x$ is found in the disordered
microgels, in all cases showing very similar exponents $x \approx 0.7$. The apparent stretched exponential growth of the domains in the diamond microgel
can be seen as the result of two processes: i) an early regime analogous to that of the disordered microgels (and therefore independent of the microstructure),
and ii) a late regime where all domains fastly merge into the final collapsed globule. This specific behavior is unusual
and is tentatively related to the regular structure of the diamond network, which leads to the formation of an homogeneous spatial distribution of nucleation centers that grow in a similar fashion. As a consequence, all them merge together in a narrow time window.

Following a similar procedure to studies of coarsening in other systems, as Ising systems, phase-separating mixtures or collapsing linear polymers \cite{Majumder2010,Midya2015,Majumder2017}, we have invoked the Ising model to establish a scaling relation of the domain dynamic correlations with the growing length scale $L$. In this case the specific microstructure (regular or disordered) has no significant effect on the scaling behavior. The faster domain growth in the late stage of the coarsening of the
diamond microgels is compensated by a similarly faster late relaxation of the dynamic self-correlations.
A universal power-law $\sim L^{-y}$ is observed, with $y \approx 1$, irrespective of the microstructure.
Though we do not expect qualitative differences in the observed coarsening scenario, quantitative differences in the scaling exponents might arise in the presence of hydrodynamic interactions, which have been neglected in this investigation. 
Work in this direction is in progress.

Finally, our results contribute to highlight the importance of developing realistic 
simulation models for microgels that incorporate the synthesis procedure, an aspect that is often skipped 
in the simulation literature by assuming regular networks. 
The analysis presented in our work reveals that regular networks can be inadequate for investigating some dynamic aspects of microgels as the deswelling kinetics. On the other hand, we believe that our characterization of the coarsening process
during the microgel collapse will motivate future experimental works by super-resolution microscopies, which are currently experiencing
noteworthy advances. 

\begin{acknowledgement}
We acknowledge financial support from the projects 
MAT2015-63704-P (MINECO-Spain and FEDER-UE) and IT-654-13 (Basque Government, Spain).
Computation/simulation for the work described in this paper was supported in part by the DeiC National HPC Centre, SDU (ABACUS 2.0).
We thank L. Rovigatti, E. Zaccarelli, D. Truzzolillo, J.A. Pomposo, E. Minina, S. Kantorovich, R. Blaak and C.N. Likos for useful discussions.
\end{acknowledgement}

%\bibliographystyle{achemso}
%\bibliography{microgels}
%\bibliography{microgels}
%\bibliography{biblioMorenoLoVerso}

\providecommand{\latin}[1]{#1}
\providecommand*\mcitethebibliography{\thebibliography}
\csname @ifundefined\endcsname{endmcitethebibliography}
  {\let\endmcitethebibliography\endthebibliography}{}

%\newpage

%\begin{center}
%FOR TABLE OF CONTENTS USE ONLY\\
%\vspace{1 cm}
%Computational Investigation of Microgels: \\  Synthesis and Effect of the Microstructure on the Deswelling Behavior\\
%\vspace{1 cm}
%Angel J. Moreno and Federica Lo Verso
%\end{center}
%\vspace{2 cm}
%\begin{figure}
%\begin{center}
%\includegraphics[width=.75\textwidth]{TOC-eps-converted-to.pdf}
%\end{center}
%\label{fig:TOC}
%\end{figure}

\end{document}